\documentclass[3p,compress]{elsarticle}

\usepackage[colorlinks]{hyperref}
\usepackage{amsmath,amssymb}
\usepackage{upgreek}

\usepackage{xcolor}

\journal{Progress in Particle and Nuclear Physics}
\usepackage[modulo,mathlines]{lineno}

\begin{document}

% increase tabular spacing 
\renewcommand{\arraystretch}{1.2}

\begin{frontmatter}

\title{Rare $B$ Decays \\ as \\ Tests of the Standard Model}

\author{Thomas Blake}
\address{Department of Physics, University of Warwick, Coventry, CV4 7AL, United
Kingdom}

\author{Gaia Lanfranchi}
\address{Laboratori Nazionali di Frascati  - INFN \\
via E. Fermi 40 - 00044 Frascati (Rome) Italy}

\author{David M. Straub}
\address{Excellence Cluster Universe, TUM, Boltzmannstr.~2, 
85748~Garching, Germany}

\begin{abstract}
One of the most interesting puzzles in particle physics today is that new physics 
is expected at the TeV energy scale to solve the hierarchy problem, and stabilise the Higgs mass, but so far
no unambiguous signal of new physics has been found. 
Strong constraints on the energy scale of new physics can be derived from precision tests of the electroweak theory
and from flavour-changing or $C\!P$-violating processes in strange, charm and beauty hadron decays.
Decays that proceed via flavour-changing-neutral-current processes are forbidden at the lowest perturbative order in the Standard Model and are, therefore, {\it rare}.
Rare $b$ hadron decays are playing a central role in the understanding of the underlying patterns of Standard Model physics and 
in setting up new directions in model building for new physics contributions. 
In this article the status and prospects of this field are reviewed.
\end{abstract}


\end{frontmatter}

\clearpage

\tableofcontents

\clearpage

\section{Introduction}
\label{sec:introduction}

In the Standard Model (SM) of particle physics the only flavour-violating interaction is the weak charged current.
The other fundamental interactions -- electromagnetism, the strong nuclear force and the weak neutral current -- are all flavour-conserving.
The probability for changes between the different quark flavours is controlled by the Cabibbo-Kobayashi-Maskawa (CKM) quark mixing matrix~\cite{Cabibbo:1963yz}.
The CKM matrix is almost flavour diagonal with off-diagonal
elements suppressed by powers of $\lambda\simeq 0.22$.

Flavour-changing-neutral-current (FCNC) processes that take a down ($d$, $s$, $b$) or up ($u$, $c$, $t$) type quark and transform it into another 
quark of the same type but of a different flavour are forbidden at 
tree level.
These processes are suppressed by the Glashow-Iliopoulos-Maiani mechanism~\cite{Glashow:1970gm} and must involve at least one off-diagonal element of the CKM matrix.
This in turn makes the processes rare.
In this review we focus primarily on rare decays of $b$ hadrons to final states involving dilepton pairs ($e^+e^-$, $\mu^+\mu^-$ or $\tau^+\tau^-$) and photons.
The short-distance contributions to the $b \to s$ and $b \to d$ transition in the SM come from loop order Feynman diagrams involving the exchange of a virtual $W$ boson.
In many extensions of the SM, new TeV scale particles can enter in competing diagrams and lead to measurable effects in the rate or other properties of the $b \to q$ decay.
A wide range of different observables are considered throughout this review.

The experimental study of rare $b$ hadron decays began with the observation of radiative $b \to s \gamma$ transitions by the CLEO experiment in 1994~\cite{Alam:1994aw} and was continued by the BaBar experiment at SLAC and the Belle experiment at KEK~\cite{Bevan:2014iga}.
By operating at the $\Upsilon(4{\rm S})$ centre-of-mass energy, BaBar and Belle were able to collect large samples of coherently produced $B^0\bar{B}^0$ and $B^+B^-$ meson pairs.
The BaBar experiment completed data taking in 2008, collecting a dataset of 467\,M $B\bar{B}$ pairs.
The Belle experiment completed data taking in 2010, collecting a dataset of 772\,M $B\bar{B}$ pairs.
The combined dataset of BaBar and Belle corresponds to an integrated luminosity of about $1\,{\rm ab}^{-1}$, for an $e^+e^-$ machine operating at the $\Upsilon(4{\rm S})$.

With the data collected during the first period of data taking (Run\,1) at the Large Hadron Collider (LHC) we now have unprecedented samples of rare decay processes due to the large $b\bar{b}$ production cross-section in high-energy $pp$ collisions.
In  $pp$ collisions at $\sqrt{s} = 7\,{\rm TeV}$, the $b\bar{b}$ production cross-section is about  $300\,\upmu{\rm b}$~\cite{Aaij:2010gn}.
This is expected to grow to $\sim 500\,\upmu{\rm b}$ at $\sqrt{s} = 14\,{\rm TeV}$.
This large production cross-section results in more than $10^{11}$ $b$ hadrons being produced in the LHC experiments in a dataset corresponding to $1\,{\rm fb}^{-1}$.
The large centre-of-mass collision energy also produces $B$ mesons with a large boost in the LHC detectors.
This enables the fast oscillation of the $B_{s}^0$--$\bar{B}_{s}^{0}$ system to be resolved and provides a clean experimental signature that can be used to identify the $b$ hadron decays.

There are three experiments at the LHC contributing to measurements of rare $b$ hadron decays; the ATLAS, CMS, and LHCb experiments.
The LHCb experiment is a dedicated experiment for studying the production and decay of $b$ and $c$ hadrons at the LHC.
It occupies the forward region, where the $b$ and $c$ hadron production is largest at the LHC.
The ATLAS and CMS experiments cover the central region of pseudorapidity at large angles to the LHC beam line. They are primarily designed to study massive particles produced with large transverse momentum.

Run\,1 of the LHC took place between 2010 and 2012 at $pp$ centre-of-mass energies of $\sqrt{s} = 7\,{\rm TeV}$ and $8\,{\rm TeV}$.
In Run\,1, ATLAS and CMS collected a dataset corresponding to about $25\,{\rm fb}^{-1}$ of integrated luminosity each.
The LHCb experiment collected data at a lower instantaneous luminosity (to reduce pile-up in the detector), recording a dataset corresponding to $3\,{\rm fb}^{-1}$ of integrated luminosity.
The ATLAS and CMS experiments can only trigger on $b$ hadron decays to final states with a dimuon pair.
By operating at a lower instantaneous luminosity, the LHCb experiment can also trigger rare decays to dielectron final-states or photons, and fully hadronic final-states.
The second period of LHC data taking (Run\,2) has just started with ATLAS and CMS recording approximately $5\,{\rm fb}^{-1}$ each of integrated luminosity in 2015 and LHCb a further $0.3\,{\rm fb}^{-1}$.

During Run\,2 of the LHC (2015-2018), the LHCb experiment expects to collect an additional 5\,fb$^{-1}$, 
and ATLAS and CMS 100\,fb$^{-1}$ of integrated luminosity each.
This, together with an increased centre-of-mass energy of $\sqrt{s}= 13$\,TeV for the LHC $pp$ collisions, 
will allow the experiments to collect datasets that are a factor of four (LHCb) and eight (ATLAS and CMS) larger than 
those collected in Run\,1, if the trigger thresholds will be kept the same.
On a longer term, an upgrade is foreseen for the LHCb experiment in 2019--2020 that will allow a 
dataset of $50\,{\rm fb}^{-1}$ to be collected in about five years of operation~\cite{Bediaga:1443882}. 
Major upgrades of the ATLAS and CMS detectors are also scheduled in 2023--2026. 
The ultimate aim for ATLAS and CMS is to reach an integrated luminosity of about 3000\,fb$^{-1}$ by around 2035.

Away from the LHC, the Belle II experiment at KEK is expected to start data taking with its full detector in 2018 and 
aims to collect an integrated luminosity of  $50\,{\rm ab}^{-1}$ by 2024. 
This will provide a dataset that is about 
a factor of 50 times larger than the combined dataset of the BaBar and Belle experiments.

This review summarises the state of rare $b$ hadron decay measurements at the start of Run\,2 of data taking at the LHC and provides
experimental prospects wherever is possible.
By ``rare'' decays, in this review we refer to decays that are mediated by a
pure FCNC transition, noting that these are not necessarily the decays with the
smallest branching fractions.
The review is structured as follows:
In Sec.~\ref{sec:framework} the theoretical framework for studying rare $B$ meson decays is introduced;
Recent experimental progress on leptonic ($B \to \ell^+ \ell^-$) decays by the LHC experiments is discussed in  Sec.~\ref{sec:leptonic};
Recent measurements of radiative $b \to s\gamma$ transitions performed by BaBar, Belle, and LHCb are introduced in Sec.~\ref{sec:radiative};
Recent  measurements of semileptonic $b \to s\ell^+\ell^-$ decays by BaBar, Belle, CDF, and the LHC experiments are discussed in Sec.~\ref{sec:semileptonic};
Searches for new light particles being produced in $b$ hadron decays are summarised in section~\ref{sec:hidden};
Searches for lepton number and lepton flavour violating processes are described in Sec.~\ref{sec:lfv};
In Sec.~\ref{sec:globalfits} a global analysis of experimental results on $b\to s$ transitions is presented;
and finally an outlook for the coming years is presented in Sec.~\ref{sec:outlook}.

\section{Theoretical framework}
\label{sec:framework}

The theory of rare $B$ decays is challenging due to the multitude of
physical scales involved. The weak interaction responsible for the flavour
change is governed by the electroweak scale set by $M_W\approx80\,$GeV; the
strong interactions responsible for the dynamics of the external states,
which are bound states of QCD, are governed by the scale
$\Lambda_\text{QCD}\approx 0.2\,$GeV, where QCD is in its non-perturbative
regime; the $b$ quark mass of roughly 4\,GeV defines an intermediate scale that
is small compared to the electroweak scale, but large compared to the QCD one.
This multi-scale problem can be tackled with the help of effective field theory
(EFT) methods. The first ingredient in the EFT method is a local operator product expansion (OPE)
that treats the weak interactions as point-like from the point of view of the
hadronic scales $m_b$ and $\Lambda_\text{QCD}$. Then, the FCNC interactions are encoded in Wilson coefficients of
dimension-six operators that are made out of the light SM fields, i.e. leptons,
the five lightest quarks, photons and gluons. This not only
allows to calculate QCD corrections to weak decays perturbatively but is also
useful when studying physics beyond the SM. Here, it can also be assumed that the new states are heavy
compared to the $b$ quark and any new physics effect can be described by a
modification of a Wilson coefficient or by the appearance
of a new operator not present in the SM.
This process is analogous to the Fermi theory of weak interactions, where the full theory is expressed in terms of the Fermi constant, $G_{F}$,
and a contact interaction between four fermions, in other words
a local operator.

With the help of the OPE, a $B$ decay amplitude to a final state $f$ can be
written schematically as
\begin{equation}
A(B\to f) = \langle f | \mathcal H_\text{eff} | B \rangle
= \frac{G_F}{\sqrt{2}} \sum_{i} \lambda \, C_i(\mu_b) \langle f |
Q_i(\mu_b) | B \rangle \,,
\end{equation}
where $\mathcal H_\text{eff}$ is the weak effective Hamiltonian, $\lambda$
is a CKM factor, $C_i$ are the Wilson coefficients, and
$\langle f | Q_i(\mu_b) | B \rangle$ are the matrix elements of the dimension-six
operators at the $b$ quark scale $\mu_b$. Since physical observables are independent of the renormalisation
scale $\mu$, this dependence cancels between the Wilson coefficients and the
matrix elements. While the Wilson coefficients can be computed in perturbation
theory at the electroweak scale $(\mu_0 \sim M_{W,t})$ where QCD is perturbative, the matrix
elements involve non-perturbative strong interactions. The methods to determine
these matrix elements depend strongly on the type of decay considered.
\begin{itemize}
\item In \emph{inclusive} radiative or semi-leptonic decays, where the hadronic
part of the final state is summed over, one can make use of the fact that $m_b$
(and to some extent even $m_c$)
is a hard scale compared to $\Lambda_\text{QCD}$ to write the decay rate in terms of the quark level transition
\begin{equation}
\Gamma(B\to X (\ell\ell, \gamma)) = \Gamma(b\to q(\ell\ell, \gamma))
+ O\left(\frac{\Lambda_\text{QCD}}{m_{b,c}}\right)
+\ldots
\label{eq:duality}
\end{equation}
This Heavy Quark Expansion
\cite{Georgi:1990um,Bigi:1993ex} (see \cite{Neubert:1993mb,Manohar:2000dt} for reviews)
allows to effectively compute the
matrix elements in perturbation theory.
The ellipsis in \eqref{eq:duality} refers to non-perturbative and other
long-distance corrections that are not formally  suppressed by $m_b$ or $m_c$.
\item In purely leptonic decays, all matrix elements are proportional to the
$B_q$ (with $q=d,s$) meson decay constant. Schematically, writing the semileptonic operators
as a product of a leptonic and a quark current, $Q=j_\ell\cdot j_q$,
\begin{align}
\langle \ell\ell | j_\ell\cdot j_q | B_q \rangle
& = 
\langle \ell\ell | j_\ell |0\rangle \cdot \langle 0| j_q | B_q \rangle \\ &
\sim \langle \ell\ell | j_\ell |0\rangle \cdot f_{B_q} \,.
\end{align}
In other words, the amplitude \emph{factorises} into a (trivial) leptonic part
and a hadronic part\footnote{This factorisation only holds up to QED corrections.}
that contains a single unknown. Decay constants are now
computed with lattice QCD (LQCD) methods to a precision of few percent
(see \cite{Aoki:2013ldr} for a review of lattice results).
\item In semi-leptonic decays with neutrinos in the final state, the decay
amplitude factorises as well,
\begin{align}
\langle \nu\nu M | j_\nu\cdot j_q | B \rangle
& =  
\langle \nu\nu | j_\nu |0\rangle \cdot \langle M| j_q | B \rangle \\ &
\sim \langle \nu\nu | j_\nu |0\rangle  \cdot F(q^2) \,.
\end{align}
The hadronic quantities are now \emph{form factors} that are functions of
the dineutrino invariant mass squared $q^2$. Depending on the hadronic
transition, several form factors can be relevant for a single decay. The most
important methods to determine form factors at present are LQCD and QCD sum
rules on the light cone (LCSR; see \cite{Khodjamirian:1998ji,Braun:1999dp,Colangelo:2000dp} for reviews). The two methods are complementary as they are
valid in different kinematical limits: LQCD is restricted to the region where
the hadronic recoil is low, corresponding to large $q^2$; and LCSR relies
on an expansion in the mass over the large energy of the final state meson and
is thus valid on the low side of the $q^2$ spectrum.
\item Semi-leptonic decays involving charged leptons are similar in first
approximation to the decays with neutrinos in the final state in that they
require the knowledge of the $B\to M$ form factors. However, the ``naive''
factorisation is now no longer exact as the charged lepton pair can also
originate from a photon emanating from a purely hadronic 
%flavour-changing
flavour-conserving interaction. Schematically,
\begin{equation}
\langle \ell\ell M | j_\ell\cdot j_q | B \rangle
\sim \langle  \ell\ell | j_\ell |0\rangle  \cdot F(q^2)  +
\text{non-factorisable corrections}\,.
\end{equation}
\end{itemize}
Decay-specific details of the determination of matrix elements will be
discussed in Sec.~\ref{sec:leptonic} and following.

\subsection{Effective Hamiltonian and Wilson coefficients}

\begin{figure}[tbp]
\centering
\includegraphics[width=0.9\linewidth]{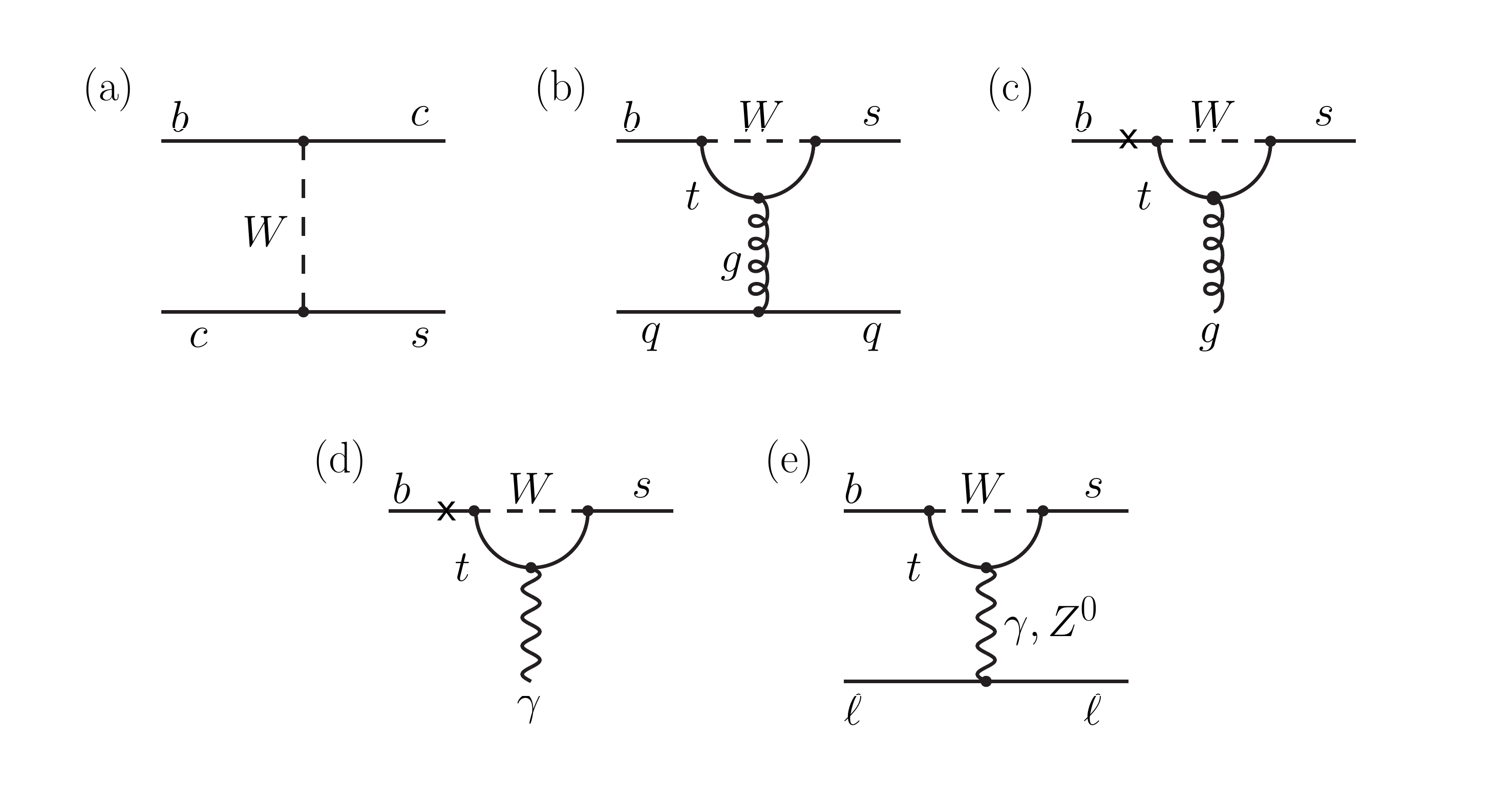}
\caption{Example tree and penguin diagrams in the
matching calculation of the Wilson coefficients for $b\to s$ transitions in the SM. The {\textsf x} indicates a chirality flip of the external quark line.}
\label{fig:heff}
\end{figure}

The effective Hamiltonian governing rare $B$ decays in the SM
contains the operators contributing to $b\to q\gamma$, $b\to q\ell^+\ell^-$,
and $b\to q\nu\bar\nu$ transitions (with $q=s$ or $d$) at the quark level.
The Wilson coefficients of these operators are determined by calculating
Feynman diagrams as the ones shown in Fig.~\ref{fig:heff} in the full SM and
\textit{matching} the result onto the effective theory.
QCD corrections lead to a \textit{mixing} of operators under renormalisation,
%so the effective Hamiltonian not only contains operators with leptons or a photon,
as the relevant effective Hamiltonian not only contains operators with leptons or a photon,
but also four-quark operators that mix into the former.
The effective Hamiltonian reads\footnote{In the following, we neglect
four-quark electroweak penguin operators which enter semi-leptonic,
leptonic and radiative decays only through
subleading QED effects.}
\begin{equation}\label{eq:heffsm}
\mathcal H_\text{eff}^{b\to q} =
\frac{4 G_F}{\sqrt{2}} \left(
\lambda_u^{(q)} \sum_{i=1}^2 C_i Q_i^u
+
\lambda_c^{(q)} \sum_{i=1}^2 C_i Q_i^c
-
\lambda_t^{(q)} \sum_{i=3}^{10} C_i Q_i
-
\lambda_t^{(q)}  C_\nu Q_\nu
+\text{h.c.}
\right)
\end{equation}
where $\lambda_p^{(q)}=V_{pb}V_{pq}^*$.
The sums contain:
the current-current operators (Fig.~\ref{fig:heff}(a)),
\begin{align}
Q_1^p  & = (\bar{q}_L \gamma_\mu T^a p_L) (\bar{p}_L \gamma^\mu T^a
b_L)
\,, &
Q_2^p & = (\bar{q}_L \gamma_\mu p_L) (\bar{p}_L \gamma^\mu b_L) \,;
\end{align}
QCD penguin operators (Fig.~\ref{fig:heff}(b)),
\begin{align}
Q_3 & = (\bar{q}_L \gamma_\mu b_L) \sum_p (\bar{p} \gamma^\mu
p) \, ,  &
Q_4 & = (\bar{q}_L \gamma_\mu T^a b_L) \sum_p (\bar{p}
\gamma^\mu T^a p) \, ,  \\
Q_5 & = (\bar{q}_L \gamma_\mu \gamma_\nu \gamma_\rho b_L)
\sum_p (\bar{p} \gamma^\mu \gamma^\nu \gamma^\rho p) \, ,
 &
Q_6 & = (\bar{q}_L \gamma_\mu \gamma_\nu \gamma_\rho T^a b_L)
\sum_p (\bar{p} \gamma^\mu \gamma^\nu \gamma^\rho T^a p) \, ,
\end{align}
where the sum runs over $p=u,d,s,c,b$;
electromagnetic and chromomagnetic dipole operators\footnote{
Also the chirality-flipped counterparts of $Q_7$ and $Q_8$
are generated in the SM, but their Wilson coefficients are suppressed
by $m_s/m_b$ in $b\to s$ transitions and $m_d/m_b$ in $b\to d$ transitions.} (Figs.~\ref{fig:heff}(c,~d)),
\begin{align}
Q_7 & = \frac{e}{16 \pi^2}
m_b (\bar{q}_L \sigma^{\mu\nu} b_R) F_{\mu\nu}
\,,  &
Q_8 & = \frac{g_s}{16 \pi^2}
m_b (\bar{q}_L \sigma^{\mu\nu} T^a b_R) G_{\mu\nu}^a
 \,;
\end{align}
and semi-leptonic operators (Fig.~\ref{fig:heff}(e)),
\begin{align}
Q_9 & = \frac{e^2}{16 \pi^2}
(\bar{q}_L \gamma_\mu b_L) \sum_\ell (\bar{\ell} \gamma^\mu \ell)
\,, &
Q_{10} & = \frac{e^2}{16 \pi^2}
(\bar{q}_L \gamma_\mu b_L) \sum_\ell (\bar{\ell} \gamma^\mu \gamma_5 \ell)
\,, \\
Q_\nu & = \frac{e^2}{8 \pi^2}
(\bar{q}_L \gamma_\mu b_L) \sum_\ell (\bar{\nu_\ell}_L \gamma^\mu \nu_{\ell L})
\,.
\end{align}
The $L$ and $R$ indices refer to left- and right-handed chiralities of the fermions.
Defined as they are in Eq.~\eqref{eq:heffsm}, with the CKM elements factored out,
all the Wilson coefficients are the same for $b\to s$ and $b\to d$ transitions in the SM.

The calculation of the Wilson coefficients at the low scale $\mu_b \sim m_b$,
which are needed for the prediction of physical observables, proceeds in two
steps. First, the matching calculation is used to determine the initial conditions
at the high scale $\mu_0\sim M_{W,t}$. Second, the calculation of the
\emph{anomalous dimension} matrix $\gamma$ needed for the solution of the
renormalisation group equation (RGE)
\begin{equation}
\mu \frac{d}{d\mu} \vec{C} = \gamma^T \vec{C} \,,
\end{equation}
which describes the mixing of the different operators and the evolution to the low-scale.
Both the initial conditions
and the anomalous dimensions
are now known for the
full set of Wilson coefficients $C_{1\text{--}10}$ at the
next-to-next-to-leading order (NNLO) in QCD and at the NLO in the electroweak
interactions\footnote{An exception is the
Wilson coefficient $C_9$, where NLO electroweak corrections are not fully known
yet.} 
\cite{Buras:1989xd,Buras:1991jm,Buras:1992tc,
Ciuchini:1992tj,Fleischer:1992gp,Buchalla:1992zm,
Buras:1993dy,Buchalla:1993bv,
Ciuchini:1993vr,Adel:1993ah,Misiak:1994zw,
Chetyrkin:1996vx,Chetyrkin:1997fm,Chetyrkin:1997gb,
Greub:1997hf,Buras:1997bk,Buras:1997xf,Ciuchini:1997xe,
Buchalla:1998ba,Baranowski:1999tq,Misiak:1999yg,Bobeth:1999mk,Bobeth:1999ww,Gambino:2003zm,
Bobeth:2003at,Gorbahn:2004my,Misiak:2004ew,Gorbahn:2005sa,Czakon:2006ss,Bobeth:2013tba}
(see \cite{Buras:2011we} for a review and historical account).
The Wilson coefficient $C_\nu$ is known at NLO in QCD and NLO in electroweak interactions
and has vanishing QCD and QED anomalous dimensions
\cite{Buchalla:1993wq,Buchalla:1997kz,Brod:2010hi}.

For the dipole operators, ``effective'' Wilson coefficients $C_{7,8}^\text{eff}$
are usually defined that, in contrast to $C_{7,8}$, are regularisation 
%scheme independent.
scheme-independent at the LO.
They are given by
\begin{align}
C_7^\text{eff}(\mu) &= C_7(\mu) + \sum_{i=1}^6 y_i C_i(\mu) \,,
&
C_8^\text{eff}(\mu) &= C_8(\mu) + \sum_{i=1}^6 z_i C_i(\mu) \,,
\end{align}
where
$y=(0,0,-\frac{1}{3},-\frac{4}{9},-\frac{20}{3},-\frac{80}{9})$,
$z=(0,0,1,-\frac{1}{6},20,-\frac{10}{3})$,
%in naive dimensional regularisation.
in the so-called naive dimensional regularisation scheme.
Moreover, it often proves convenient to define ``effective'' coefficients that
contain additional contributions of four-quark
operators that always appear in matrix elements in a fixed combination with
other Wilson coefficients. This applies in particular to the Wilson coefficient
$C_9^\text{eff}$. Different conventions as to which contribution to include
in the effective coefficient exist in the literature.

Numerical values and theoretical uncertainties of all the Wilson coefficients in \eqref{eq:heffsm},
using the state of the art for the initial conditions and anomalous dimensions
(including QED corrections to the running and the mixing with four-quark electroweak
penguin operators)
are shown in Table~\ref{tab:wc} for four different choices of the scale $\mu_b$.
The numerical renormalisation group evolution was performed using
the flavio package \cite{flavio}.

\begin{table}[tbp]
\renewcommand{\arraystretch}{1.2}
\centering
\begin{tabular}{lrrrrrrrrr}
\hline
 & \multicolumn{4}{c}{$\mu_b$ [GeV]} & \multicolumn{2}{c}{$\delta
C_i(\mu_b=4.2\,\text{GeV})$} \\
 & $2.0$ & $4.2$ & $4.8$ & $5.0$ & $\mu_0$ & $M_t$, $\alpha_s$\\
\hline
$C_1$ & $-0.492$ & $-0.294$ & $-0.264$ & $-0.255$ & $0.009$ & $0.002$\\
$C_2$ & $1.033$ & $1.017$ & $1.015$ & $1.014$ & $0.001$ & $0.000$\\
$C_3$ & $-0.0133$ & $-0.0059$ & $-0.0051$ & $-0.0048$ & $0.0002$ & $0.0001$\\
$C_4$ & $-0.147$ & $-0.087$ & $-0.080$ & $-0.078$ & $0.000$ & $0.001$\\
$C_5$ & $0.0009$ & $0.0004$ & $0.0004$ & $0.0003$ & $0.0000$ & $0.0000$\\
$C_6$ & $0.0030$ & $0.0011$ & $0.0009$ & $0.0009$ & $0.0001$ & $0.0000$\\
$C_7^\mathrm{eff}$ & $-0.3189$ & $-0.2957$ & $-0.2915$ & $-0.2902$ & $0.0002$ &
$0.0005$\\
$C_8^\mathrm{eff}$ & $-0.1780$ & $-0.1630$ & $-0.1606$ & $-0.1599$ & $0.0005$ &
$0.0004$\\
$C_9$ & $4.349$ & $4.114$ & $4.053$ & $4.033$ & $0.012$ & $0.007$\\
$C_{10}$ & $-4.220$ & $-4.193$ & $-4.189$ & $-4.187$ & $0.000$ & $0.033$\\
\hline
\end{tabular}
\caption{Numerical values of the $\Delta B=1$ Wilson coefficients at the
$b$ quark scale $\mu_b$ for four different choices of $\mu_b$, to NNLO accuracy.
The last two columns show the uncertainty for $\mu_b=4.2$\,GeV stemming from
the variation of the matching scale $\mu_0$ between 80 and 320\,GeV (roughly
$M_t/2$ and $2 M_t$), which gives an indication of the size of
neglected higher-order corrections, as well as from parametric uncertainties
that are dominated by the strong coupling for $C_{1\text{--}6}$ and by $M_t$
for $C_{9,10}$ (an additional uncertainty due to electroweak scheme dependence in $C_9$
is not shown).}
\label{tab:wc}
\end{table}

\subsection{CKM hierarchies}\label{sec:VtdVts}

The GIM mechanism \cite{Glashow:1970gm}, which ensures the cancellation of
FCNCs for degenerate quark masses, is broken
strongly by the large top quark mass in the SM. This is reflected by the fact
that the FCNC operators in the effective Hamiltonian in Eq.~\eqref{eq:heffsm} are
multiplied by the combination of CKM elements $\lambda_t^{(q)}=V_{tb}V_{tq}^*$.
Since the values of the Wilson coefficients are equal for $b\to s$ and $b\to d$
transitions up to CKM factors, the rates of $b\to d$ transitions are
expected to be suppressed by a factor $|V_{td}/V_{ts}|^2$ compared to $b\to s$
transitions and measurements of these rates can be interpreted as
determinations of $|V_{td}/V_{ts}|$.

While the scaling with $|V_{td}/V_{ts}|$ is true at leading order,
there are important corrections to this ratio from matrix elements of the current-current operators
$Q_{1,2}^{u,c}$ that are multiplied by
the CKM combinations $\lambda_u^{(q)}$ and $\lambda_c^{(q)}$
in Eq.~\eqref{eq:heffsm}.
CKM unitarity implies $\lambda^{(q)}_u+\lambda^{(q)}_c+\lambda^{(q)}_t=0$, so
one of the three quantities can be eliminated. In $b\to s$ transitions, there
is a hierarchy $\lambda^{(s)}_{t,c}\sim \lambda^2\gg\lambda^{(s)}_u\sim
\lambda^4$, where $\lambda$ is the sine of the Cabibbo angle. Consequently,
$Q_{1,2}^{(u)}$ can usually be neglected and $\lambda_c^{(s)}\approx-\lambda_t^{(s)}$ can be assumed.
An important consequence is that direct $C\!P$ asymmetries in $b\to s$ transitions
are tiny in the SM.
In $b\to d$ transitions,
the CKM hierarchies are instead such that all three quantities are comparable,
$\lambda^{(d)}_t\sim\lambda^{(d)}_c\sim\lambda^{(d)}_u\sim \lambda^3$.
As a result, many $b\to d$ transitions receive additional hadronic contributions
involving $Q_{1,2}^u$ that lead to larger theoretical uncertainties compared
to the corresponding $b\to s$ transition and need to be taken into account when
extracting $|V_{td}/V_{ts}|$.

\subsection{Effective Hamiltonian beyond the Standard Model}\label{sec:heffbsm}

\noindent
In extensions of the SM involving new \textit{heavy} particles, new physics can
\begin{enumerate}
 \item modify the Wilson coefficients of the operators present in the SM;
 \item generate new operators not present in the SM.
\end{enumerate}
While there is a large number of FCNC four-quark operators, these only enter
into the semi-leptonic and radiative decays discussed in this review through
higher-order corrections. Moreover, several
of the four-quark operators present in the SM receive sizeable contributions from
renormalisation group mixing with the current-current operator $Q_2$ that is
generated at tree level, diluting possible new physics effects. Consequently,
in a large class of models it is sufficient to consider  BSM contributions only to the 
dipole operators and operators with two quark and two lepton fields.
In full generality, 
BSM contributions only to the the dipole operators can be different between $b\to s$ and
$b \to d$ transitions and can have quark chirality opposite to the SM (indicated by primes below),

\begin{align}
(Q_7^{(\prime)})_{q} & = \frac{e}{16 \pi^2}
m_b (\bar{q}_{L(R)} \sigma^{\mu\nu} b_{R(L)}) F_{\mu\nu}
\,,  &
(Q_8^{(\prime)})_{q} & = \frac{g_s}{16 \pi^2}
m_b (\bar{q}_{L(R)} \sigma^{\mu\nu} T^a b_{R(L)}) G_{\mu\nu}^a
\,,
\end{align}
Beyond the SM, lepton flavour universality could also be violated, resulting
in a dependence of the operators involving leptons on the lepton flavour. In general,
\begin{align}
(Q_9^{(\prime)})_q^\ell & = \frac{e^2}{16 \pi^2}
(\bar{q}_{L(R)} \gamma_\mu b_{L(R)}) (\bar{\ell} \gamma^\mu \ell)
\,, &
(Q_{10}^{(\prime)})_q^\ell & = \frac{e^2}{16 \pi^2}
(\bar{q}_{L(R)} \gamma_\mu b_{L(R)}) (\bar{\ell} \gamma^\mu \gamma_5 \ell)
\,, \nonumber\\
(Q_\nu^{(\prime)})_q^\ell & = \frac{e^2}{8 \pi^2}
(\bar{q}_{L(R)} \gamma_\mu b_{L(R)}) (\bar{\nu_\ell}_L \gamma^\mu \nu_{\ell L})
\,,
\label{eq:sl-operators-np}
\end{align}
where the corresponding Wilson coefficients can take different values for different flavours of lepton.
Beyond the SM basis there can also be contributions from scalar ($Q_S$), pseudoscalar ($Q_P$), and tensor ($Q_T$) operators involving two quark fields and two leptons,
\begin{align}
(Q_S^{(\prime)})_q^\ell &= \frac{e^2}{16 \pi^2} m_b
(\bar{q}_{L(R)} b_{R(L)}) (\bar{\ell} \ell)
\,, &
(Q_P^{(\prime)})_q^\ell &= \frac{e^2}{16 \pi^2} m_b
(\bar{q}_{L(R)} b_{R(L)}) (\bar{\ell}  \gamma_5 \ell)
\,, \nonumber\\
(Q_T^{(\prime)})_q^\ell &= \frac{e^2}{16 \pi^2}
(\bar{q}_{R(L)} \sigma^{\mu\nu} b_{L(R)}) (\bar{\ell}_{R(L)} \sigma_{\mu\nu}
\ell_{L(R)})
\,, &
\end{align}

In theories that violate lepton flavour,  semi-leptonic operators with
different lepton flavour can also be present, e.g.
\begin{align}
(Q_9^{(\prime)})_q^{\ell_i\ell_j} & = \frac{e^2}{16 \pi^2}
(\bar{q}_{L(R)} \gamma_\mu b_{L(R)}) (\bar{\ell_i} \gamma^\mu \ell_j)
\,, &
\end{align}
where $i\neq j$, and analogously for $Q_{10,S,P,T}^{(\prime)}$.

There are several symmetry-motivated cases where relations between the Wilson
coefficients of these operators are expected.
\begin{itemize}
\item Minimal flavour violation (MFV) in the quark sector
\cite{D'Ambrosio:2002ex}  implies $(C_i)_d = (C_i)_s$ and $C_i'\approx0$. The same
relations are obtained in models with a minimally broken $U(2)^3$ symmetry.
\item Constrained MFV \cite{Buras:2000dm} implies $C_{S,P,T}^{(\prime)}=0$ in
addition to the MFV relations, i.e. there are no operators beyond the SM ones.
Up-to small differences between the hadronic systems involved, in MFV models decay rates are expected to follow the CKM hierarchy i.e. the rate of $b \to d$ processes is suppressed by $|V_{td}/V_{ts}|^2$ with respect to $b \to s$ processes.
\item Lepton flavour universality implies
$(C_i^{(\prime)})_q^e=(C_i^{(\prime)})_q^\mu=(C_i^{(\prime)})_q^\tau$.
\item Lepton flavour conservation implies
$(C_k^{(\prime)})_q^{\ell_i\ell_j}=0$.
\item Any weakly coupled
heavy new physics implies \cite{Alonso:2014csa}
$(C_S)_q^\ell=-(C_P)_q^\ell$, $(C_S')_q^\ell=(C_P')_q^\ell$, and
$(C_T^{(\prime)})_q^\ell=0$.\footnote{This assumes the dimension-6 Standard Model
Effective Field Theory (SMEFT) with a linearly realised Higgs boson
\cite{Buchmuller:1985jz, Grzadkowski:2010es}, i.e. is not valid for a
strongly-interacting Higgs sector \cite{Cata:2015lta}. Corrections to these relations are of dimension eight in the SMEFT.}
\end{itemize}

\section{Leptonic decays}
\label{sec:leptonic}

\subsection{Standard Model predictions}

The leptonic decays $B_q\to \ell^+\ell^-$, where $q=d,s$ and $\ell=e,\mu,\tau$,
are especially rare in the SM.
In addition to the SM loop and CKM suppression, the two spin-$1/2$ muons originate from a pseudoscalar $B$ meson and the decay is helicity suppressed.
The branching fractions of these decays can be written
\begin{equation}
\text{BR}(B_q\to\ell^+\ell^-)_\text{SM} =
\tau_{B_q}
\frac{G_F^2\alpha_\text{em}^2}{16\pi^2}
f_{B_q}^ 2
m_\ell^2 m_{B_q} \sqrt{1-\frac{4 m_\ell^2}{m_{B_q}^2}}
|V_{tb}V_{tq}^*|^2 |C_{10}^\text{SM}|^2 \,,
\label{eq:br-bll-sm}
\end{equation}
where $f_{B_q}$ is the $B_q$ meson decay constant.
In the SM the branching fraction only depends on the Wilson coefficient $C_{10}$, whose contribution is suppressed (due to helicity suppression) by $m_{\ell}^{2}/m_{B_q}^2$.
The helicity suppression is consequently particularly strong for $B_q \to e^+ e^-$ and $B_q \to \mu^+ \mu^-$ due to the smallness of the electron and muon masses.

The branching fractions of these decays, accounting for NLO electroweak corrections \cite{Bobeth:2013tba} and NNLO
QCD corrections \cite{Hermann:2013kca} to $C_{10}$, is known to better than 10\% precision~\cite{Bobeth:2013uxa}.
The SM branching fraction uncertainty is shared almost evenly between our knowledge of the $B^0$ and $B^0_s$ meson decay constants from lattice QCD~\cite{Witzel:2013sla, Na:2012kp, Bazavov:2011aa}, and the CKM matrix elements.

For the $B_s$ decays, the sizeable life-time difference between the heavy and
light $B_s$ mass eigenstates leads to a difference between the ``prompt''
branching fraction (before $B_s$--$\bar B_s$ mixing) given by \eqref{eq:br-bll-sm}
and the time-integrated one that is measured by experiments. Denoting the
latter as $\overline{\text{BR}}$, one has \cite{DeBruyn:2012wk}
\begin{equation}
\overline{\text{BR}}(B_s\to\ell^+\ell^-)
=
\left[
\frac{1+\mathcal A_{\Delta\Gamma} \, y_s}{1-y_s^2}
\right]
\text{BR}(B_s\to\ell^+\ell^-) \,,
\end{equation}
where $y_s=\Delta\Gamma_s/(2\Gamma_s)=0.063(5)$ \cite{Amhis:2014hma}.
In the SM, $\mathcal A_{\Delta\Gamma}=+1$, so the time-integrated branching
fraction is
roughly 6\% larger than the prompt one. Adopting the parameters in Table~\ref{tab:leptonic:inputs}
one finds the following SM predictions,
\begin{align}
\overline{\text{BR}}(B_s\to e^+ e^-)_{\rm SM} &= (8.24 \pm 0.36)\times 10^{-14}
,&
\text{BR}(B^0\to e^+ e^-)_{\rm SM} &= (2.63 \pm 0.32)\times 10^{-15}
,\\
\overline{\text{BR}}(B_s\to\mu^+\mu^-)_{\rm SM} &= (3.52 \pm 0.15)\times 10^{-9}
,&
\text{BR}(B^0\to\mu^+\mu^-)_{\rm SM} &= (1.12 \pm 0.12)\times 10^{-10}
,\label{eq:bsmumu_SM} \\
\overline{\text{BR}}(B_s\to\tau^+\tau^-)_{\rm SM} &= (7.46 \pm 0.30)\times 10^{-7}
,&
\text{BR}(B^0\to\tau^+\tau^-)_{\rm SM} &= (2.35 \pm 0.24)\times 10^{-8}
.
\end{align}
These predictions differ slightly from those of Ref.~\cite{Bobeth:2013uxa} by the choice of renormalisation scheme used for $C_{10}$ and in the values used for $f_{B_q}$.

\begin{table}
\centering
\begin{tabular}{ccc}
\hline
Parameter & Value & Ref. \\
\hline
$\alpha_\text{em}(M_Z)$ & $1/127.940(14)$ & \cite{Agashe:2014kda}\\
$\alpha_{s}(M_Z)$ & $0.1185(6)$ & \cite{Agashe:2014kda}\\
$\tau_{B^0}$ & $1.520(4)~\text{ps}$ & \cite{Amhis:2014hma} \\
$\tau_{B_s^{H}}$ & $1.604(10)~\text{ps}$& \cite{Amhis:2014hma}\\
$f_{B^0}$ & $190.5(4.2)~\text{MeV}$ &  \cite{Aoki:2013ldr}\\
$f_{B_s}$ & $226.0(2.2)~\text{MeV}$ &  \cite{Rosner:2015wva}\\
$|V_{cb}|$ & $4.221(78)\times10^{-2}$  & \cite{Agashe:2014kda}\\
$|V_{ub}|$ & $3.72(16)\times10^{-3}$  & \cite{Lattice:2015tia}\\
$|V_{tb}V^{*}_{ts}/V_{cb}|$ & $0.980(2)$ & \\
$|V_{tb}V^{*}_{td}/V_{ub}|$ & $2.45(15)$ & \\
\hline
\end{tabular}
\caption{
Numerical inputs for the SM calculation of ${\rm BR}(B_q \to \ell^+\ell^-)$.
\label{tab:leptonic:inputs}
}
\end{table}

Assuming the validity of the SM, the ratio\footnote{Here, a factor
$\sqrt{1-4m_\mu^2/m_{B^0}^2}/\sqrt{1-4m_\mu^2/m_{B_s}^2}$ has been neglected,
as it differs from 1 by less than $0.01\%$.}
\begin{equation}
\frac{\text{BR}(B^0\to \mu^+ \mu^-)}{\overline{\text{BR}}(B_s\to \mu^+ \mu^-)}
= \frac{\tau_{B^0}}{\tau_{B_s}}\frac{|V_{td}|^2}{|V_{ts}|^2} \frac{m_{B^0}}{m_{B_s}} \frac{f_{B^0}^2}{f_{B_s}^2}(1-y_s)
\end{equation}
allows the extraction of the ratio of CKM elements $|V_{td}/V_{ts}|$ with the
smallest theoretical uncertainty in rare $B$ decays. Using the HFAG averages
for the $B^0$ and $B_s$ lifetimes in Table~\ref{tab:leptonic:inputs},
where
\cite{Amhis:2014hma}
\begin{align}
\tau_{B_s^{\rm H}}&=\frac{\tau_{B_s}}{1-y_s} \,,
\end{align}
and the recent HPQCD lattice calculation of the ratio of the $B^0$ and $B_s$ decay
constants \cite{Dowdall:2013tga},
\begin{equation}
f_{B_s}/f_{B^0} = 1.205 \pm 0.007 \,,
\end{equation}
one obtains
\begin{equation}
\frac{|V_{td}|}{|V_{ts}|}
=
\sqrt{\frac{\text{BR}(B^0\to \ell^+ \ell^-)}{\overline{\text{BR}}(B_s\to \ell^+ \ell^-)}}
\times (1.238 \pm  0.007 \pm 0.004) \,,
\end{equation}
with the first error due to the decay constants and the second one due to
the lifetimes. The total theoretical uncertainty is below one percent. This can be compared to the
extraction from the ratio of $B^0$ and $B_s$ mass differences,
\begin{equation}
\frac{\Delta M_d}{\Delta M_s} = \frac{|V_{td}|^2}{|V_{ts}|^2} \frac{m_{B^0}}{m_{B_s}}\frac{1}{\xi^2} \,.
\end{equation}
Using the recent lattice calculation of $\xi$ from the FNAL/MILC collaborations
\cite{Bazavov:2016nty}, this leads to a relative uncertainty of $1.5\%$ in $|V_{td}/V_{ts}|$
(while the experimental uncertainty is already negligible).

\subsection{New physics sensitivity}

The important role of the branching fraction of the $B_s \to \mu^+ \mu^-$ and $B^0 \to \mu^+ \mu^-$
decays in models with extended Higgs sectors has been widely discussed in the literature.
Beyond the SM, the (prompt) branching fraction is modified as
\begin{equation}
\frac{\text{BR}(B_q\to\ell^+\ell^-)}{\text{BR}
(B_q\to\ell^+\ell^-)_\text{SM}}
=
\frac{|S|^2\left(1-\frac{4m_\ell^2}{m_{B_q}^2}\right)+|P|^2}{|C_{10}^\text{SM}
|^2 }
,
\label{eq:brbqll-np}
\end{equation}
where
\begin{align}
P &= \left[(C_{10})_q^\ell-(C_{10}')_q^\ell\right] +  \frac{m_{B_q}^2}{2m_\ell}
\left[(C_{P})_q^\ell-(C_{P}')_q^\ell\right]
\,,&
S &= \frac{m_{B_q}^2}{2m_\ell} \left[(C_{S})_q^\ell-(C_{S}')_q^\ell\right]
\,,
\end{align}
while $\mathcal A_{\Delta\Gamma}$, relevant for $B_s\to\ell^+\ell^-$, is given by
\begin{equation}
 \mathcal A_{\Delta\Gamma} =
\frac{\text{Re}\left(P^2-S^2\right)}{|P|^2+|S|^2} \,.
\label{eq:ADG}
\end{equation}
Equations\ \eqref{eq:brbqll-np} and \eqref{eq:ADG} show the complementary
dependence of the branching ratio and $\mathcal A_{\Delta\Gamma}$ on the
Wilson coefficients, so measurements of both quantities would provide independent information
on new physics.
The value of $\mathcal A_{\Delta\Gamma}$ can be determined separately from the branching
fraction by measuring the $B_s\to\ell^+\ell^-$ effective lifetime \cite{DeBruyn:2012wk}
\begin{equation}
\tau_{\ell^+\ell^-} =
\frac{\int_0^\infty t \langle \Gamma(B_s(t)\to\ell^+\ell^-)\rangle dt}
{\int_0^\infty \langle \Gamma(B_s(t)\to\ell^+\ell^-)\rangle dt}
\end{equation}
that satisfies the relation
\begin{equation}
\mathcal A_{\Delta\Gamma} y_s =
\frac{(1-y_s^2)\tau_{\ell^+\ell^-}-(1+y_s^2)\tau_{B_s}}
{2\tau_{B_s}-(1-y_s^2)\tau_{\ell^+\ell^-}} \,.
\end{equation}
In the presence of new physics, the mixing-induced $C\!P$ asymmetry
$S_{\ell^+\ell^-}$ in $B_s\to\ell^+\ell^-$ decays could be non-zero.
It is given in terms of the Wilson coefficients as
\cite{DeBruyn:2012wk,Buras:2013uqa}
\begin{equation}
S_{\ell^+\ell^-} =
\frac{\text{Im}\left(P^2-S^2\right)}{|P|^2+|S|^2} \,.
\end{equation}
The value of $S_{\ell^+\ell^-}$ can be extracted from the time-dependent tagged rate asymmetry,
\begin{equation}
\frac{\Gamma(B_s(t)\to\ell^+\ell^-)-\Gamma(\bar B_s(t)\to\ell^+\ell^-)}
{\Gamma(B_s(t)\to\ell^+\ell^-)+\Gamma(\bar B_s(t)\to\ell^+\ell^-)}
=\frac{S_{\ell^+\ell^-}\sin(\Delta M_s t)}
{\cosh(y_s t/\tau_{B_s})+\mathcal A_{\Delta\Gamma}\sinh(y_s t/\tau_{B_s})}
\,.
\end{equation}

Unlike the SM contribution from $Q_{10}$, contributions from scalar and pseudoscalar operators are not helicity suppressed
and models with non-zero $C_S^{(\prime)}$ and $C_P^{(\prime)}$ can result in large enhancements of the branching fraction for these decays.
A prime example of this are models with an extended Higgs sector such
as the MSSM. 
In the MSSM, even for flavour-diagonal soft-terms, chargino loops
lead to a contribution to $(C_{S,P})_q^\ell$ that is proportional to the ratio of the two vacuum expectation values of the two Higgs doublets cubed, i.e. $\tan^3\beta$~\cite{Babu:1999hn,Isidori:2001fv,Buras:2002wq}.
This contribution is sizeable in scenarios motivated by Grand Unification.
Apart from scalar exchange, scalar and pseudoscalar operators can also be generated
by the tree-level exchange of the vector leptoquark states with quantum numbers
$U_1=(\mathbf{3},\mathbf{1},2/3)$ and $V_2=(\overline{\mathbf{3}},\mathbf{2},5/6)$
under the SM gauge group (see \cite{Dorsner:2016wpm} for a recent review).
In the presence of pseudoscalar operators, the branching fractions could also
be suppressed.
The strong constraints on tree-level scalar exchange from the measurement
of $B_s \to \mu^+ \mu^-$ is in principle sensitive to new physics scales
close to 1000~TeV \cite{Buras:2014zga}.

Even in the absence of scalar or pseudoscalar operators, the $B_q\to\ell^+\ell^-$ decay provides
an important test of models predicting deviations in $(C_{10})_q^\ell$ or
a non-zero $(C_{10}')_q^\ell$. While these Wilson coefficients also contribute
to semi-leptonic $b\to q\ell^+\ell^-$ transitions, $B_q\to\ell^+\ell^-$ decays are subject
to smaller hadronic uncertainties and allow the cleanest theoretical extraction of the $(C_{10}^{(\prime)})_q^\ell$  Wilson
coefficients.
In the case of $B_s\to\ell^+\ell^-$, a new physics effect in one of these Wilson
coefficients implies a similar new physics effect either in $B\to K\ell^+\ell^-$
or $B\to K^*\ell^+\ell^-$, as a cancellation with other Wilson coefficients hiding
the effect in all semi-leptonic observables simultaneously is not possible
due to the large number of observables with different dependences on the
coefficients.
There are two qualitatively different effects generating
contributions to $(C_{10}^{(\prime)})_q^\ell$: effective flavour-changing $Z$ couplings
and short-distance semi-leptonic operators.

Effective flavour-changing $Z$ couplings can either be generated at the loop
level (e.g. in the MSSM \cite{Buchalla:2000sk}) or at tree level (e.g. in models with partial
compositeness \cite{Straub:2013zca} or in the closely related Randall-Sundrum models
\cite{Blanke:2008yr,Bauer:2009cf}). They generate mostly
$(C_{10}^{(\prime)})_q^\ell$ rather
than $(C_9^{(\prime)})_q^\ell$ due to the accidentally suppressed vector
coupling of the
$Z$ to charged leptons in the SM. An important feature of these
$Z$-mediated effects is that they are lepton flavour universal and thus imply
the same relative enhancement or suppression of $B_q\to\mu^+\mu^-$ and
$B_q\to\tau^+\tau^-$;
moreover, they also enter in $B\to K^{(*)}\nu\bar\nu$ decays \cite{Buras:2014fpa}.
Models with MFV also imply effects
in the decays $K^+\to\pi^+\nu\bar\nu$ and $K_L\to \pi^0\nu\bar\nu$.

Short-distance semi-leptonic operators can be generated at loop level
or at tree level by the exchange of a new heavy neutral vector boson ($Z'$,
see e.g. \cite{Buras:2012jb}) or
various types of scalar or vector leptoquarks \cite{Dorsner:2016wpm}.
In the $Z'$ case, the $Z'$-mediated contribution often competes with a
$Z$-mediated contribution due to $Z$-$Z'$ mixing \cite{Buras:2014yna}.
In the leptoquark case, one expects an effect for only a single lepton flavour
(if the leptoquark only couples to one lepton flavour) or the presence of
a lepton flavour-violating decay $B_q\to \ell_1^+ \ell_2^-$ (if the leptoquark
couples to several lepton flavours).

Measuring the decays $B_q\to\tau^+\tau^-$, in addition to $B_q\to\mu^+\mu^-$,
would allow independent and complementary tests of the SM
and upper bounds on the decay rate will constrain extensions of the SM that violate
lepton flavour universality.
While the relative enhancement or suppression of these modes is the same%
\footnote{In the scalar case, a small difference arises due to the factor $(1-4m_\ell^2/m_{B_q}^2)$ in \eqref{eq:brbqll-np}.}
in models where new physics enters through scalar or pseudoscalar operators
or flavour-changing $Z$ couplings, $B_q\to\tau^+\tau^-$ would probe models
with new physics dominantly coupling to the third generation, e.g. leptoquark
models motivated by the $B\to D^{(*)}\tau\nu$ anomalies (see Sec.~\ref{sec:semileptonic:universality}).

\subsection{Experimental aspects}

\begin{figure}[htb]
\begin{center}
\includegraphics[width=0.65\linewidth]{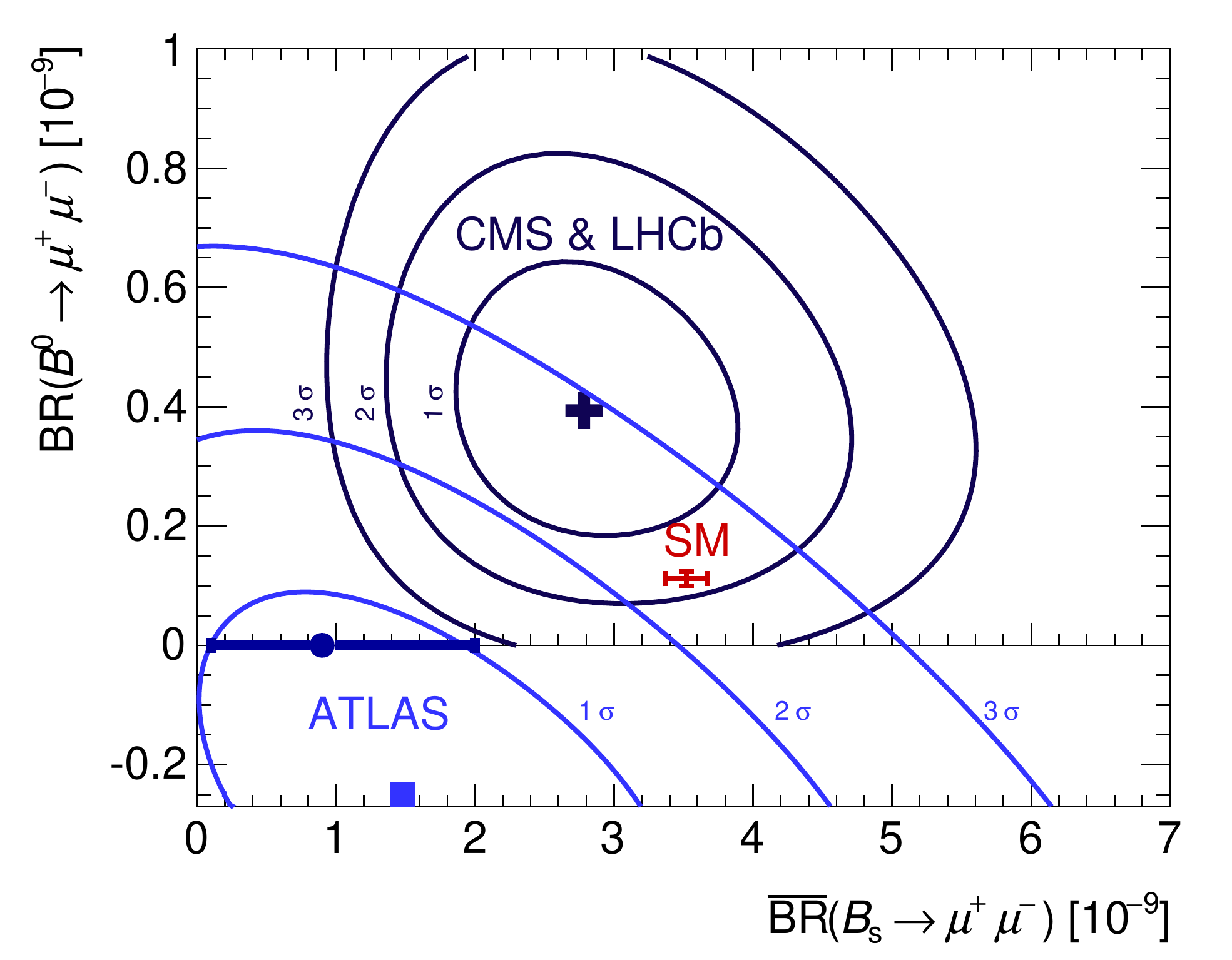}
\caption{\small Likelihood contours  for the $B_s \to \mu^+ \mu^-$ and $B^0 \to \mu^+ \mu^-$ branching fractions
from a fit to the combined LHCb and CMS datasets (solid black lines) and ATLAS dataset (solid blue line)
corresponding to intervals of $1, 2, 3 \,\sigma$. In all cases no constraint of non-negative branching fractions has been imposed.
The cross indicates the best fit point for the combined datasets of LHCb and CMS in the two-dimensional plane of the branching fractions.
The solid bullet corresponds to the maximum of the likelihood of the ATLAS dataset within
the boundary of non-negative branching fractions, with the error bars covering the 68.3\% confidence range for
$\overline{\text{BR}}(B_s \to \mu^+ \mu^-)$. The contours are reproduced from Refs.~\cite{CMS:2014xfa}  and \cite{Aaboud:2016ire}.
The SM prediction is also shown (see text).}
\label{fig:bsmumu}
\end{center}
\end{figure}

At the start of data taking at the LHC, no experimental evidence for either the $B^0 \to \mu^+ \mu^-$ or  $B_s \to \mu^+ \mu^-$ decay had been found.
Upper limits on the branching fractions were more than one order of magnitude above the SM predictions, the best limits being provided by the CDF collaboration~\cite{Aaltonen:2011fi}: ${\rm BR}(B^0 \to \mu^+ \mu^-) < 6.0 \times 10^{-9}$  and BR$(B_s \to \mu^+ \mu^-)< 4.0 \times 10^{-8}$ at 95\% CL.
In 2013 LHCb published the first evidence of the $B^0_s \to \mu^+ \mu^-$ decay based on a dataset corresponding to 2.0\,${\rm fb}^{-1}$ of integrated luminosity collected in 2011 and the first part of 2012~\cite{Aaij:2012nna}.
In 2014 the CMS and LHCb collaborations  performed a joint analysis of their datasets
collected during Run\,1 of the LHC, obtaining a statistical significance of  $6.2 \,\sigma$ for the $B^0_s$ mode and $3.0 \,\sigma$ for the
$B^0$ mode~\cite{CMS:2014xfa, Chatrchyan:2013bka, Aaij:2013aka}.
The combined fit leads to
\begin{align}
\begin{split}
\overline{\text{BR}}(B_s \to \mu^+ \mu^-) & = (2.8^{+0.7}_{-0.6}) \times 10^{-9} \,, \\
{\rm BR}(B^0 \to \mu^+ \mu^-) & = (3.9 ^{+1.6}_{-1.4}) \times 10^{-10} \,,
\end{split}
\end{align}
where the
uncertainties include both statistical and systematic sources of uncertainty. Systematic uncertainties
constitute 35\% and 18\% of the total uncertainty for the $B^0_s$
and $B^0$ decays, respectively.
The two dimensional interval resulting from the fit is shown in Fig.~\ref{fig:bsmumu}.
The result is compatible with the SM
branching fraction prediction for the $B^0_s \to \mu^+ \mu^-$ and $B^0 \to  \mu^+ \mu^-$ decays at the level of
1.2$\sigma$ and 2.2$\sigma$, respectively.
This result concludes a search that started more than three decades ago
and starts the era of precision measurements of the properties of this decay.

The ATLAS collaboration has also recently published the results of its search for the $B_q \to \mu^+ \mu^-$ decays, using the dataset collected during Run\,1 of the LHC~\cite{Aaboud:2016ire}.  For the $B^0$ decay, ATLAS sets an upper limit on the
branching fraction of  ${\rm BR}(B^0 \to \mu^+ \mu^-) <  4.2 \times 10^{-10}$ at 95\% CL.
For the $B^0_s$ decay, the ATLAS central value is
$\overline{\text{BR}}(B^0_s \to \mu^+ \mu^-) = (0.9^{+1.1}_{-0.8})\times 10^{-9}$.
The two dimensional contour of the ATLAS result is also shown in Fig.~\ref{fig:bsmumu}.
The ATLAS result is consistent with the Standard Model expectation (and with the combined CMS + LHCb result) at about two standard deviations.

The channels with electrons in the final state ($B^0 \to e^+ e^-$ and $B^0_s \to e^+ e^-$) are much more strongly helicity
suppressed than the dimuon modes.  To observe these decays in the SM we would require a sample of $B$ mesons that is significantly larger than the one accumulated during Run\,1 of the LHC.  The best existing limits on the branching fractions of these decays are from the CDF collaboration \cite{Aaltonen:2009vr}, who sets limits of
\begin{align}
{\rm BR}(B^0 \to e^+ e^-)  &< 8.3 \times 10^{-8}\,,\\
\overline{{\rm BR}}(B_s \to e^+ e^-) &< 2.8 \times 10^{-7}~\text{at~90\%~CL}\,.
\end{align}
These limits are still about seven orders of magnitude above the SM predictions.
Note, the two limits are not independent, when setting limits on ${\rm BR}(B^0 \to e^+ e^-)$ it is assumed that $\overline{{\rm BR}}(B_s \to e^+ e^-)$ is zero and vice versa.

The branching fraction of $B_{q} \to  \tau^+ \tau^-$ decays is much larger, due to the large tau lepton mass.
However, due to the fact that there are multiple neutrinos among the tau decay products 
(between two and four depending on the tau final states),
the $\tau^+ \tau^-$  final state is considerably more difficult to access experimentally than the $e^+ e^-$ and $\mu^+ \mu^-$ modes.
The only existing published limit\footnote{The LHCb collaboration presented this summer the preliminary upper limits 
$B^0 \to \tau^+ \tau^- <1.3 \times 10^{-3}$ and $B^0_s \to \tau^+ \tau^- <2.4 \times 10^{-3}$ at 90\% CL based 
on the full Run I dataset~\cite{LHCb-CONF-2016-011}.}
 comes from the BaBar collaboration~\cite{Aubert:2005qw}
\begin{align}
{\rm BR}(B^0 \to \tau^+ \tau^-) < 4.1 \times 10^{-3}\,.
\end{align}
This limit is still about five orders of magnitude above the SM predictions.
Indirect constraints of ${\rm BR}(B_s \to \tau^+ \tau^-) < 3\%$  can be placed on the $B_s$ meson decay using the width difference in the neutral $B_s$ meson
system~\cite{Dighe:2010nj}.
A limit of ${\rm BR}(B_s \to \tau^+ \tau^-) < 5\%$ can also derived from data from the ALEPH experiment~\cite{Grossman:1996qj}.

\subsection{Experimental prospects}
The upgrade of the
LHCb experiment in 2019-2020 will allow the collaboration to collect a data sample of 50 fb$^{-1}$ in about five years of data taking~\cite{Bediaga:1443882}, while the upgrades foreseen for ATLAS and CMS experiments in 2023-2026 will allow the two collaborations to collect about 3000 fb$^{-1}$ each by 2035.
With such a huge data sample, ATLAS and CMS will measure the branching fraction of the $B_s \to \mu^+ \mu^-$ decay with an accuracy
of about 10\% and the ratio of the branching fractions ${\rm BR}(B^0\to \mu^+\mu^-)/\overline{{\rm BR}}(B_s \to \mu^+ \mu^-)$ with 
an accuracy of about 20\%~\cite{CMS:2015iha}.

A preliminary upper limit has been set by LHCb  on the $B_s \to \tau^+ \tau^-$ decay, and both LHCb and Belle II are expected to improve it in the next years.
With a data sample of 5 ab$^{-1}$ collected at the $\Upsilon(5S)$, Belle II should be able to 
set a limit of $2 \times 10^{-3}$ at 90\% CL \cite{BELLE2-NOTE-0021}.
The best upper limit on the $B^0 \to e^+ e^-$ decay from the B-factory experiments 
is ${\rm BR}(B^0 \to e^+ e^-) < 1.13 \times 10^{-7}$ at 
90 \% CL, obtained by BaBar~\cite{Aubert:2007hb} with 384$\times 10^6$ $B\overline{B}$ pairs which correspond roughly to  350 fb$^{-1}$ of integrated luminosity. With 50 ab$^{-1}$ collected by the end of 2024,  Belle II should be able to reach an upper limit on  ${\rm BR}(B^0 \to e^+ e^-)$ of $\sim 10^{-8}$ at 90 \% CL, 
assuming a background contamination similar to BaBar.

Finally, extrapolating from the LHCb measurement of the lifetime of $B_s \to K^+K^-$~\cite{Aaij:2014fia}, a meaningful 
constraint on $S$ and $P$ from the $B_s \to \mu^+ \mu^-$ lifetime will not be possible with the data collected by 
the ATLAS, CMS and LHCb experiments during Run\,2 of the LHC. More interesting constraints are possible towards 
the end of the LHC data-taking, when the uncertainty on the measured lifetime will be better than the difference 
between the lifetimes of the heavy- and light-mass eigenstates.

\section{Radiative decays}
\label{sec:radiative}

\subsection{Introduction}

The electromagnetic radiative decays based on the quark-level transition
$b \to s \gamma$ or $ b \to d \gamma$ play a crucial role in testing the
Standard Model.
At leading order in the SM, these processes are generated by the
electromagnetic dipole operator $(Q_7)_q$,
while the contribution of the chirality-flipped operator $(Q_7')_q$,
leading to an opposite final-state photon helicity,
is suppressed by $m_q/m_b$.
New physics can manifest itself by
modifying their corresponding Wilson coefficients, potentially breaking the universality between $b\to
s\gamma$ and $b\to d\gamma$ processes.
Apart from measuring the rates of the radiative decays to test the agreement
with the SM, it is of particular interest to probe the helicity of the $b\to
q\gamma$ transition and to look for possible new sources of $C\!P$ violation.

The radiative quark decay is probed in the following processes:
\begin{itemize}
 \item The \textit{inclusive} decays  $\bar B \to  X_s \gamma$ and $ \bar B \to  X_d
\gamma$, where $B (\overline{B})$ stands for $B^{0,+} (\overline{B}^{0,-})$, respectively,
and $X_s$ and $X_d$ are defined as any final state not containing
charmed hadrons and with
$s \overline{q}$ or  $d \overline{q}$ flavour quantum numbers, where $q$
is the spectator from the $\bar B$ meson. As discussed in
Sec.~\ref{sec:framework}, inclusive branching ratios are theoretically
exceptionally clean. In
addition to the branching fractions,  direct $C\!P$ asymmetries of these
inclusive decays are of interest.
\item The \textit{exclusive} decays $B^{+,0}\to K^*\gamma$ and
$B_s\to\phi\gamma$ probing the $b\to s\gamma$ transition as well as
$B^{+,0}\to\rho^{+,0} \gamma$, $B^{0}\to\omega\gamma$ and $B_s\to K^*\gamma$ probing
the $b\to d\gamma$ transition. These decays are theoretically challenging as
they require not only the knowledge of the heavy-to-light form factors at
zero momentum transfer (which is far away from the kinematical regime
accessible to lattice QCD), but also of additional hadronic matrix elements
that are not contained in the form factors. The exclusive radiative decays of
the neutral $B^0$ and $B_s$ mesons offer additional sensitivity to new physics
via time-dependent $C\!P$ asymmetries, giving access to $C\!P$ violation in the
interference between mixing and decay.
\item \textit{Exclusive} decays with a three-body hadronic final state such
as $B\to K_1(1400)(\to K\pi\pi)\gamma$. In these decays, angular correlations
between the decay products could be used to determine the photon helicity
\cite{Gronau:2001ng,Gronau:2002rz} (see also \cite{Kou:2010kn,Kou:2016iau}).
\item Baryonic decays of $\Lambda_b\to \Lambda^{(*)}\gamma$, where the photon
polarisation could be determined from the angular decay distribution, exploiting
the spin-$\frac{1}{2}$ nature of the $\Lambda_b$ baryon.
\end{itemize}
Complementary sensitivity to the $b\to q\gamma$ transition is also given by
the semi-leptonic decays discussed in Sec.~\ref{sec:semileptonic}.

\subsection{Inclusive radiative decays}
\label{ssec:inclusive_radiative}

\subsubsection{Standard Model predictions}
\label{sssec:inclusive_radiative:predictions}

The inclusive decays $\bar B\to X_q\gamma$ ($q=d,s$) are theoretically
appealing as the matrix elements can be computed in perturbation theory at
leading order in the heavy quark limit.
Schematically the rate can be written as
\begin{equation}
 \Gamma(\bar B\to X_q\gamma) = \Gamma(b\to X_q\gamma)\,,
 + \delta \Gamma_\text{non.}
\end{equation}
where the first term on the right-hand side is the perturbatively calculable rate
and the last term a non-perturbative contribution.
A theoretical challenge is the fact that experiments can only resolve photons
down to a minimum energy $E_0$.
While experiments at the $B$ factories involved measurements with $E_0 \geq 1.7\,{\rm GeV}$,
theory predictions are conventionally given at $E_0 = 1.6$ GeV where non-perturbative
uncertainties are under better control.
For $E_0=1.6$~GeV, the non-perturbative uncertainty on
the $\bar B\to X_s\gamma$ rate was estimated at 5\% \cite{Benzke:2010js}. For
$\bar B \to  X_d\gamma$, these uncertainties are even larger since $\lambda_u^{(d)}$ is
not Cabibbo-suppressed in the $b\to d$ transition, in contrast to the $b\to s$
transition \cite{Asatrian:2013raa}.

Concerning the perturbative part of the calculation, it amounts to performing the
matching of the Wilson coefficients at the electroweak scale, the
renormalisation group evolution down to the scale of the $b$ quark mass, and
the computation of the on-shell matrix elements of all contributing operators
at that scale. The first two steps have already been discussed in
Sec.~\ref{sec:framework}. The calculation of the matrix elements has seen
significant progress in recent years
\cite{Gorbahn:2005sa, Asatrian:2006rq,  Czakon:2006ss, Ewerth:2008nv, Boughezal:2007ny,
Misiak:2010tk, Misiak:2010sk,
Asatrian:2010rq, Ferroglia:2010xe, Kaminski:2012eb,
Huber:2014nna, Czakon:2015exa}, allowing an improved NNLO prediction of the
$\bar B \to X_s \gamma$ branching fraction for a
photon energy $E_{\gamma} > 1.6 $ GeV \cite{Misiak:2015xwa}:
\begin{equation}
\text{BR}(\bar B \to X_s \gamma)_\text{SM} = (3.36 \pm 0.23) \times 10^{-4}
\quad(E_\gamma > 1.6\,{\rm GeV})\,.
\label{eq: rad_inclusive}
\end{equation}
The total uncertainty is given by the sum in quadrature of non-perturbative
(5\%), higher-order (3\%), interpolation (3\%) and parametric (2\%) uncertainties.
For $\bar B \to  X_d\gamma$, taking the contribution of  $b\to du\bar u\gamma$ tree-level process into account, one arrives at the
SM prediction for $E_{\gamma} > 1.6$\,GeV~\cite{Misiak:2015xwa}
\begin{equation}
\text{BR}(\bar B \to X_d \gamma)_\text{SM} = \left(1.73^{+0.12}_{-0.22}\right) \times 10^{-5}
\quad(E_\gamma > 1.6\,{\rm GeV})\,.
\end{equation}

In addition to the branching fractions, new physics can also affect the direct
$C\!P$ asymmetry between $\bar{B}$ and $B$ decays, defined as
\begin{equation}
A_{C\!P}^{X_q\gamma} =
\frac{\Gamma(\bar B\to X_q\gamma)-\Gamma(B\to X_{\bar q}\gamma)}{\Gamma(\bar B\to X_q\gamma)+\Gamma(B\to X_{\bar q}\gamma)}
\label{eq: acp}
\end{equation}
The short-distance contribution to $A_{C\!P}^{X_s\gamma}$ is CKM suppressed
and therefore small in the SM,
while it is sizeable for $A_{C\!P}^{X_d\gamma}$
\cite{Kagan:1998bh, Ali:1998rr, Hurth:2003dk}. However, it was
pointed out in Ref.~\cite{Benzke:2010tq} that $A_{C\!P}^{X_q\gamma}$ are actually dominated by
a non-perturbative long-distance contribution that mar the sensitivity of $A_{C\!P}^{X_q\gamma}$ to new
physics. The SM predictions are estimated to lie in the ranges \cite{Benzke:2010tq}
\begin{align}
-0.6\%&<\left(A_{C\!P}^{X_s\gamma}\right)_\text{SM}<2.8\%\,, &
-62\%&<\left(A_{C\!P}^{X_d\gamma}\right)_\text{SM}<14\%\,.
\end{align}
While the SM predicts quite different asymmetries
for $\bar B \to  X_s \gamma$ and $\bar B \to  X_d \gamma$, the combined asymmetry
\begin{align}
A^{X_{s+d}\gamma}_{C\!P}  =
\frac{\left[\Gamma(\bar B\to X_s\gamma)+\Gamma(\bar B\to X_d\gamma)\right]-\left[\Gamma(B\to X_{\bar s}\gamma)+\Gamma(B\to X_{\bar d}\gamma)\right]}
{\left[\Gamma(\bar B\to X_s\gamma)+\Gamma(\bar B\to X_d\gamma)\right]+\left[\Gamma(B\to X_{\bar s}\gamma)+\Gamma(B\to X_{\bar d}\gamma)\right]}
\label{eq:acp_comb}
\end{align}
is ${\cal O}(10^{-6})$ in the SM, with nearly exact cancellation of the opposite sign asymmetries for
$\bar B \to  X_s\gamma$ and $\bar B \to  X_d \gamma$ decays.
The combined $C\!P$ asymmetry, $A_{C\!P}^{X_{s+d}\gamma}$, is also sensitive to different
new physics scenarios than $A_{C\!P}^{X_s \gamma}$ ~\cite{Hurth:2003dk}.
Thus measurements of this joint asymmetry complement
those of $A_{C\!P}^{X_s \gamma}$ to constrain new physics models.
In addition, the $C\!P$ asymmetry difference between the charged and neutral $B$ decay,
\begin{equation}
\Delta A_{C\!P}^{X_q\gamma} =
A_{C\!P}^{X_q^-\gamma}-A_{C\!P}^{X_q^0\gamma} \,,
\end{equation}
is expected to vanish in the SM and potentially receives a non-zero contribution
in the presence of non-standard $C\!P$ violation in $(C_7)_q$ or $(C_8)_q$.
Measurements of  $A^{X_{s+d}\gamma}_{C\!P}$ and $\Delta A_{C\!P}^{X_q\gamma} $ therefore constitute much cleaner test of the SM than measuring  $A_{C\!P}^{X_q\gamma}$ directly.

\subsubsection{Experimental aspects}

Measurements of the $\bar B \to  X_s \gamma$ branching fraction have been performed by the B-factory experiments using both inclusive approaches
and by summing together many exclusive modes.
The fully inclusive measurement is performed by detecting a high energy photon with $E_{\gamma}$ close to half of the $b$ quark mass.
Without at least partially reconstructing the $X_s$ or $X_d$ system it is not possible to separate the contributions from $\bar B \to  X_s \gamma$ and $\bar B \to  X_d \gamma$ decays. The latter contribution is subtracted by the B-factory experiments assuming%
\footnote{As mentioned above, this relation receives corrections from tree-level
contributions to $B\to X_d\gamma$ that are of order 10\% with a
sizeable uncertainty \cite{Asatrian:2013raa,Misiak:2015xwa}.}
\begin{equation}
\frac{{\rm BR}(\bar B \to  X_d \gamma) }{{\rm BR} (\bar B \to  X_s \gamma)  }
\propto |V_{td}/V_{ts}|^2 = 0.044 \pm 0.003.
\label{eq:vtd_vts}
\end{equation}
The photon energy is determined by a direct measurement of  a neutral cluster in the experiments electromagnetic calorimeter
and therefore depends on the calorimeter energy resolution.
The statistical uncertainty and the minimum $E_{\gamma}$ is driven by experimental backgrounds.
The Belle experiment uses $E_{\gamma}$ as low as $1.7\,{\rm GeV}$~\cite{Limosani:2009qg}. Below $1.7\,{\rm GeV}$, the signal is expected to be small and the background becomes very large.
The main backgrounds are from continuum processes of $e^+ e^- \to q \overline{q}$ or $\tau^+ \tau^-$, where $q=u,d,s,c$.
To control the level of the background, the event can be tagged as containing a $B\overline{B}$ pair by reconstructing a high-momentum electron or muon ({\it lepton tag})
from the semileptonic decay of the non-signal $B$.
By using the lepton tag approach it is possible to measure the $C\!P$-flavour tag for the combined $\bar B \to  X_{s+d} \gamma$ decays and thus can be used to measure the
direct $C\!P$ asymmetry.

In addition to the lepton-tag approach, BaBar~\cite{Aubert:2007my} has also used the {\it recoil-B} technique,
in which the signal ({\it recoil}) $B$ meson is tagged by fully reconstructing the non-signal $B$ meson in a hadronic
decay mode. This technique (along with the reconstruction in semileptonic decays) has been broadly used at the
B-factory experiments to study rare decays with multiple neutrinos, e.g., $B  \to  \tau \nu$ and $B \to  K \nu \overline{\nu}$.
Although this method is currently statistically limited, it is very promising for the future, i.e. at a high-luminosity B-factory operating at the $\Upsilon(4S)$.

The inclusive branching fraction can be also measured by reconstructing the $X_s$ as the sum of as many exclusive final states as possible.
In this case the photon energy in the $B$ rest frame is obtained using the mass of the $X_s$ system,
\begin{equation}
E_{\gamma} = \frac{m^2 _{B} - m^2_{X_S} }{ 2 m_B }
\end{equation}
to an accuracy that is much better than can be achieved by measuring the photon energy directly with the experiment's calorimeter.
This method is systematically limited by uncertainties in the $X_s$ hadronisation (which is simulated using JETSET~\cite{Sjostrand:1995iq}),
which influence both the efficiency for the selected decay modes and the estimate of the contribution to $\Gamma(\bar B \to  X_s\gamma)$ from the unmeasured modes.
Using this method the $b \to s \gamma$ and the $b \to d \gamma$ decays can be experimentally separated and the $b \to d \gamma$ transition measured.
This method also determines the  flavour and charge of the $b \to s \gamma$ decay, allowing measurements
of direct $C\!P$ and isospin asymmetries in inclusive $b \to s \gamma$ decays.

\begin{figure}[tbp]
\centering
\includegraphics[width=0.6\linewidth]{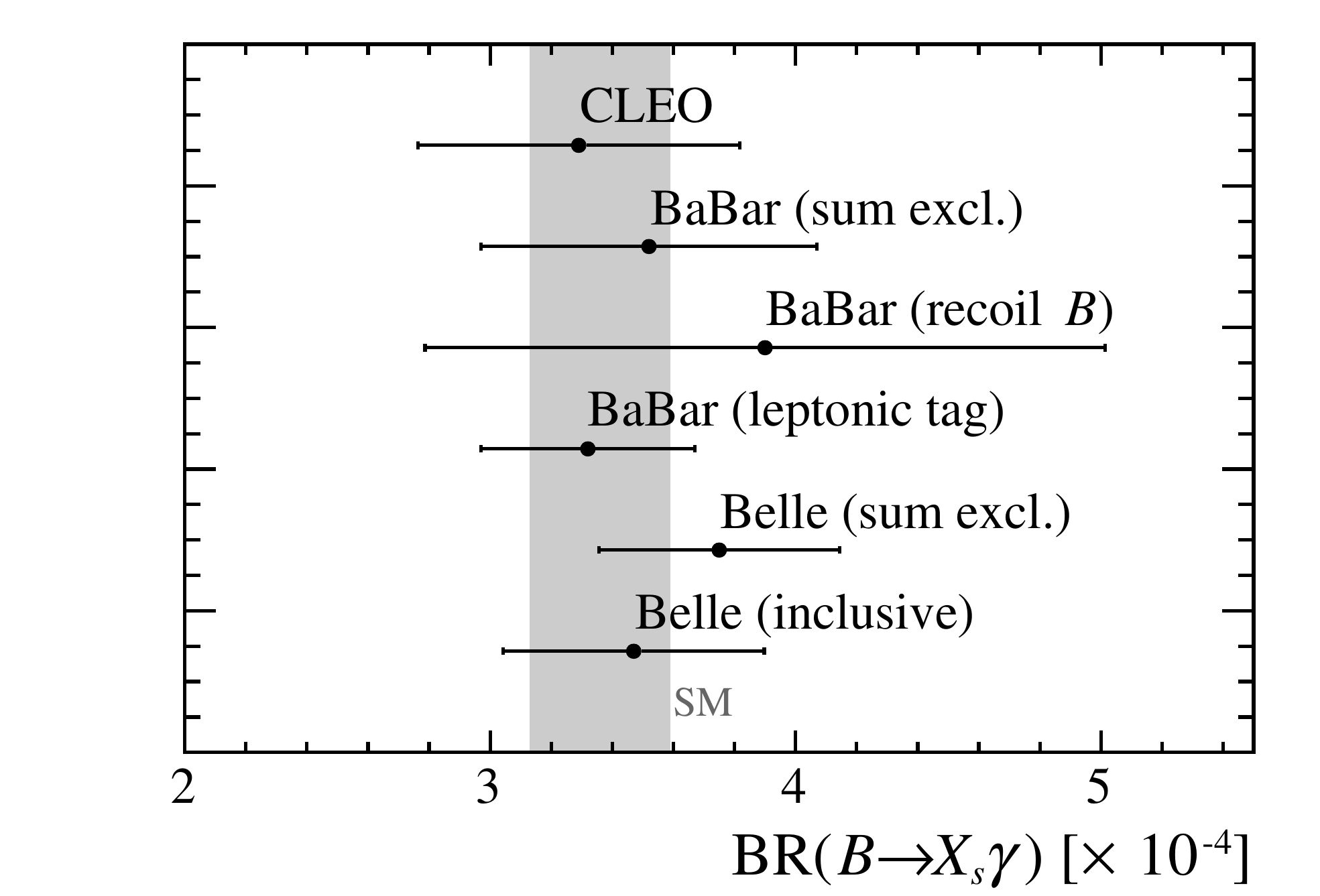}
\caption{
Inclusive branching fraction of $\bar B \to  X_s \gamma$ measured by BaBar, Belle and CLEO~\cite{Chen:2001fja}. The measurements have been extrapolated to $E_{\gamma} > 1.6\,{\rm GeV}$ from larger $E_{\rm T}$ cut values using the HFAG prescription (based on HQET)~\cite{Amhis:2014hma}. For BaBar three semi-independent measurements are included; summing a large number of exclusive decays as an approximation of the fully exclusive rate~\cite{Lees:2012wg}; a fully inclusive measurement of $X_s\gamma$ and $X_d\gamma$ after fully reconstructing the other $B$ meson in the event~\cite{Aubert:2007my}; and a fully inclusive measurement of $X_s\gamma$ and $X_d\gamma$ after using a high-$p_{\rm T}$ lepton to tag the event as a $B\bar{B}$ event~\cite{Lees:2012ym}. For Belle two measurements are presented; a fully inclusive approach with and without a high-$p_{\rm T}$ lepton tag~\cite{Limosani:2009qg}; and a sum over exclusive final states~\cite{Saito:2014das}.
\label{fig:btosgamma:inclusive}
}
\end{figure}

The latest measurements of the inclusive $\bar B \to  X_s \gamma$ branching fraction from the B-factory experiments\footnote{This summary describes the situation before the recent result from Belle collaboration~\cite{Belle:2016ufb}.} are summarised in Fig.~\ref{fig:btosgamma:inclusive}.
The measurements have $E_{\gamma}$ thresholds ranging from 1.7 to over 2.0\,GeV, while theoretical predictions are usually made
with a minimum $E_{\gamma}$ of 1.6\,GeV.
The approach adopted by HFAG and the PDG is to perform the extrapolation of
the experimental results down to 1.6\,GeV from the lowest measured experimental threshold using HQET.

The untagged and lepton tagged fully inclusive approaches give by far the best accuracy on the branching fraction
above any given energy threshold.
The quoted world average of extrapolated values\footnote{See recent update of the HFAG averages for the ICHEP2016 conference, http://www.slac.stanford.edu/xorg/hfag/.} is % of extrapolated values is~\cite{Amhis:2014hma}
% ----
\begin{equation}
{\rm BR}(\bar B \to  X_s \gamma) = (3.43 \pm 0.19) \times 10^{-4} \;\;\; (E_{\gamma} > 1.6 \;{\rm GeV})\,,
\end{equation}
%
%where the first uncertainty is the combination of the statistical and systematic uncertainties, and the second uncertainty
%depends on the model used in the extrapolation.
The world-average extrapolated branching fraction is in excellent agreement with theoretical predictions
and has a comparable uncertainty.

The tag of the other $B$ of the event (via lepton tag or recoil method) allows to know the flavour of the signal $B$ and therefore to measure the
$C\!P$ asymmetry of the inclusive $\bar B \to  X_{s+d} \gamma$ processes. Although the SM predicts quite different asymmetries
for $\bar B \to  X_s \gamma$ and $\bar B \to  X_d \gamma$, the $X_s$ and $X_d$ final states
cannot be distinguished in the fully inclusive technique and only the combination can be measured. As discussed in Sec.~\ref{sssec:inclusive_radiative:predictions}, the asymmetry of the combination is expected to be ${\cal O}(10^{-6})$ in the SM.
The results obtained by BaBar for two different tagging methods are
\begin{align}
A_{C\!P}^{X_{s+d}\gamma} &= 0.06 \pm 0.06_{\rm stat} \pm 0.02_{\rm syst} &&\text{(lepton tag \cite{Lees:2012ufa})},\\
A_{C\!P}^{X_{s+d}\gamma} &= 0.10 \pm 0.18_{\rm stat}  \pm 0.05_{\rm syst}&&\text{(recoil technique \cite{Aubert:2007my})},
\end{align}
respectively.

The only method which allows to measure the $A_{C\!P}^{X_s \gamma}$ is the sum of the exclusive modes in the final state.
Results exist from CLEO~\cite{Coan:2000pu}, Belle~\cite{Nishida:2003paa} and BaBar~\cite{Aubert:2008be} and are reported in Table~\ref{tab:rad_inc_acp}.
The results are in good agreement with theoretical predictions.

\begin{table}[tbp]
\centering
\renewcommand{\arraystretch}{1.2}
\begin{tabular}{cc}
\hline
$A_{C\!P}^{X_s \gamma}$ & Experiment \\ \hline
$-0.011 \pm 0.030_{\rm stat} \pm 0.014_{\rm syst}$ & BaBar~\cite{Aubert:2008be} \\
$\phantom{-}0.002 \pm 0.050_{\rm stat} \pm 0.030_{\rm syst}$ & Belle~\cite{Nishida:2003paa} \\
$-0.079 \pm 0.108_{\rm stat} \pm 0.022_{\rm syst} $ & CLEO~\cite{Coan:2000pu} \\
\hline
\end{tabular}
\caption{Measurements of the direct $C\!P$ asymmetry in $\bar B\to X_s\gamma$ by CLEO, BaBar and Belle.}
\label{tab:rad_inc_acp}
\end{table}

The importance of a measurement of $|V_{td}/V_{ts}|$ has been outlined in Sec.~\ref{sec:VtdVts}.
The measurement of the inclusive $\bar B \to  X_d \gamma$ process allows us to extract the ratio $|V_{td}/V_{ts}|$ with a better sensitivity
with respect to the exclusive modes $B \to \rho \gamma$ and $B \to K^* \gamma$ that are limited by
the theoretical uncertainty of the form factors.
With the sum-of-exclusive mode approach it is possible to
discriminate between $\bar B \to  X_d \gamma$ and $\bar B \to  X_s \gamma$ final states by identifying a kaon amongst the decay products of the $X_{s,d}$ system.
The BaBar collaboration has studied the ratio of ${\rm BR}(\bar B \to  X_d \gamma)/ {\rm BR}(\bar B \to  X_s \gamma)$ using the sum of seven
exclusive final states with masses of up to $2.0\,{\rm GeV}/c^{2}$~\cite{delAmoSanchez:2010ae}.
After correcting for unobserved decay modes, they obtain a ratio of branching fractions
${\rm BR}(\bar B \to  X_d \gamma)/ {\rm BR}(\bar B \to  X_s \gamma) = 0.040 \pm 0.009_{\rm stat} \pm 0.010_{\rm syst}$ from which they determine
$|V_{td}/V_{ts}| = 0.199 \pm 0.022_{\rm stat} \pm 0.024_{\rm syst} \pm 0.002_{\rm th}$\footnote{We note that the theory uncertainties due to the
tree-level contribution to $B\to X_d\gamma$ mentioned above have not
been taken into account in this determination.}.
The measured ratio is completely consistent with the value of $|V_{td} / V_{ts}|$ derived from $B_{(s)}^0$-$\overline{B}_{(s)}^0$ mixing.

\subsection{Exclusive radiative decays}\label{sec:rad-excl}

In exclusive radiative decays the hadronic system is fully reconstructed.
The SM predictions for the rate of exclusive processes tend to come with larger uncertainties but by fully reconstructing the final state there are additional observables that can be measured.
The observables accessible in $B\to V\gamma$ decays, where $V$ is a vector meson, include:
branching fractions; isospin asymmetries, defined as
\begin{equation}
A_I^{V\gamma} = \frac{c_V^2 \Gamma(\bar B^0\to V\gamma) - \Gamma(B^-\to V\gamma)}{c_V^2 \Gamma(\bar B^0\to V\gamma) + \Gamma(B^-\to V\gamma)}
\end{equation}
where $c_{K^*} = 1$ and $c_\rho = c_\omega = \sqrt{2}$;
and $C\!P$ asymmetries.
The $C\!P$ asymmetries are particularly interesting in the case of neutral $B^0$ and
$B_s$ decays due to the interference with mixing.
A similar set of observables exist for $B \to A \gamma$ decays, where $A$ is an axial-vector meson.
Exclusive decays can also be used to test the polarisation of photons produced in the $b \to q\gamma$ transition.
In the SM the polarisation of the photons is almost entirely
left-hand polarised due to the chiral nature of the weak charged current
interaction.
The right-handed component is suppressed by the ratio of quark masses
$m_s/m_b$ at leading order in the heavy quark limit
but also receives contributions of order $\Lambda_\text{QCD}/m_b$ \cite{Grinstein:2004uu}.
There are several methods that can be used to test the photon polarisation, these include:
\begin{itemize}
\item time dependent interference between neutral $B$ and $\bar{B}$ mesons
decaying to a common final-state;
\item so-called up-down asymmetries in $B \to K \pi \pi \gamma$ decays;
\item the production polarisation of $\Lambda_b$ baryons in $\Lambda_b \to
\Lambda^{(*)}\gamma$ decays;
\item and the angular distribution of $B \to V \ell^+\ell^-$ decays at low
$\ell^+\ell^-$ masses (discussed in Sec.~\ref{sec:semileptonic}).
\end{itemize}

\subsubsection{Standard Model predictions}

The computation of observables in exclusive $B\to V\gamma$ decays requires the determination of:
\begin{itemize}
 \item the Wilson coefficients $C_{1}$ through $C_8$ at the
 scale of the $b$ quark mass (see Sec.~\ref{sec:framework});
 \item the $B\to V$ tensor form factors at $q^2=0$;
 \item and ``non-factorisable'' hadronic effects that do not correspond to form
 factors.
\end{itemize}

Concerning the form factors, the kinematical constraint $q^2=0$ corresponds to
maximal hadronic recoil, which makes it hard to simulate in lattice QCD, which is
most effective at low hadronic recoil. QCD sum rules on the light cone (LCSR)
exploit the fact that the energy of the vector meson is largest for $q^2=0$ and
is thus best suited for determining the relevant form factors.
State-of-the-art numerical results for the $B\to V$ form factors from LCSR
are given in Ref.~\cite{Straub:2015ica} as an update of Ref.~\cite{Ball:2004rg}.
This calculation uses a sum rule with an interpolating current for the $B$
meson and a light-cone distribution amplitude for the vector meson.
Alternatively, LCSR can be formulated with $B$ meson light-cone distribution
amplitudes and an interpolating current for the vector meson
\cite{Khodjamirian:2006st}; currently, this results in larger uncertainties due to
the limited experimental knowledge on the $B$ meson light-cone distribution amplitudes.

An understanding of the ``non-factorisable'' effects is crucial as it determines
the size of the strong phases as well as the size of isospin-breaking effects.
They have been calculated within the QCD factorisation framework
\cite{Bosch:2001gv,Beneke:2001at,Beneke:2004dp}.
Partial results exist within LCSR
\cite{Ball:2006eu,Khodjamirian:2010vf,Dimou:2012un,Lyon:2013gba}
that go beyond the
heavy quark limit and avoid endpoint divergences encountered in QCD
factorisation \cite{Kagan:2001zk}.
Using a hybrid QCDF/LCSR approach, the following SM predictions for the isospin
asymmetries have been obtained\footnote{Strictly speaking, these predictions refer
to the \textit{CP-averaged} isospin asymmetries. In the SM, this difference is
negligible.} \cite{Lyon:2013gba}
\begin{align}
\begin{split}
 A_I^{K^*\gamma} &= (4.9 \pm 2.6)\% \,,\\
 A_I^{\rho\gamma} &= (5.2 \pm 2.8)\% \,.
 \end{split}
\end{align}

\subsubsection{Time dependent analyses}

\begin{figure}
\centering
\includegraphics[width=0.5\linewidth]{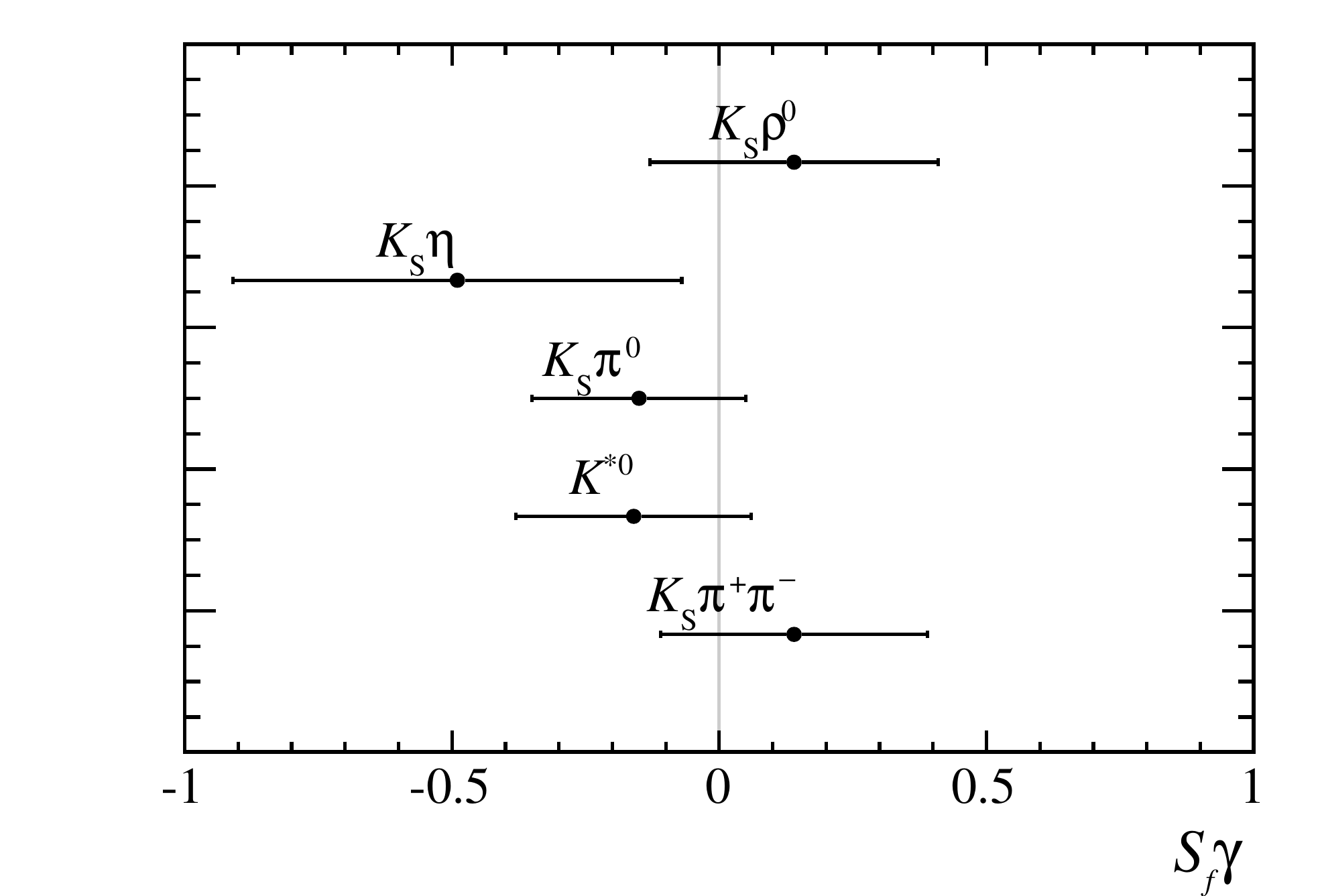}
\caption{
Time dependent $C\!P$ asymmetry of $B \to f \gamma$ decays. This asymmetry is non-zero only if the photon from the $b \to s \gamma$ transition is not pure left-hand polarised.
The data-points are averages from  the HFAG collaboration~\cite{Amhis:2014hma}.
Note, the $K_{\rm S}\pi^0$ and $K^{*0}$ data points are highly correlated as are the $K_{\rm S} \pi^+\pi^-$ and the $K_{\rm S} \rho^0$ data points.
\label{fig:radiative:timedependent}
}
\end{figure}

If a neutral $B$ and $\overline{B}$ meson decay to a final state with a common hadronic system $f$, the time dependent $C\!P$ asymmetry for the system can be written~\cite{Atwood:1997zr,Atwood:2004jj}
\begin{align}
{\cal A}(B_q \to f\gamma) = \frac{S_f \sin \left( \Delta M_q t \right) - C_f \cos \left( \Delta M_q t \right) }{\cosh \left( {\Delta\Gamma_{q}} t/2 \right) - {\cal A}_{\Delta\Gamma,f} \sinh \left( {\Delta\Gamma_{q}}t/2 \right) }~.
\end{align}
Here, $\Delta M_q$ and $\Delta \Gamma_{q}$ are the mass and width difference between the heavy and light mass-eigenstates of the $B^0_{q}$-$\overline{B}^0_{q}$ system,
with $q=d$ or $s$.
For the $B^0$ system, $\Delta \Gamma_{d} \approx 0$ such that
\begin{align}
{\cal A}(B_q \to f\gamma) =  S_f \sin \left( \Delta M_q t \right) - C_f \cos \left( \Delta M_q t \right)~.
\end{align}
The term $S_f$ provides sensitivity to the photon polarisation,
\begin{align}
S_f \propto \sin (2 \psi) \sin (2\beta)~,
\end{align}
where $\beta \equiv {\rm arg}(-V_{cd} V_{cb}^* / V_{td} V_{tb}^* )$ and $\tan\psi$ is the ratio of the amplitudes for the left- and right-handed polarisations.
The angle $\beta$ is one of the angles of the CKM triangle and is measured to be $\sin (2\beta) = 0.679 \pm 0.020$.
In order to determine $S_f$ it is necessary to determine the flavour of the $B$ meson at production using flavour tagging techniques.
This time-dependent technique has been exploited by the BaBar and Belle experiments to determine $S_{K^* \gamma}$ (using $K^* \to K^0 \pi^0$)~\cite{Aubert:2008gy,Ushiroda:2006fi}, $S_{K_S \eta \gamma}$~\cite{Aubert:2008js} and $S_{K_S \rho^0\gamma}$~\cite{Li:2008qma,Sanchez:2015pxu}.
A summary of the results is presented in Fig.~\ref{fig:radiative:timedependent}.
The different measurements are all consistent with the SM expectation which is expected to be ${\cal O}(0.01)$.
The uncertainty on the $S_{K^*\gamma}$ measurement from BaBar and from Belle is $\sim 0.20$.
This type of measurement is challenging for the LHC experiments, where it is much more difficult to tag the initial flavour of the $B$ meson.
Furthermore, the efficiency to reconstruct $K_{\rm S}$ mesons is typically much smaller at the LHC than the B-factory experiments.

In the $B_{s}^0$--$\overline{B}_{s}^0$ system, $\Delta \Gamma_s$ is non-zero and there is also sensitivity to the photon polarisation through the ${\cal A}_{\Delta \Gamma}$ parameter.
This has a nice experimental advantage that it can be measured through the effective lifetime of the system without the need for flavour-tagging.
The parameter
\begin{align}
{\cal A}_{\Delta\Gamma} \propto \sin(2\psi) \cos(2 \beta_s)
\end{align}
is also expected to be close to zero due to the dependence on $\sin(2\psi)$.
The most promising channel to measure ${\cal A}_{\Delta\Gamma}$ is through the decay $B_s \to \phi\gamma$~\cite{Muheim:2008vu}.
The LHCb experiment has recently measured~\cite{Aaij:2016ofv}
\begin{align}
{\cal A}_{\Delta\Gamma} = -0.98 {}^{+0.46}_{-0.52} {}^{+0.23}_{-0.20}\,,
\end{align}
which is consistent with the SM expectation of $\sim 0.05$ at the level of two standard deviations.

\subsubsection{Up-down asymmetry in \texorpdfstring{$B \to K \pi \pi \gamma$}{B->Kππγ} decays}

The photon polarisation can also be determined from the photon direction in $B \to K \pi \pi \gamma$ decays.
The basic idea is to compute the angle between the direction of the photon and the plane containing the pions in the $K\pi\pi$ rest-frame.
This quantity is odd under parity and the average value of this triple product has one sign if the photon is left-hand polarised and a different sign if the photon is right-hand polarised.
The up-down asymmetry, $A_{\rm u,d}$, is proportional to the photon polarisation
\begin{align}
\lambda_{\gamma} = \frac{|C_7|^2 - |C'_7|^2}{|C_7|^2 + |C'_7|^2}~.
\end{align}
An unpolarised photon would have no preferred direction  and zero up-down asymmetry.

\begin{figure}
\begin{center}
\includegraphics[width=0.65\linewidth]{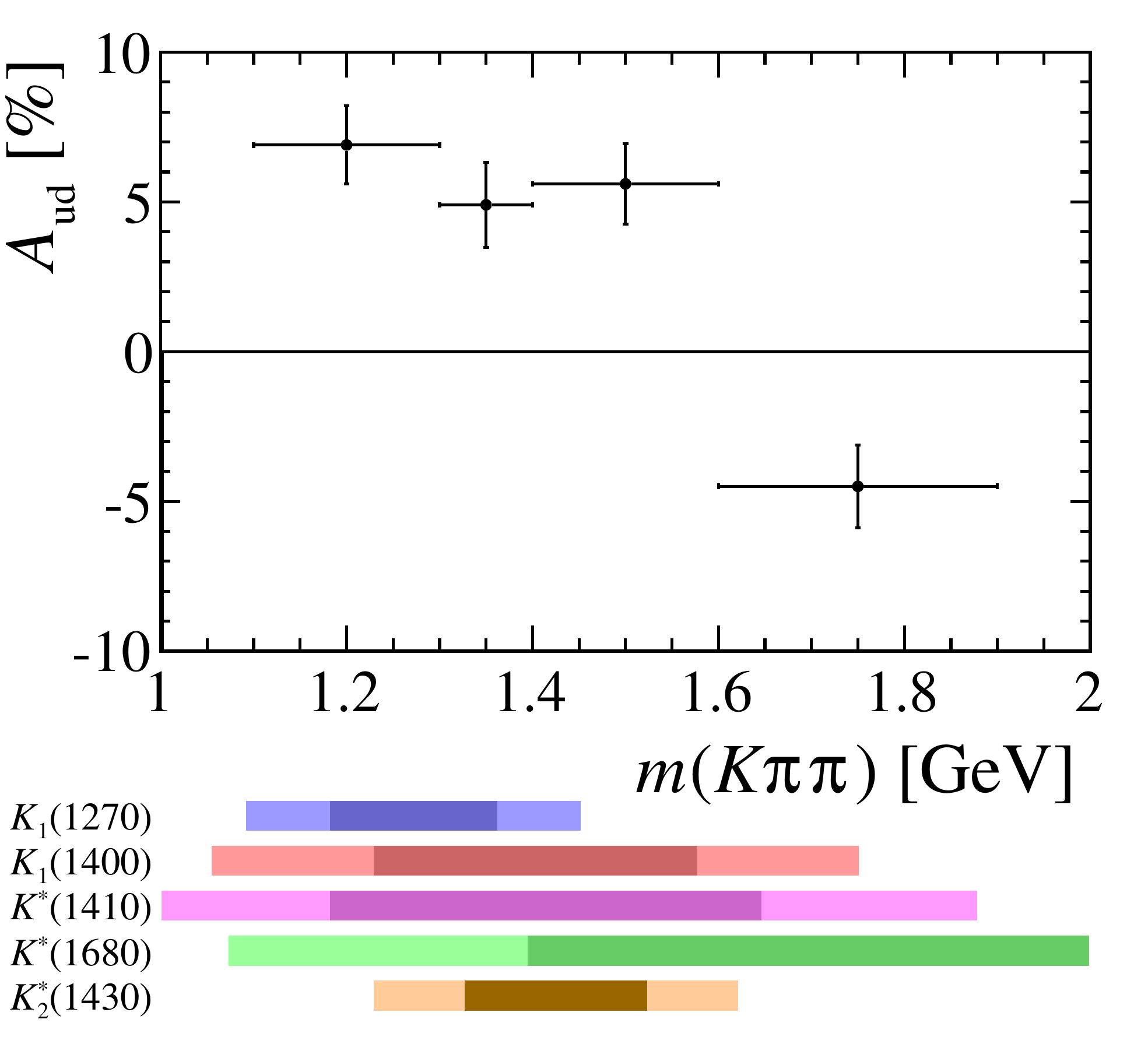}
\end{center}
\caption{
Up-down asymmetry of photons produced in $B \to K\pi\pi\gamma$ system with respect to the decay plane of the $\pi\pi$ pair measured by the LHCb experiment in Ref.~\cite{Aaij:2014wgo}.
The $K\pi\pi$ mass of the $K_1$ and $K^*$ resonances contributing to the up-down asymmetry is indicated by the shaded regions. These correspond to windows of one and two natural widths about the pole mass of the resonance.
\label{fig:radiative:updown}
}
\end{figure}

The up-down asymmetry has been measured by the LHCb experiment in four different regions of the mass of the $K\pi\pi$ system~\cite{Aaij:2014wgo}.
The result of this analysis is shown in Fig.~\ref{fig:radiative:updown}.
The measurement is inconsistent with zero polarisation at more than five standard deviations.
However, from Fig.~\ref{fig:radiative:updown} it is evident that the polarisation has a different sign in different mass regions.
The value of the up-down asymmetry and its sign depend on the hadronic resonances contributing to the $K\pi\pi$ system.
The prediction for the $K_{1}(1400)\gamma$ decay is $0.33\pm 0.05$~\cite{Gronau:2002rz}.
This prediction is valid only for a pure $K_{1}(1400)\gamma$ decay, while in Fig.~\ref{fig:radiative:updown}
the region around 1400 MeV contains contributions also from other measurements.
A recent measurement by the BaBar experiment provides additional information on the composition of the $K\pi\pi$ system that can be used to help determine $\lambda_{\gamma}$~\cite{Sanchez:2015pxu}.
In the first  $K\pi\pi$ mass region, the data predominantly comprises the $K_1(1270)$ meson.
At larger masses, the data comprises a mixture of the two $K_{1}$ states, the $K^*(1410)$ and the $K^*(1680)$.

\subsubsection{Baryonic decays}

The photon polarisation can also be determined from the angular distribution of radiative $\Lambda_b$ decays, exploiting the spin-$\tfrac{1}{2}$ nature of the $\Lambda_b$ baryon.
The final state of $\Lambda_b \to \Lambda^{(*)} \gamma$ decays can be described by two angles.
The angle $\theta_\gamma$ denotes the angle between the direction of the photon and the $\Lambda_b$-spin.
The angle $\theta_p$ denotes the angle between the proton from the $\Lambda^{(*)}$ baryon and the $\Lambda^{(*)}$ direction in the $\Lambda^{(*)}$ rest-frame.
The differential decay for $\Lambda_b \to \Lambda(1115)\gamma$, where the spin-$\tfrac{1}{2}$ $\Lambda(1115)$ decays weakly, can be described by~\cite{Hiller:2001zj,Legger:2006cq}
\begin{align}
\frac{{\rm d}\Gamma}{{\rm d}\cos\theta_{\gamma}} \propto 1 - \lambda_{\gamma} P_{\Lambda_b} \cos\theta_\gamma \quad,\quad  \frac{{\rm d}\Gamma}{{\rm d}\cos\theta_{p}} \propto 1 - \alpha_{p,1/2} \cos \theta_p \,.
\end{align}
Here, $\alpha_{p,1/2}$ is the well known $\Lambda$ asymmetry parameter, $\alpha_{p,1/2} = 0.642\pm 0.013$, and $P_{\Lambda_b}$ is the production polarisation of the $\Lambda_b$ baryon.
The situation is different for the strongly decaying $\Lambda(1520)$ baryon, where
\begin{align}
\frac{{\rm d}\Gamma}{{\rm d}\cos\theta_{\gamma}} \propto 1 - \alpha_{\gamma,3/2} P_{\Lambda_b} \cos\theta_\gamma \quad,\quad  \frac{{\rm d}\Gamma}{{\rm d}\cos\theta_{p}} \propto 1 - \alpha_{p,3/2} \cos^2\theta_p \,.
\end{align}
In this case, $\alpha_{\gamma,3/2}$ is proportional to $\lambda_{\gamma}$ but is diluted by the mixture of different $\Lambda$ helicity states.
Unfortunately the production polarisation of $\Lambda_b$ baryons in LHC collisions is now known to be small, $P_{\Lambda_b} = 0.06 \pm 0.07 \pm 0.02$~\cite{Aaij:2013oxa}.
The decay involving the $\Lambda(1115)$ is also experimentally challenging due to the long-lifetime of the $\Lambda$, which has $c\tau = 7.8$\,cm. In the laboratory frame, a sizeable fraction of the $\Lambda$ baryons decay outside the acceptance of the tracking systems of the LHC experiments and cannot be reconstructed.
No measurement of the rate of $\Lambda^{(*)}\gamma$ decays exists.
The large $\Lambda_b$ baryon production rates at the LHC should enable these measurements to be performed by the LHCb experiment with its LHC Run\,2 dataset.

\subsection{Experimental prospects}

With regards to inclusive $\bar B \to  X_s \gamma$ decays, the Belle II experiment will be able to measure ${\rm BR}(\bar B \to  X_s \gamma)$ to $\sim 6\%$. This is a conservative estimate of the achievable uncertainty based on the systematic uncertainty on
the current Belle branching fraction measurement at $E_{\gamma} > 1.7\,{\rm GeV}$~\cite{Aushev:2010bq}.
The recoil-tag method has an efficiency of $\sim 0.3\%$ but the
larger Belle II dataset should enable a competitive measurement of BR$(\overline{B} \to X_s \gamma)$
to be performed using this method with a statistical precision close to Belle's current measurement.
This measurement would have the same uncertainty as current theoretical predictions.
Time dependent $C\!P$ violation measurement are not statistically limited and the Belle II experiment should ultimately reach a precision of $0.05$ on $S_{K^* \gamma}$ with its full dataset of $50\,{\rm ab}^{-1}$ of integrated luminosity~\cite{Aushev:2010bq}. The Belle II experiment will also reach a precision of $0.15$ on measurements of $S_{\rho \gamma}$ from $b \to d$ radiative transitions.

In the $B_s$ system, an upgrade of the LHCb experiment will be able to measure a SM-like value of $A_{\Delta \Gamma}$ in $B_s \to \phi\gamma$ decays to a precision of around  0.03 with a dataset of $50\,{\rm fb}^{-1}$ of integrated luminosity~\cite{LHCb:2011dta}.
The LHCb experiment will also be able to make the first tests of the photon polarisation using baryonic $b$-hadron decays with its Run\,2 dataset.

\section{Semileptonic decays}
\label{sec:semileptonic}

Semi-leptonic rare decays based on the $b \to q \ell^+ \ell^-$ transition,
where $\ell$  is an electron or
a muon, provide important insight into the SM structure and are sensitive
to new physics contributions to the operators $Q_7^{(\prime)}$, $Q_9^{(\prime)}$ and $Q_{10}^{(\prime)}$.
Compared to leptonic or radiative decays, which are only sensitive to effects in $Q_{10}^{(\prime)}$ or $Q_{7}^{(\prime)}$, respectively, they are sensitive to a more
diverse range of new physics effects.
Both inclusive and exclusive semileptonic decays have the virtue that they
give access to angular observables sensitive to physics beyond the SM.
The large number of such observables and their complementary dependences on
Wilson coefficients make semileptonic decays indispensable in global analyses
of new physics in rare $B$ decays.

However, theoretically, the hadronic uncertainties are more challenging as the
lepton pair can also originate from a photon
originating from flavour-conserving QED vertices.
This is particularly evident when the dilepton invariant mass
squared $q^2$ is close to the mass of the charmonium resonances $J\!/\!\psi(1S)$
and $\psi(2S)$, when the rate
is enhanced by orders of magnitude.
In addition, \textit{exclusive} decays $B\to M\ell^+\ell^-$ to a meson $M$ require the
knowledge of the $B\to M$ form factors in the full kinematic range
$4m_\ell^2 < q^2 < (m_B-m_M)^2$.
Experimentally, one challenge is represented by the small branching fractions
of ${\cal O}(10^{-6})$ for $b \to s \ell^+ \ell^-$ decays and $10^{-8}$
for $b \to d \ell^+ \ell^-$ decays, since the rate of the decays is suppressed relative to $b \to q \gamma$ decays
by an additional factor of $\alpha_{\rm em}$.

\subsection{Inclusive decays}
\label{sec:semileptonic:inclusive}

The inclusive semi-leptonic modes $B \to X_q \ell^+  \ell^-$ are not limited by  theoretical uncertainties
on the hadronic form-factors for the $B \to M$ transitions of exclusive $B \to  M \ell^+ \ell^-$ decay modes. They are, therefore, predicted
with higher accuracy than the exclusive decay modes.

\subsubsection{Observables in inclusive semi-leptonic decays}

The angular distribution of  inclusive $B \to X_q \ell^+ \ell^-$ decays can be described by three independent observables,
$H_{\rm T}$ , $H_{\rm A}$ and $H_{\rm L}$, from which the short-distance
Wilson coefficients can be determined~\cite{Lee:2006gs}.
The observables $H_{\rm T}$ and $H_{\rm L}$ correspond to the transverse and longitudinal polarisation of the hadronic system. The observable $H_{\rm A}$ generates a forward-backward asymmetry in the dilepton system, $A_{\rm FB} = \tfrac{3}{4} H_{\rm A}(q^2)$.
This forward-backward asymmetry is a feature of rare $b \to s\ell^+\ell^-$ decays and arises  because there are contributions from both vector ($\bar{\ell}\gamma^{\mu}\ell$) and axial-vector ($\bar{\ell}\gamma^{\mu}\gamma_5\ell$) currents in the dilepton system.

In the absence of QED corrections, the double differential decay rate as a function of the dilepton invariant mass squared, $q^2$, and $\cos\theta_{\ell}$,
where $\theta_{\ell}$ is the angle between the $\ell^-$ and the $B^0$ (or $B^+$) meson momentum vectors
in the $\ell^+ \ell^-$ centre-of-mass frame, is:
\begin{equation}
\frac{ {\rm d}^2 \Gamma}{{\rm d} q^2 \,{\rm d} \cos\theta_{\ell}} = \frac{3}{8} \left[(1+ \cos^2\theta_{\ell}) H_{\rm T}(q^2) + 2 H_{\rm A} (q^2) \cos\theta_{\ell}  + 2 (1-\cos^2\theta_{\ell}) H_{\rm L}(q^2) \right]~.
\label{dg_over_dqdz}
\end{equation}
Here, the functions $H_i (q^2)$ contain the $q^2$ dependence of the rate and are independent of $\cos\theta_{\ell}$.
Integrating over $\cos\theta_{\ell}$ yields
\begin{align}
\frac{{\rm d}\Gamma}{{\rm d}q^2}  = H_{\rm T}(q^2) + H_{\rm L}(q^2)~.
\label{eq:dgamma-inclusive}
\end{align}

In the presence of QED corrections, an important difference with respect to the
\textit{exclusive} semi-leptonic decays regards the treatment of final-state
photon radiation and the definition of the dilepton
invariant mass squared $q^2$. In exclusive decays (including the leptonic decays),
the dilepton system is usually defined to be \textit{fully inclusive} of (soft)
bremsstrahlung, while direct (hard) emission can be suppressed experimentally
by cuts on the invariant masses of the initial and final state mesons.
In the \textit{inclusive} case, the dilepton system is usually defined without
any final state photon radiation, which is instead included in the hadronic
invariant mass $m_{X_q}$. In contrast to the exclusive decays, this invariant mass
is an independent kinematical variable.
As a consequence of final-state photon emission, the angular distribution
is distorted and \eqref{dg_over_dqdz} has to be modified to
include terms of higher powers in $\cos\theta_{\ell}$ \cite{Huber:2015sra}.

\subsubsection{Standard Model predictions}

\noindent
As sketched in the introduction,
the experimentally measured decay rate is related to the partonic decay through
\begin{equation}
\Gamma(B\to X_q \ell^+\ell^-) = \Gamma(b\to q\ell^+\ell^-)
+ O\left(\frac{\Lambda_\text{QCD}}{m_{b,c}}\right) +\ldots\,.
\label{eq:inclusivebxll}
\end{equation}
The perturbative calculation of the partonic decay rate $\Gamma(b \to q\ell^+\ell^-)$
requires
the Wilson coefficients discussed already in Sec.~\ref{sec:framework},
but also the calculation of the matrix elements of the contributing operators.
The state of the art is NNLL (next-to-next-to-leading logarithmic) accuracy for
the the decay rate, which corresponds to one-loop matrix elements for the operators
$Q_{7\ldots10}$ and two-loop matrix elements for $Q_{1,2}$
\cite{Asatrian:2001de,Asatryan:2001zw,Asatryan:2002iy,Ghinculov:2002pe,Asatrian:2002va,Ghinculov:2003qd,Greub:2008cy}.
Power corrections to the heavy quark limit including powers $1/m_b^2$, $1/m_b^3$,
and $1/m_c^2$,
corresponding to the last term in \eqref{eq:inclusivebxll},
have been computed in \cite{Falk:1993dh,Ali:1996bm,Chen:1997dj,Buchalla:1997ky,Buchalla:1998mt,
Bauer:1999kf,Ligeti:2007sn}.
Additional non-perturbative corrections have been estimated to be around 5\%
in Ref.~\cite{Huber:2015sra}.

In the experimental analyses of  the $B\to X_s\ell^+\ell^-$ decay, a cut on the hadronic
invariant mass $m_{X_q}<m_{X_q}^\text{cut}$ is imposed.
This affects in particular the low-$q^2$ region, leading to a significant suppression
of the rate, but also introducing additional uncertainty
\cite{Lee:2005pk,Lee:2005pwa,Lee:2008xc,Bernlochner:2011di,Bell:2010mg}.
If only a restricted range in $q^2$ is considered, collinear photon emission
leads to numerically important logarithmic QED corrections
\cite{Huber:2005ig,Huber:2008ak,Huber:2015sra}.

Including the current state of the art, the SM branching fractions
\begin{align}
\begin{split}
\text{BR}(B\to X_s e^+e^-)_\text{SM} &= (1.67\pm 0.10)\times 10^{-6} \quad(q^2 \in [1,6]\,{\rm GeV^2})\,,\\
\text{BR}(B\to X_s e^+e^-)_\text{SM} &= (2.20\pm 0.70)\times 10^{-7} \quad(q^2 > 14.4\,{\rm GeV^2})\,,\\
\text{BR}(B\to X_s\mu^+\mu^-)_\text{SM} &= (1.62\pm 0.09)\times 10^{-6} \quad(q^2 \in [1,6]\,{\rm GeV^2})\,,\\
\text{BR}(B\to X_s\mu^+\mu^-)_\text{SM} &= (2.53\pm 0.70)\times 10^{-7} \quad(q^2 > 14.4\,{\rm GeV^2})\,
\end{split}
\end{align}
have been obtained, along with predictions for the angular observables and
additional bins, in Ref.~\cite{Huber:2015sra}. The uncertainties
of the high-$q^2$ predictions can be reduced by normalizing the branching
fraction to the $B\to X_u\ell\nu$ decay \cite{Ligeti:2007sn,Huber:2015sra}.

The inclusive decay $B\to X_d\ell^+\ell^-$ as well as inclusive decays with tau
leptons in the final state have not been studied to the same level of detail
as $B\to X_s e^+ e^-$ and $B \to X_s \mu^+ \mu^-$ decays.

\subsubsection{Experimental aspects}

The inclusive $B \to X_s \ell^+  \ell^-$ processes have been measured at BaBar and Belle
following the sum-of-exclusive states method, summing up as many exclusive final states as possible.
In a similar fashion to the $B \to X_s \gamma$  analysis, the $X_s$ state is reconstructed as one kaon plus a number of additional pions.
In practice measurements are limited to at most two pions.
The presence of large background from semileptonic $B$ decays prevents to use in this case a fully inclusive approach.
Backgrounds from $B \to D (\to X_s \ell \nu) \ell \nu$ decays are particularly problematic at the B-factory experiments because the $B$ and $D$ vertices are not easily resolved.
The {\it reconstructed-B-tag} (recoil) approach has not been pursued because millions of $B$ tags are needed in order to measure
an inclusive branching fraction of a few ${\cal O}(10^{-6})$.

The latest measurements of the inclusive branching fractions by BaBar and Belle are reported in Table~\ref{tab:semileptonic_inc:range}.
The systematic uncertainty on the measurements is dominated by the
the modelling of the $X_s$ system (as in $B \to X_s \gamma$ measurements, see Sec.~\ref{ssec:inclusive_radiative}) and the estimation of background from semileptonic decays.

In contrast to the inclusive $b \to s\gamma$ branching fraction measurements, which only determine the magnitude of $C_7$ and not its sign, inclusive semi-leptonic decays are sensitive to the sign of the Wilson coefficient $C_7$.
The data is consistent with the SM sign-convention, \emph{i.e.} that $C_7$ has the opposite sign to $C_9$.
This generates destructive interference between the contributions from $C_7$ and $C_9$, which is required to explain the observed branching fraction~\cite{Gambino:2004mv}.

%The results for the inclusive $b \to s \ell^+ \ell^-$ branching fraction are consistent with the SM prediction
%when interpolated to account for the the veto region~\cite{Ali:2002jg}.

%In all cases the dilepton mass regions $q^2 < 0.2\,{\rm GeV}^2$ and around the $J\!/\!\psi$ and $\psi'$ masses are
%not considered. The excluded regions around the $J\!/\!\psi$ and $\psi'$ are interpolated
%assuming the SM with no interference~\cite{Ali:1996bm, Ali:2002jg}.

It is also conventional for experiments to report branching fraction measurements that encompass the ``full'' $q^2$ range above a minimum threshold.
For example the  BaBar experiment reports a measurement of ${\rm BR}( B \to X_s \ell^+ \ell^-)$, based on 471 million  $B \overline{B}$ events of~\cite{Lees:2013nxa}
\begin{align}
{\rm BR}( B \to X_s \ell^+ \ell^-) = (6.73 {}^{+0.70}_{-0.64} {}^{+0.34}_{-0.35}  \pm 0.50 ) \times 10^{-6}\quad(q^2 > 0.1\,{\rm GeV^2})~.
\end{align}
These analysis remove the $q^2$ regions around the $J\!/\!\psi$ and $\psi'$ masses and correct for the excluded region using a phenomenological model for the shape of the differential decay distribution as a function of $q^2$ (in the absence of the narrow charmonimum states)~\cite{Ali:1996bm, Ali:2002jg}.
The final uncertainty on the BaBar measurement comes from both the model used to extrapolate the branching fraction to include the removed $q^2$ regions and the extrapolation to the full $X_s$ system from the ten exclusive states that were measured.

Restricting  the analysis exclusively to final states from which a decaying $B$ meson's flavour can be inferred,
BaBar also reports a measurement of the direct $C\!P$ asymmetry, $A_{C\!P}$, in bins of dilepton mass.
Integrated over the full dilepton mass range and averaging over decays to $e^+e^-$ and $\mu^+\mu^-$ they find~\cite{Lees:2013nxa}
\begin{align}
A_{C\!P}(B \to X_s \ell^+ \ell^-) = 0.04 \pm 0.11 \pm 0.01\,,
\end{align}
which is consistent with SM expectations~\cite{Ali:1998sf}.

A measurement of the $B \to X_s \ell^+ \ell^-$ forward-backward asymmetry has been performed by the Belle collaboration in Ref.~\cite{Sato:2014pjr}.
The forward-backward asymmetry is extrapolated from the sum of 10 exclusive $X_s$
states with an invariant mass $M_{X_s} < 2.0\,{\rm GeV}$, corresponding to
50\% of the inclusive rate.
For $q^2>10.2\,{\rm GeV}^2$, the wrong sign $A_{\rm FB}$ is excluded at the 2.3 $\sigma$ level.
For $q^2<4.3\,{\rm GeV}^2$, the result is within $1.8\; \sigma$ of the SM expectation.

\begin{table}
\centering
\begin{tabular}{lcc}
\hline
Mode & BaBar $[1,6]$ & Belle $[1,6]$  \\
\hline
${\rm BR}(B \to X_s e^+ e^-)$ & $(1.93^{+0.47}_{-0.45} {}^{+0.21}_{-0.16} \pm 0.18)\times 10^{-6}$ & -- \\
${\rm BR}(B \to X_s \mu^+ \mu^-)$ & $(0.66^{+0.82}_{-0.76} {}^{+0.30}_{-0.24} \pm 0.07 )\times 10^{-6}$ & -- \\
${\rm BR}(B \to X_s \ell^+ \ell^-)$ & $(1.60^{+0.41}_{-0.39} {}^{+0.17}_{-0.13} \pm 0.18) \times 10^{-6}$ & $(1.49 \pm 0.50^{+0.41}_{-0.32}) \times 10^{-6}$\\
\hline
Mode & BaBar $[>14.2]$ & Belle $[>14.4]$  \\
\hline
${\rm BR}(B \to X_s e^+ e^-)$ & $(0.56^{+0.19}_{-0.18} {}^{+0.03}_{-0.03} \pm 0.00)\times 10^{-6}$ & -- \\
${\rm BR}(B \to X_s \mu^+ \mu^-)$ & $( 0.60^{+0.31}_{-0.29}{}^{+0.05}_{-0.04} \pm 0.00)\times 10^{-6}$ & -- \\
${\rm BR}(B \to X_s \ell^+ \ell^-)$ & $(0.57^{+0.16}_{-0.15} {}^{+0.03}_{-0.02} \pm 0.00) \times 10^{-6}$ & $(0.42 \pm 0.12 ^{+0.06}_{-0.07}) \times 10^{-6}$\\
\hline
\end{tabular}
\caption{
Measurements of inclusive $B \to X_s \ell^+\ell^-$ branching fractions in the $q^2$ ranges $1 < q^2 < 6\,{\rm GeV}^{2}$ and $q^2 > 14\,{\rm GeV}^{2}$ by the BaBar~\cite{Lees:2013nxa} and Belle~\cite{Iwasaki:2005sy} experiments.
The systematic uncertainty from the BaBar experiment is split into two parts; experimental uncertainties and the uncertainty associated to the hadronisation of the $X_s$ system (and the extrapolation of the observed modes to the full $m_{X_s}$ range).
The $X_s$ hadronisation uncertainty is tiny at large $q^2$ because there is no phase-space to produce higher $K^*$ states.
\label{tab:semileptonic_inc:range}
}
\end{table}

\subsubsection{Experimental prospects}

The latest experimental measurements of ${\rm BR}(B \to X_s \ell^+ \ell^-)$ from BaBar and Belle have systematic uncertainties that are comparable in size to the statistical uncertainty on the measurements. The systematic uncertainties are limited by the modelling of the background (from semileptonic decays) and the knowledge of the hadronisation of the $X_s$ system. To improve the branching fraction measurement at Belle II it will be necessary to reduce both of these sources of uncertainty.
Measurements of $A_{\rm FB}(X_s \ell^+ \ell^-)$ and $A_{C\!P}(X_s \ell^+ \ell^-)$ are not systematically limited and significant improvements are expected with the Belle II dataset.

\begin{figure}[!hbt]
\centering
\includegraphics[width=0.8\linewidth]{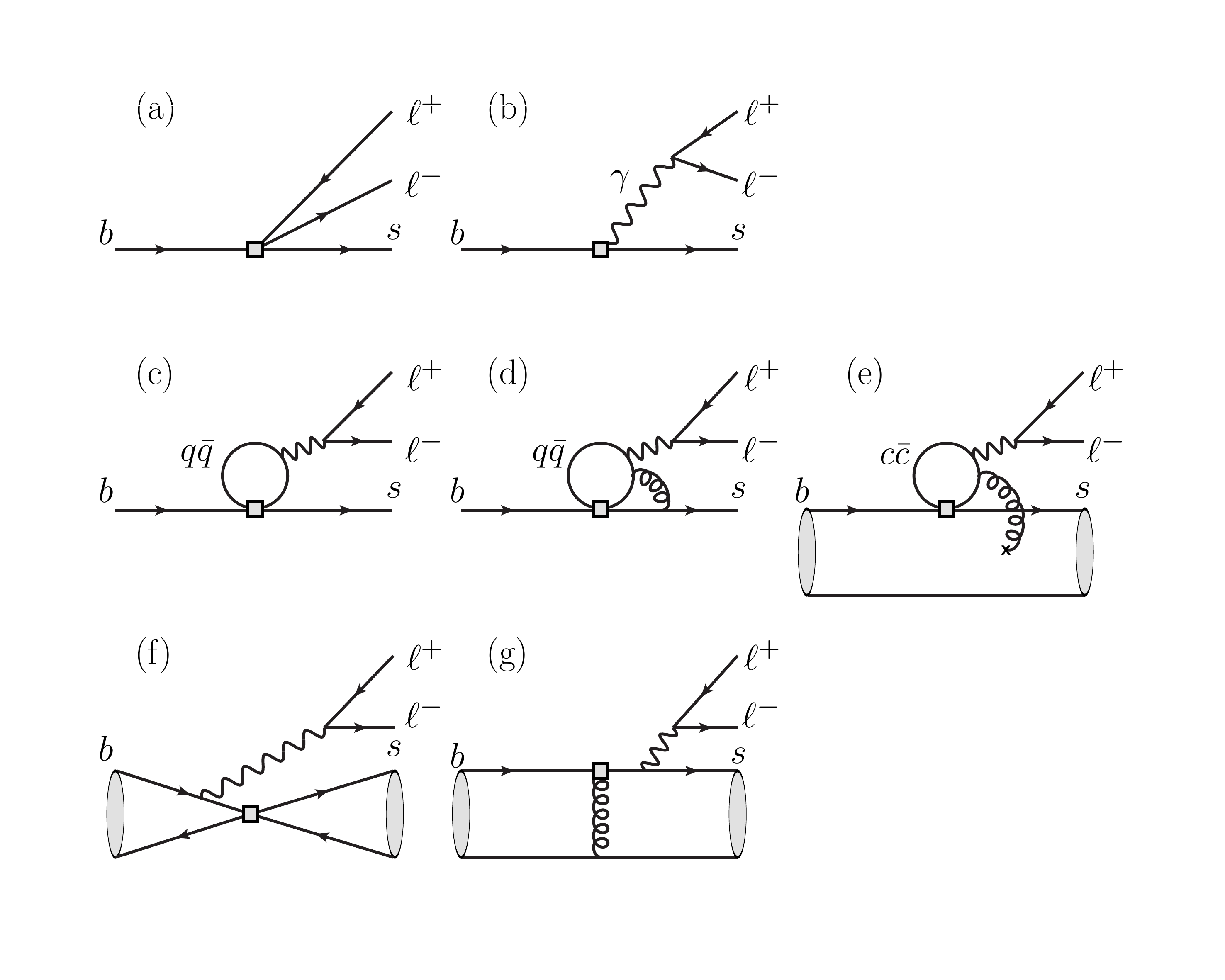}
\caption{
Schematic representation of local-operators and non-factorisable corrections in $b\to s\ell^+\ell^-$ decays.
The semileptonic operators $Q_{9,10}$ correspond to diagram (a) and the virtual photon contribution from  $Q_7$ to diagram (b).
The short-distance contribution involving particles at mass scales above $m_b$ are integrated out and represented by the shaded box.
Quark-loop contributions (c), including QCD corrections (d) are
calculable perturbatively, except in the charmonium resonance region
where the $c\bar c$ loop goes on shell and except for soft gluon corrections
(e). Weak annihilation (f) and hard spectator scattering (g) can be calculated
in QCD factorisation at low dilepton invariant  mass.}
\label{fig:semileptonic:btoslldiagrams}
\end{figure}

\begin{figure}[!htb]
\centering
\includegraphics[width=0.85\linewidth]{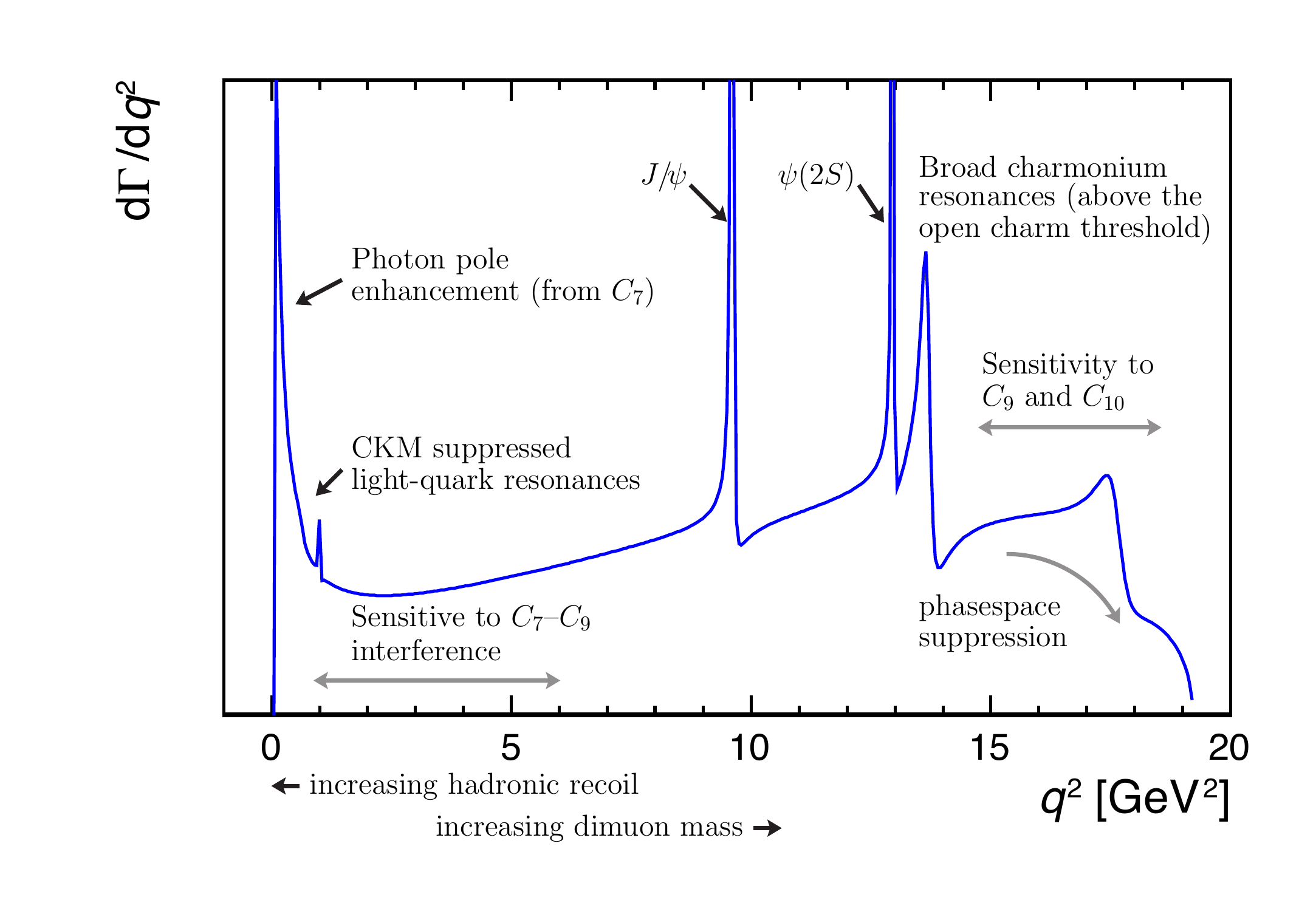}
\caption{
Cartoon illustrating the dimuon mass squared, $q^2$, dependence of the differential decay rate of $B \to K^{*}\ell^+\ell^-$ decays. The different contributions to the decay rate are also illustrated. For $B \to K\ell^+ \ell^-$ decays there is no photon pole enhancement due to angular momentum conservation.
\label{fig:semileptonic:cartoon}
}
\end{figure}

\subsection{Exclusive decay rates}
\label{sec:semileptonic:exclusiverates}

The most commonly studied exclusive decays are $B^{\pm,0}\to K\ell^+\ell^-$,
$B^{\pm,0}\to K^*\ell^+\ell^-$, $B_s\to \phi\ell^+\ell^-$  and $\Lambda_b \to \Lambda \ell^+ \ell^-$ probing the
$b\to s$ transition, as well as $B^{\pm,0}\to \rho \ell^+\ell^-$, $B^0\to \omega \ell^+\ell^-$,
and $B^{\pm,0}\to \pi \ell^+\ell^-$ probing the $b\to d$ transition.
These processes are discussed in detail below.

\subsubsection{Standard Model predictions}

In a similar fashion to the exclusive radiative decays, the prediction of exclusive
semi-leptonic decay rates requires the knowledge of Wilson coefficients as well
as hadronic form factors and non-factorisable hadronic effects that are not
contained in the form factors.

Concerning the form factors, at low $q^2$ the most precise predictions come from
LCSR, as already discussed in Sec.~\ref{sec:rad-excl} for the $B$ to vector
form factors and the situation is analogous for the $B$ to pseudoscalar
form factors \cite{Ball:2004ye,Bharucha:2012wy,Imsong:2014oqa}.
At high $q^2$, where
the hadronic recoil is small, the form factors can be simulated in lattice QCD
and significant progress in this direction has been made recently.
The uncertainties are smallest for the
$B$ to pseudoscalar transitions
$B\to K$ \cite{Bailey:2015dka}
and
$B \to \pi$ \cite{Flynn:2015mha,Lattice:2015tia,Bailey:2015nbd}.
In these transitions, there are only three independent form factors, one of
which does not contribute to the SM prediction in the limit of massless leptons
(which is a good approximation for electrons and muons). Further precision can
be gained by performing combined fits of the lattice results valid at high $q^2$
and LCSR results valid at low $q^2$
\cite{Bharucha:2010im,Imsong:2014oqa,Altmannshofer:2014rta}.
For the $B$ to vector form factors, only a single lattice computation exists so
far, comprising the $B\to K^*$ and $B_s\to \phi$ form factors
\cite{Horgan:2013hoa,Horgan:2015vla}. With vector
mesons in the final state, a challenge is their short lifetime -- in contrast
to the pseudoscalar mesons $\pi$ and $K$, $K^*$ and $\phi$ are not stable under
the strong interactions. The finite lifetime is neglected in the lattice
simulation and represents a source of systematic uncertainty.
Overcoming this limitation is in the focus of current efforts \cite{Agadjanov:2016fbd}.
As for the
$B$ to pseudoscalar transitions, combined fits of lattice and LCSR results
valid in different kinematical regimes lead to increased precision and less
dependence on extrapolation models \cite{Straub:2015ica}.

Beyond the form-factors, the next most significant uncertainties are hadronic
uncertainties associated to non-factorisable corrections. These are illustrated in Fig.~\ref{fig:semileptonic:btoslldiagrams}.
Diagrams (a) and (b) represent the leading order short-distance contributions from the
operators $Q_{7\ldots10}$ that factorise ``naively'' into a hadronic and leptonic
current.
The size of the non-factorisable effects and the
theoretical methods required to compute them vary strongly with $q^2$
(see Fig.~\ref{fig:semileptonic:cartoon} for a cartoon of the $q^2$ dependence of the differential branching ratio and the relevant hadronic effects).

At intermediate $q^2$, around the masses of the $J/\psi$ and $\psi(2S)$, the charm
loop in diagram (c) goes on shell, the decays turn into non-leptonic decays,
e.g. $B\to K J/\psi (\to\ell^+\ell^-)$, and quark-hadron duality breaks down~\cite{Beneke:2009az}.
These regions are typically vetoed in the experimental analyses.

At low $q^2$, the relevant non-factorisable effects include
weak annihilation as in diagram (f) and hard spectator scattering as in diagram
(g).
They have been calculated for $b\to s$ and $b\to d$ transitions involving
vector mesons in QCD factorisation to NLO in QCD \cite{Beneke:2001at,Beneke:2004dp}
as well as in soft-collinear effective theory \cite{Ali:2006ew}
and shown to be negligible in $B\to K\ell^+\ell^-$ decays
\cite{Bobeth:2007dw,Bartsch:2009qp}.
Weak annihilation and spectator scattering involving $Q_8$ have been computed
also in LCSR \cite{Dimou:2012un,Lyon:2013gba}.
Diagram (c) corresponds to the contribution of four-quark operators
that is usually written as a contribution to the ``effective'' Wilson coefficient
$C_9^\text{eff}$. Perturbative QCD corrections to the matrix elements of
$Q_{1,2}$ as in diagram (d) are numerically sizeable and are known from the inclusive decay as
discussed above.
The main challenge in exclusive $b\to s$ decays at low $q^2$ is represented by
soft gluon corrections to the charm loop shown in diagram (e). These have
been estimated in LCSR \cite{Khodjamirian:2010vf,Khodjamirian:2012rm}
but remain a significant source of uncertainty.

At very low $q^2\lesssim 1\,\text{GeV}^2$, narrow resonances
due to the light unflavoured mesons $\rho$, $\omega$, $\phi$ etc.\ appear in the
spectrum. While these resonances are not described locally by the QCDF
calculation, their effect in binned observables (where the bin size is large compared
to the width of the states) is negligible for exclusive decays
based on the $b\to s\ell^+\ell^-$ transition like $B\to K^{(*)}\ell^+\ell^-$
\cite{Khodjamirian:2010vf,Jager:2012uw}.

At high $q^2$, above the open charm threshold
$q^2_\text{oc}\approx 15\,\text{GeV}^2$ broad $c\bar{c}$ resonances appear in
the differential decay distribution and local quark-hadron duality should not
be expected to hold. A local operator expansion in powers of $E/\sqrt{q^2}$
has been used however to argue that for the rate \textit{integrated} in the
entire high $q^2$ region $q^2>15\,\text{GeV}^2$, it is well approximated by
the perturbative calculation consisting of the naively factorising part and
the matrix elements of the four-quark operators including perturbative corrections
to the matrix elements \cite{Grinstein:2004vb,Beylich:2011aq}.

In the exclusive decays based on the $b\to d\ell^+\ell^-$ transition, an
additional complication is given by the fact, discussed in
Sec.~\ref{sec:framework},
that the CKM combination $\lambda^{(d)}_u$ is not Cabibbo-suppressed with
respect to $\lambda^{(d)}_t$. As a consequence, several effects that are
small in exclusive $b\to s\ell^+\ell^-$ transitions become important,
including weak annihilation and narrow light meson resonances.
A discussion of these effects for the case of $B\to\pi\ell^+\ell^-$, using
QCD factorisation, LCSR, and the hadronic dispersion relation, has recently
been presented in \cite{Hambrock:2015wka}.

Apart from $B$ meson decays, the $b\to q\ell^+\ell^-$ transition is also
probed by baryonic decays such as $\Lambda_b\to \Lambda\ell^+\ell^-$.
Progress towards robust SM predictions in this mode has been made recently
by deriving the full angular distribution and estimating hadronic effects
\cite{Boer:2014kda} as well as by computing the relevant form factors in LQCD
\cite{Detmold:2016pkz}.

\subsubsection{Branching fraction measurements}

\begin{figure}[!htb]
\begin{center}
\includegraphics[width=0.48\linewidth]{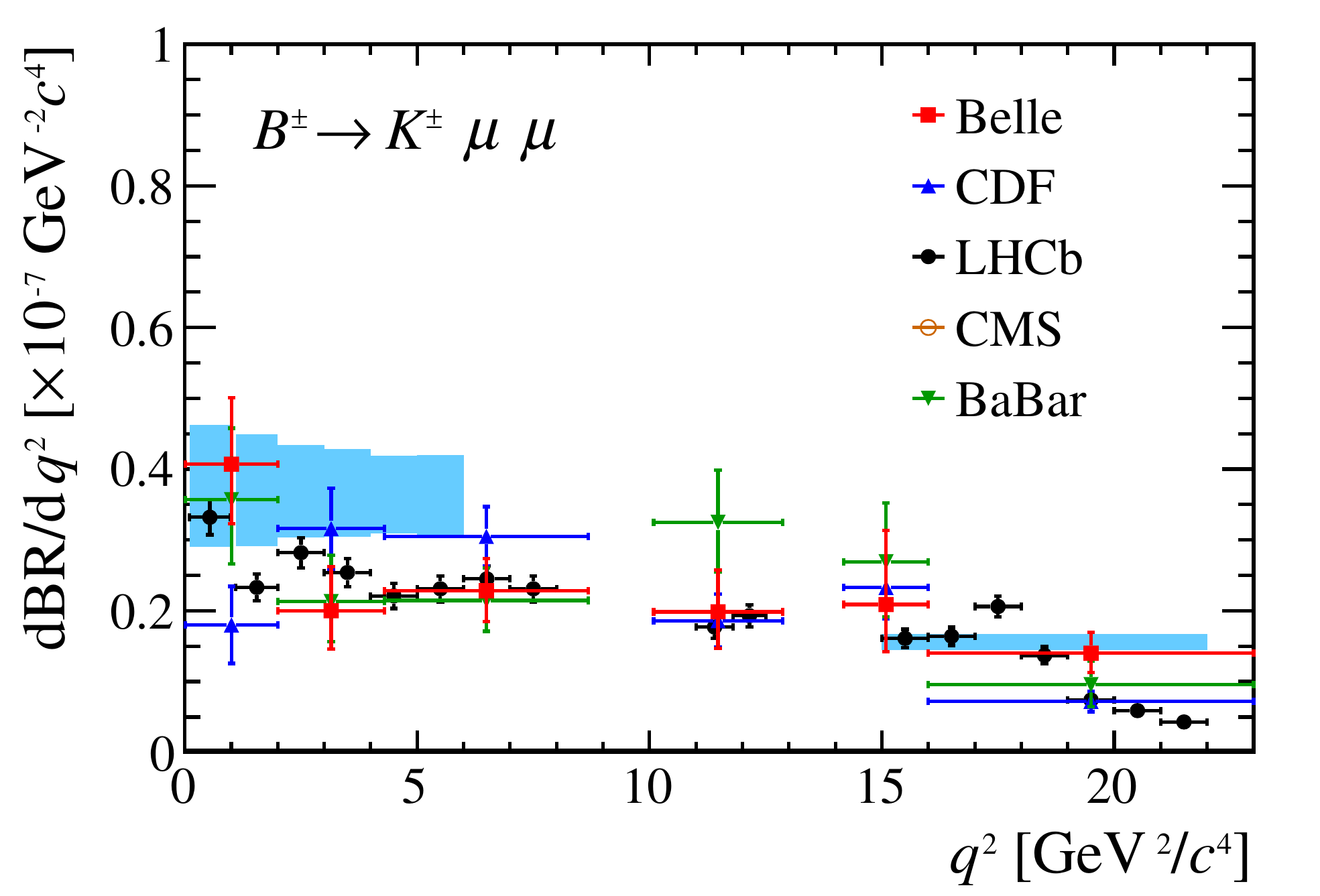}
\includegraphics[width=0.48\linewidth]{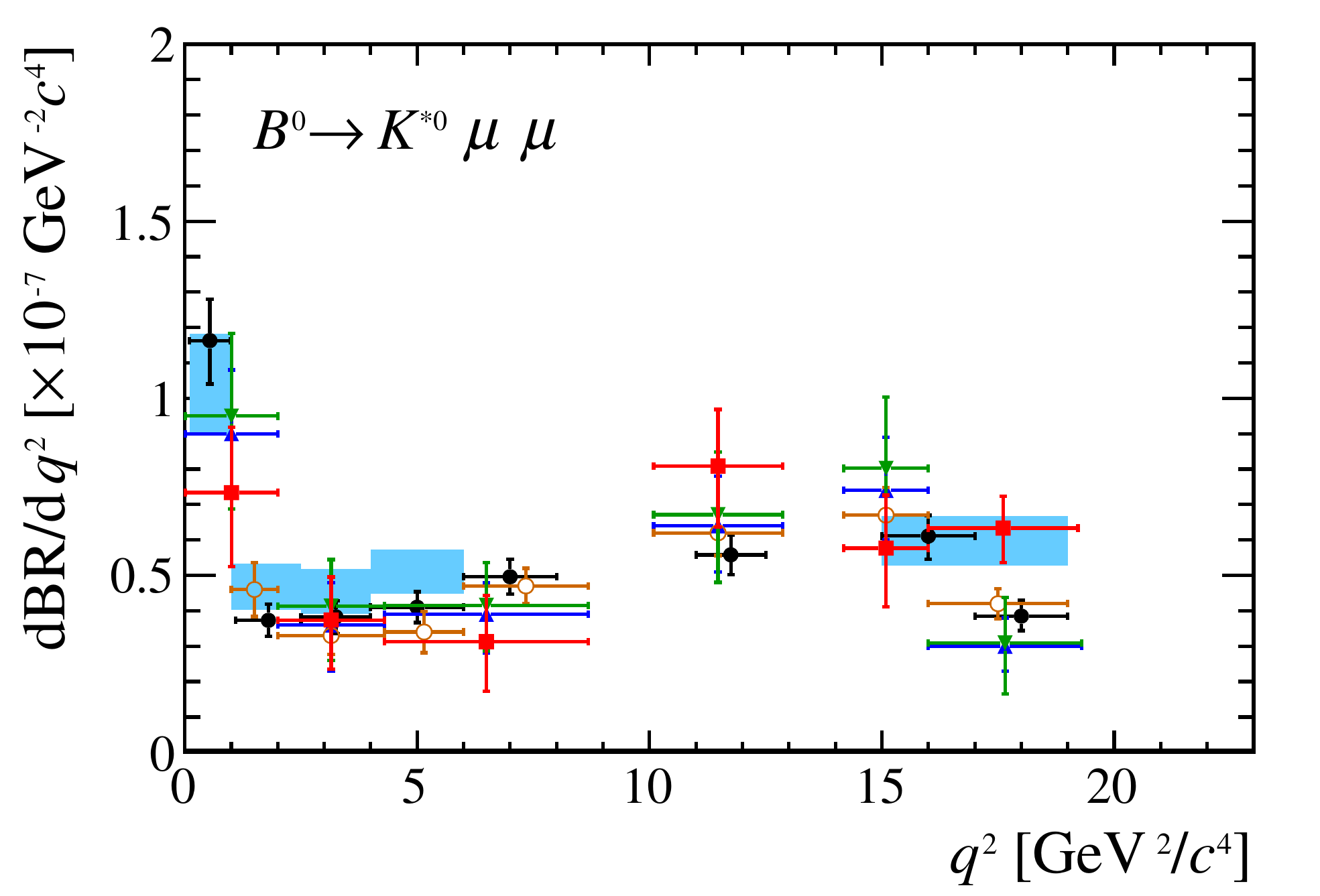}  \\
\includegraphics[width=0.48\linewidth]{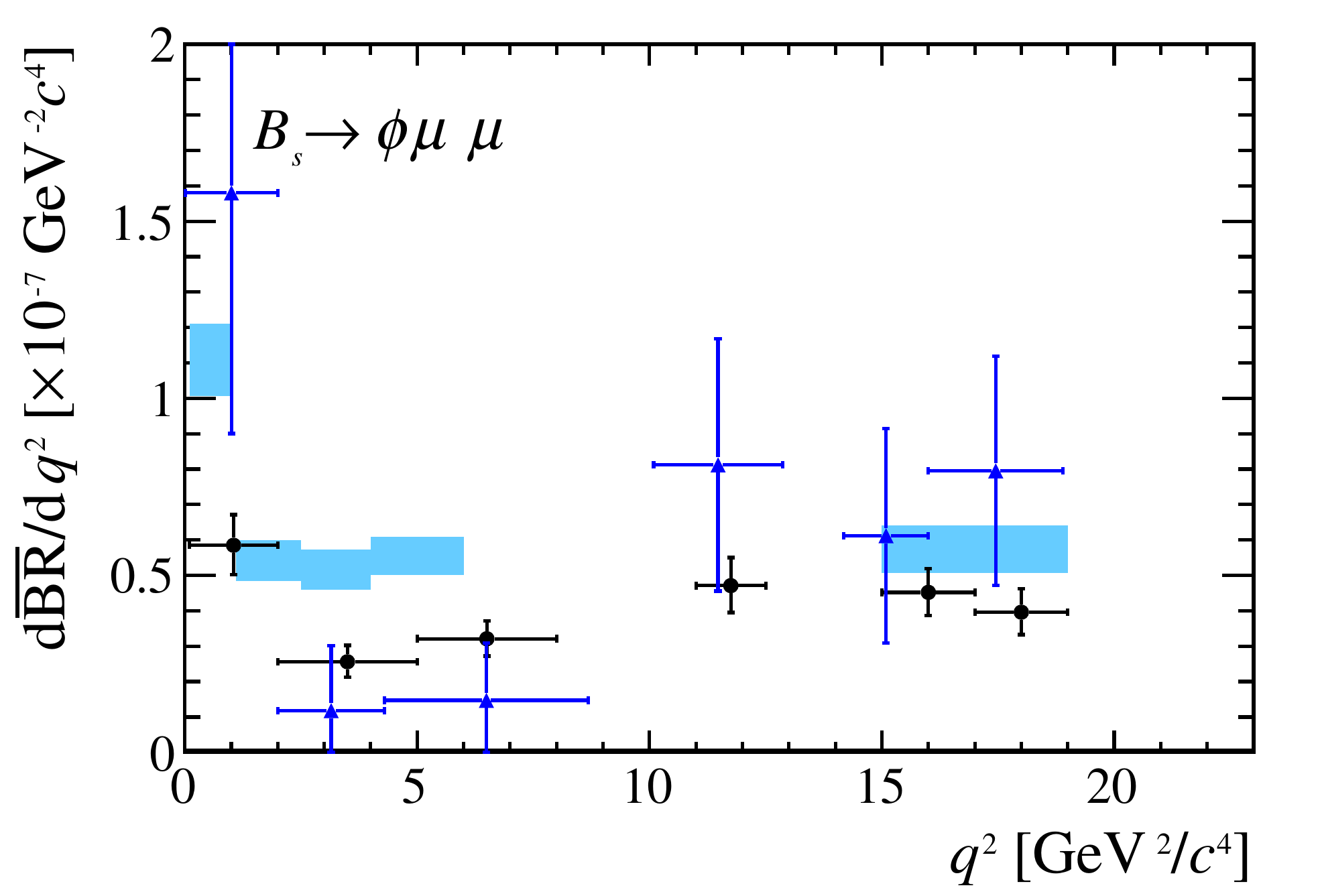}
\includegraphics[width=0.48\linewidth]{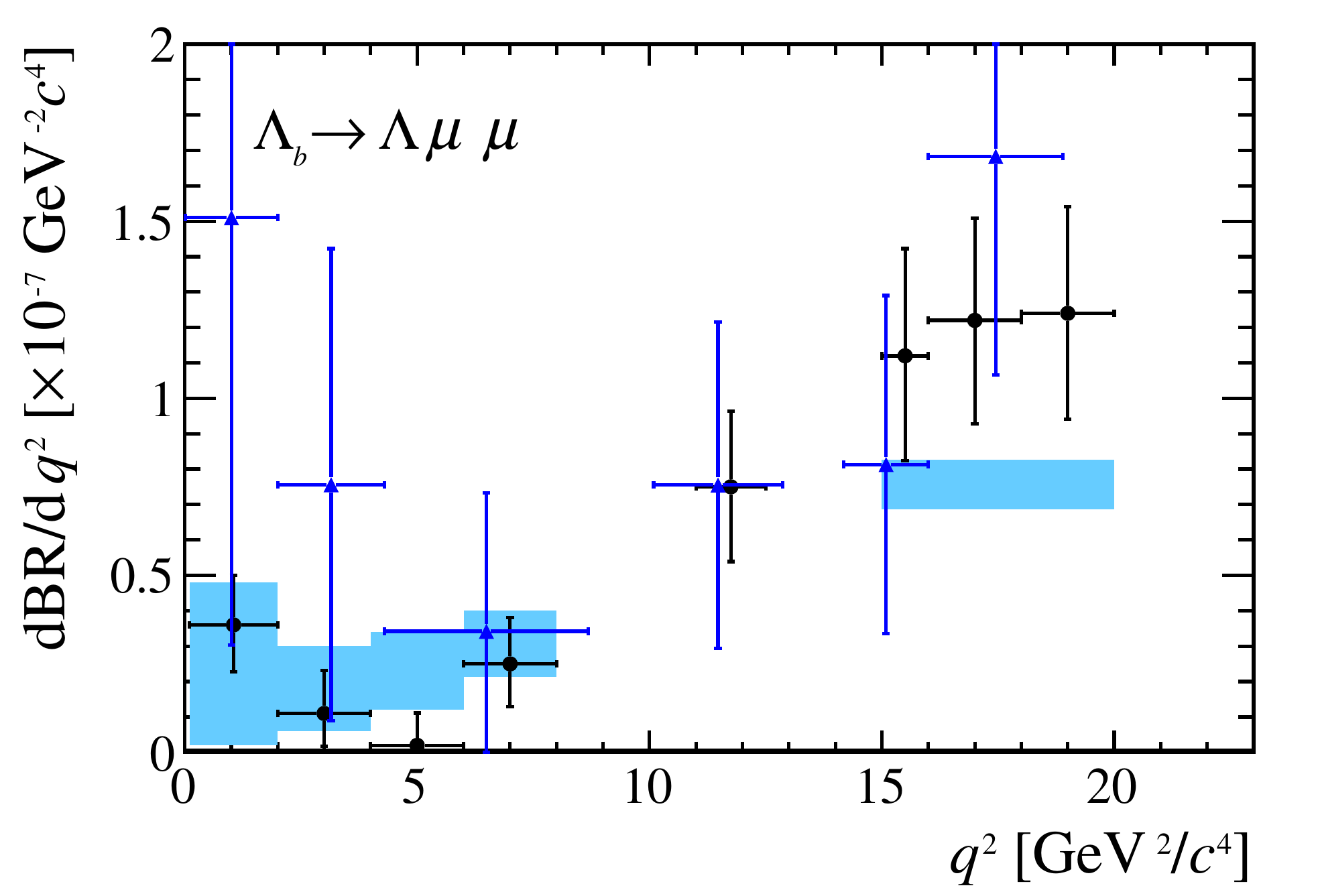}
\end{center}
\caption{
Differential decay rate as a function of the dimuon mass squared, $q^2$, for different $b \to s\mu^+\mu^-$ processes measured by the BaBar, Belle, CDF, CMS and LHCb experiments.  The experimental data are taken from Refs.~\cite{Lees:2012tva,Wei:2009zv,Aaltonen:2011qs,Khachatryan:2015isa,Aaij:2014pli,Aaij:2016flj,Aaij:2015esa,Aaij:2015xza}. The theoretical predictions for the branching fractions correspond to the shaded regions. For more details see text.
\label{fig:semileptonic:branching}
}
\end{figure}

A summary of experimental measurements of the differential branching fraction for the $b \to s\ell^+\ell^-$ processes $B \to K\ell^+\ell^-$, $B \to K^{*}\ell^+\ell^-$, $B_s\to\phi \ell^+\ell^-$ and $\Lambda_b \to \Lambda \ell^+\ell^-$, as a function of $q^2$, is provided in Fig.~\ref{fig:semileptonic:branching}.
The measurements from BaBar~\cite{Lees:2012tva} and Belle~\cite{Wei:2009zv} combine final-states with dielectron and dimuon pairs and combine final-states that are related by isospin, i.e. they combine $B^{0} \to K^{*0}\ell^+\ell^-$ and $B^+ \to K^{*+} \ell^+\ell^-$ decays which differ only by the flavour of the spectator quark in the decay.
The CDF~\cite{Aaltonen:2011qs}, CMS~\cite{Khachatryan:2015isa} and LHCb~\cite{Aaij:2014pli,Aaij:2016flj,Aaij:2015esa,Aaij:2015xza} measurements represented in the figure only concern dimuon final-states and do not include decays to final-states with neutral $K^0$ mesons or $\pi^0$.
The dimuon pair  provides a clean experimental signature that can be used to select the decays in the experiments triggers.
The LHCb experiment can also select events with electron or fully hadronic final-states but, even in LHCb, the trigger threshold for these final-states is much higher.
For example, in Run\,1 the LHCb trigger required a single electron with $E_{\rm T} > 3.6\,{\rm GeV}$ as opposed to a single muon with $p_{\rm T} > 1.76\,{\rm GeV}$.
It is also difficult for the LHC experiments to reconstruct decays involving long-lived particles ($K_{\rm S}$ or $\Lambda$) or final-states with $\pi^0$.
The $K_{\rm S}$ and $\Lambda$ typically have lifetimes of tens of centimetres in the LHC detectors and the longer lived $K_{\rm L}$ have a lifetime of hundreds of metres and decay outside the detectors acceptance.
At the B-factory experiments, the $K_{\rm L}$ can be detected in the experiments calorimeter.

With the large datasets available at the LHC, the experimental uncertainty on the branching fraction of many exclusive $b\to s\ell^+\ell^-$ decays is now much more precise than the corresponding SM predictions.
The theoretical prediction in Fig.~\ref{fig:semileptonic:branching} mostly use LCSR predictions for the form-factors at large recoil (low $q^2$) and predictions from Lattice QCD at low recoil (large $q^2$). For $\Lambda_b\to\Lambda\mu^+\mu^-$, predictions from Lattice QCD are used across the full $q^2$ range.
No predictions are provided close to the narrow $c\bar{c}$ resonances.
In this region the assumptions used to compute the SM predictions break down.
At low recoil (large $q^2$) predictions are only given averaged from $15\,{\rm GeV}^2$ to the kinematic limit over which the contribution from the broad charmonium resonances is thought to be well described by a local OPE.
For $B_s\to \phi\mu^+\mu^-$, the theory prediction takes into account the sizeable lifetime difference between the
two $B_s$ mass eigenstates that leads to a difference between the prompt and the time-integrated
branching fractions \cite{Descotes-Genon:2015hea}.

In general, the experimental measurements of the $b\to s\ell^+\ell^-$ branching fractions tend to lie below the SM expectations across the full $q^2$ range. The discrepancy is largest for the $B_s\to \phi\mu^+\mu^-$ decay in the large recoil (low $q^2$) region. The exception to this trend is the differential branching fraction of the $\Lambda_b \to \Lambda \mu^+ \mu^-$ where, at least at low recoil (large $q^2$), the measured branching fraction is above (but consistent with) the SM prediction.

The largest source of systematic uncertainty on the experimental measurements arises from the uncertainty on the rates of $B \to J\!/\!\psi (\to \ell^+ \ell^-) M$ decays that are used to normalise the $B \to M \ell^+ \ell^-$ branching fractions. For $B \to K\ell^+\ell^-$ and $B\to K^{*}\ell^+\ell^-$ decays this uncertainty is at the level of 4-6\%. For $B_s \to \phi \mu^+\mu^-$ it is at the level of 7.5\%.  This uncertainty is presently much larger for $\Lambda_b \to \Lambda\mu^+\mu^-$ decays at the level $\sim20\%$ and at large $q^2$ represents the largest source of experimental uncertainty on the LHCb measurement.

For the decay $B \to K^* \ell^+\ell^-$, due to the wide width of the $K^*$ meson, an important experimental consideration is the background from non-$K^*$ $B \to K\pi \mu^+\mu^-$ decays \emph{e.g.} from decays where the $K\pi$ is in an S-wave configuration~\cite{Lu:2011jm,Doring:2013wka}. This is an irreducible background that needs to be accounted for in the experimental analyses. These non-$K^*$ backgrounds have a different angular structure to the $B \to K^*\ell^+\ell^-$ decay. The  S-wave  contribution has been measured to be  $\mathcal{O}(5\%)$ in the mass windows used by the B-factory and LHC experiments~\cite{Aaij:2016flj}.

The only experimental constraint on the branching fraction of rare semileptonic decays to ditau final-states is a recent search for the decay $B^+ \to K^+ \tau^+\tau^-$ by the BaBar collaboration, who sets a limit~\cite{TheBaBar:2016xwe}.
\begin{align}
{\rm BR}(B^+ \to K^+ \tau^+\tau^-) < 2.3 \times 10^{-3}~\text{at~90\%~CL}.
\end{align}
This is four orders of magnitude above the SM prediction for the decay.
Decays to ditau final states are challenging to measure experimentally for two reasons.
First, there are a large number of possible decay modes for the $\tau$ lepton and experiments typically only consider a small number of these. Second, there are at least two missing neutrinos in the final state (and four if both tau leptons are reconstructed through $\tau^+ \to \mu^+ \nu_{\mu} \bar{\nu}_{\tau}$).
This makes it difficult to separate the signal from backgrounds from other $b$ hadron decays with missing final-state particles.
To reduce this background, BaBar fully reconstructs the other $B$ meson in the event (so called $B$ meson tagging technique) to determine the charge and momentum of the signal $B$.
This is not possible at the LHC experiments.

Rare $b \to d \ell^+\ell^-$ decays have very small branching fractions in the SM due to the small size of the $|V_{td}|$ CKM matrix element.
These processes have branching fractions of ${\cal O}(10^{-8})$.
In contrast to $b\to s\ell^+\ell^-$ decays light hadronic resonances that proceed via $b \to u\bar{u}d$ transitions also play an important role.
The LHCb experiment has measured the branching fraction of the $B^+ \to \pi^+\mu^+ \mu^-$ and $B^0 \to \rho^0 \mu^+\mu^-$\footnote{With its current dataset the LHCb collaboration does not separate contributions from the $\rho^0$ and other $\pi^+\pi^-$ states.}
decays using the Run\,1 dataset from the LHC~\cite{Aaij:2015nea,Aaij:2014lba}.
The branching fractions of the processes are consistent with their SM expectations. However, they once again sit on the low side of the predictions.
The LHCb dataset for $B^+ \to \pi^+\mu^+\mu^-$ is sufficient that the contribution of the $\rho$ and $\omega$ to the dimuon mass spectrum is visible.
These resonant contributions account for a significant fraction of the measured branching fraction in the region $q^2 < 1\,{\rm GeV}^{2}$.

\subsubsection{Isospin asymmetries}

In Fig.~\ref{fig:semileptonic:branching}, the BaBar and Belle measurements combine processes that only differ by the flavour (and charge) of the spectator quark in the $B$ meson.
This is a reasonable average to make in the SM, where isospin asymmetries
\begin{align}
A_{\rm I} =
\frac{\Gamma(B^0 \to K^{(*)0} \ell^+\ell^-) - \Gamma(B^+ \to K^{(*)+} \ell^+ \ell^-)}{\Gamma(B^0 \to K^{(*)0} \ell^+\ell^-) + \Gamma(B^+ \to K^{(*)+} \ell^+ \ell^-)}
\end{align}
are small.
For  $q^2 \gtrsim 4\,{\rm GeV}^{2}$ the asymmetry is expected to be $A_{\rm I} \approx -1\%$~\cite{Feldmann:2002iw}.
At small values of $q^2$ $A_{\rm I}(B \to K^* \mu^+ \mu^-)$ is expected to grow to become compatible with the isospin asymmetry seen in $B \to K^* \gamma$ decays of $A_{\rm I} = (5.2 \pm 2.6) \%$~\cite{Amhis:2014hma,Aubert:2009ak,Nakao:2004th}.
For $B \to K \mu^+\mu^-$ decays $A_{\rm I}$ remains small across the full range of $q^2$~\cite{Lyon:2013gba}.
The isospin breaking of the SM is driven by contributions from weak annihilation processes accompanied by ISR/FSR producing a virtual photon.
Measurements of $A_{\rm I}$ by LHCb~\cite{Aaij:2014pli}, BaBar~\cite{Lees:2012tva} and Belle~\cite{Wei:2009zv} are consistent with the SM expectation of a small isospin asymmetry.

\subsubsection{\texorpdfstring{$C\!P$}{CP} asymmetries}\label{sec:sl:cpa}

The experimental measurements from the BaBar and Belle experiments in Fig.~\ref{fig:semileptonic:branching} also combine charge-conjugated processes ($B$ and $\bar{B}$ decays).
Direct $C\!P$ asymmetries
\begin{align}
A_{C\!P} =
\frac{\Gamma(\overline{B} \to \overline{K} \ell^+ \ell^-) - \Gamma({B} \to {K} \ell^+ \ell^-)}
{\Gamma(\overline{B} \to \overline{K} \ell^+ \ell^-) + \Gamma({B} \to {K} \ell^+ \ell^-)}
\end{align}
are expected to be small in the SM for $b\to s$ processes because of the small size of $\lambda_u^{(s)}$ compared to  $\lambda_{t}^{(s)}$.
They can, however, be larger for $b \to d$ transitions where  $\lambda_u^{(d)}$ and $\lambda_{t}^{(d)}$ are more similar in size.
At low $q^2$, the $C\!P$ asymmetry of $B^+\to \pi^{+} \mu^+\mu^-$ decays is expected to be approximately $-10\%$, becoming larger in the region where the short-distance contribution interferes with the $\rho$ and $\omega$~\cite{Hambrock:2015wka}.
%are expected to be small in the SM for $b\to s$ processes because of the small size of $|V^{}_{ub} V^{*}_{us}|$ compared to  $|V^{}_{tb} V^{*}_{ts}|$.
%They can, however, be larger for $b \to d$ transitions where $|V^{}_{ub} V^{*}_{ud}|$ and $|V^{}_{tb} V^{*}_{td}|$ are more similar in size.
%At low $q^2$, the $C\!P$ asymmetry of $B^+\to \pi^{+} \mu^+\mu^-$ decays is expected to be approximately $-10\%$, becoming larger in the region where the short-distance contribution interferes with the $\rho$ and $\omega$~\cite{Hambrock:2015wka}.
The most precise measurements of these direct $C\!P$ asymmetries comes from the LHCb experiment~\cite{Aaij:2014bsa, Aaij:2015nea},
\begin{align}
A_{C\!P}(B^0\to K^{*0} \mu^+\mu^-) & = +0.035 \pm 0.024_{\rm stat} \pm 0.003_{\rm syst} \\
A_{C\!P}(B^+\to K^{+} \mu^+\mu^-) & = +0.012 \pm 0.017_{\rm stat} \pm 0.001_{\rm syst}  \\
A_{C\!P}(B^+\to \pi^{+} \mu^+\mu^-) & = -0.11 \pm 0.12_{\rm stat} \pm 0.01_{\rm syst} \,.
\end{align}
These measurements are consistent with SM expectations.

\subsection{Angular distribution}
\label{sec:semileptonic:angular}

The angular distribution of $B \to K \ell^+\ell^-$ decays is described by a single angle, $\theta_{\ell}$, defined as the angle between the flight direction of the $\ell^+$ and the direction of the $B$ in the dilepton rest-frame.
The differential decay rate is given by~\cite{Bobeth:2007dw}
\begin{align}
\frac{\mathrm{d}^2\Gamma(B \to K \ell^+\ell^-)}{\mathrm{d}\cos\theta_{\ell}\,\mathrm{d}q^2} = \frac{3}{4}( 1 - F_{\rm H}) ( 1- \cos^2\theta_{\ell}) + \frac{1}{2} F_{\rm H} + A_{\rm FB} \cos\theta_{\ell} \, ,
\end{align}
where both the constant term $F_{\rm H}$, and the forward-backward asymmetry, $A_{\rm FB}$, depend on $q^2$.
In the SM, $A_{\rm FB}$ arises from QED corrections and is tiny. The $F_{\rm H}$ term is also small for $\ell = e, \mu$.
In the presence of BSM physics, both $A_{\rm FB}$ and $F_{\rm H}$ can be non-zero.
The most recent measurements by LHCb using $B \to K\mu^+\mu^-$ decays are consistent with the SM expectation of $A_{\rm FB} = 0$ and $F_{\rm H} \approx 0$~\cite{Aaij:2014tfa}.

\begin{figure}[!htb]
\centering
\includegraphics[width=1.0\linewidth]{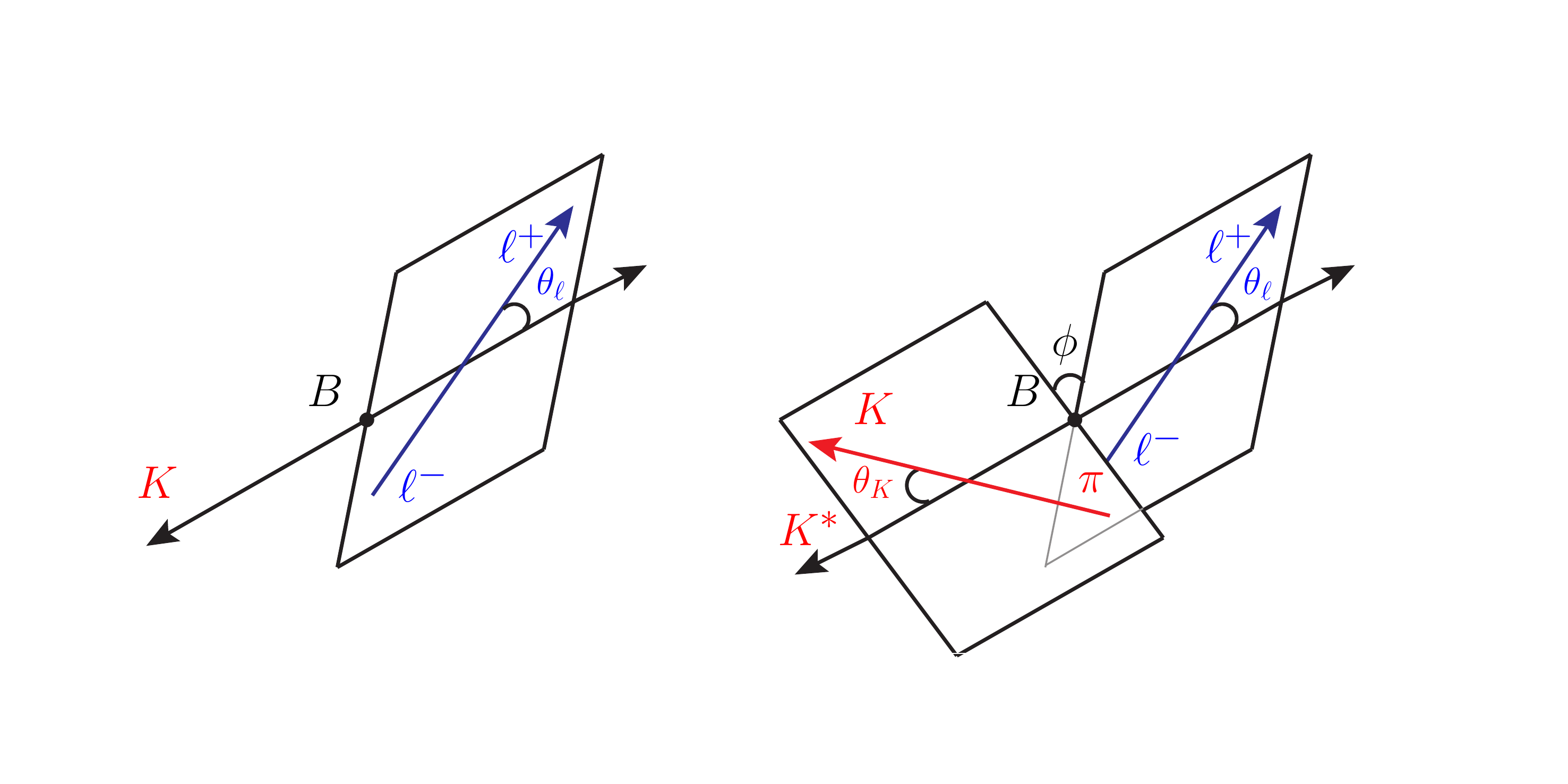}
\caption{
Cartoon illustrating the angular conventions used by experiments for $B \to K \ell^+ \ell^-$ and $B \to K^* \ell^+ \ell^-$ decays. For the $B$ decay: the angle $\theta_{\ell}$ is defined by the direction of the positive lepton in the dilepton rest frame and the flight direction of the dilepton pair in the $B$ rest-frame; the angle $\theta_{K}$ is defined as the angle between the flight direction of the kaon in the $K^*$ rest-frame and the direction of the $K^*$ in the $B$ rest-frame. The angle $\phi$ is the angle between the decay plane containing the $K\pi$ and the decay plane of the dilepton pair.  The angles $\theta_\ell$ and $\theta_K$ are defined in the range $[0,\pi]$. The angle $\phi$ is defined in the range $[-\pi,\pi]$. For $\phi$, positive angles correspond to the case where the $K\pi$ plane is in advance of the dilepton plane. The convention for $\bar{B}$ decays can be found using $C\!P$.
\label{fig:btosll:angles}
}
\end{figure}

The angular distribution of $B \to K^*\ell^+\ell^-$ decays is more complex and is described by three angles:
the angle between the flight direction of the $\ell^+$ and the $B$ in the rest frame of the dilepton system,  $\theta_\ell$;
the angle between the flight direction of the $K$ and the $B$ in the rest frame of the $K^*$ system, $\theta_{K}$;
and the angle between the plane containing the $\ell^+$ and $\ell^-$ and the plane containing the $K$ and $\pi$,  labelled $\phi$.
The angular convention adopted by the experiments is illustrated in Fig.~\ref{fig:btosll:angles}  for  $B \to K^* \ell^+ \ell^-$ and $B \to K \ell^+\ell^-$ decays.
The convention of Ref.~\cite{Aaij:2013iag} is used when going from $B$ to $\bar{B}$ decays.
A detailed discussion of this angular convention and how it relates to other angular conventions that appear in literature can be found in Ref.~\cite{Gratrex:2015hna} -- the largest ambiguity in the literature relates to  the definition of the $\phi$ angle for the $B$ and $\bar{B}$ decays.

The differential decay rate in terms of the angular variables is given by
\begin{align}
\begin{split}
\frac{\mathrm{d}^4\Gamma(\overline{B} \to \overline{K}^* \ell^+\ell^-)}{\mathrm{d}\cos\theta_{\ell}\,\mathrm{d}\cos\theta_{K}\,\mathrm{d}\phi\,\mathrm{d}q^2} = \frac{9}{32\pi} \sum_j I_{j} f_{j}(\cos\theta_\ell,\cos\theta_K,\phi)~, \\
\frac{\mathrm{d}^4\Gamma({B} \to {K}^* \ell^+\ell^-)}{\mathrm{d}\cos\theta_{\ell}\,\mathrm{d}\cos\theta_{K}\,\mathrm{d}\phi\,\mathrm{d}q^2} = \frac{9}{32\pi} \sum_j \bar{I}_{j} f_{j}(\cos\theta_\ell,\cos\theta_K,\phi)~,
\end{split}
\end{align}
where $I_j$ and $\bar{I}_{j}$ are functions of $q^2$ and depend on the $K^*$ transversity amplitudes.
The angular dependence of each term,  $f_{j}(\cos\theta_\ell,\cos\theta_K,\phi)$ , originates from the spherical harmonic functions associated with different polarisation states of the $K^*$ and dilepton system.
This angular distribution has been first discussed in \cite{Kruger:1999xa}, extended to include
right-handed currents in \cite{Kim:2000dq,Kruger:2005ep}, scalar operators in \cite{Altmannshofer:2008dz}
and tensor operators in \cite{Bobeth:2012vn}.
A general rederivation of these expressions and a comprehensive discussion of different angular conventions
has been presented in Ref.~\cite{Gratrex:2015hna}.
The same formalism can be applied to other $B \to V\ell^+\ell^-$ decays, such as $B_{s}\to\phi\ell^+\ell^-$ and $B^0 \to \rho \ell^+\ell^-$.
The self-tagging nature of the $B \to K^* \ell^+ \ell^-$ decay means that it is possible to determine both $C\!P$-averaged and $C\!P$-asymmetric quantities,
\begin{align}
S_{i} = (I_i + \bar{I}_i)\Big/\frac{{\rm d}\Gamma}{{\rm d} q^2}\quad,\quad A_i = (I_i - \bar{I}_i)\Big/\frac{{\rm d}\Gamma}{{\rm d} q^2}~,
\end{align}
using the notation of \cite{Altmannshofer:2008dz}.
The final state of the decays $B_s \to \phi \ell^+ \ell^-$ and $B^0 \to \rho \ell^+ \ell^-$ are not flavour specific and it is not possible to separate the $S_i$ and $A_i$ without performing a time-dependent flavour tagged analysis (where the initial flavour of the $B$-meson is tagged at production). In time-integrated, untagged, analyses experiments measure an admixture of the $C\!P$ conserving and $C\!P$ violating observables (depending on
how the observables transform under a $C\!P$ conjugation, but with the angles $\theta_\ell$ and $\theta_K$ associated to the same-charge particle in both decays), namely
$S_3$, $S_4$ and $S_7$ plus $A_5$, $A_{6s}$, $A_{8}$ and $A_{9}$.

The angular $C\!P$ asymmetries $A_{7,8,9}$ have the special property that they are
odd under a ``na{\"i}ve'' $T$ transformation, i.e.\ under a reflection of all momenta
and spins without actually reversing the time direction of the process,
since they correspond to kinematic triple correlations.
While non-zero $J_{7,8,9}$ do not yet signal a violation of $T$ (and $C\!P$) invariance,
since they can be generated by final state interaction phases as well,
$A_{7,8,9}$ are true measures of $C\!P$ violation. In contrast to the direct
$C\!P$ asymmetries discussed in section \ref{sec:sl:cpa}, they do not require the
presence of any strong phases to be sensitive to new sources of $C\!P$ violation
\cite{Bobeth:2008ij}.

The angular distribution of $\Lambda_b \to \Lambda^{(*)}\ell^+\ell^-$ decays can be more complex again if the spin-$\tfrac{1}{2}$ $\Lambda_b$ baryon is polarised at production.
Decays involving the $\Lambda(1115)$ provide a unique set of observables because the $\Lambda(1115)$ decays weakly (with both vector and axial vector contributions).

\subsubsection{Angular observables in \texorpdfstring{$B \to V\ell^+ \ell^-$}{B->Vll} decays}

The matrix element for the $B \to V\ell^+ \ell^-$ decay can be written as
\begin{align}
{\cal M}(m,n) \propto \varepsilon_{V}^{\mu*}(m)\,M_{\mu\nu}\,\varepsilon_{\ell^+ \ell^-}^{\nu*}(n)~,
\end{align}
where $\varepsilon_V^{\mu*}(m)$ and $\varepsilon_{\ell^+ \ell^-}^{\nu*}(n)$ are polarisation vectors for the vector meson and the dilepton system
\begin{align}
\begin{split}
\varepsilon^{\mu*}(\pm) & = (0,1,\pm i,0)/\sqrt{2} \\
\varepsilon^{\mu*}(0) & = (0,0,0,1) \\
\varepsilon^{\mu*}(t) & = (1,0,0,0) \\
\end{split}
\end{align}
Angular momentum conservation requires $(m,n) = (0,0), (0,t), (+,+), (-,-)$.
The matrix element can then be expanded in terms of seven helicity amplitudes $H_{0}^{\rm L,R}$, $H_t$, $H_{+}^{\rm L,R}$ and $H_{-}^{\rm L,R}$ or seven transversity amplitudes, which are related to the helicity amplitudes through
 \begin{align}
A_{\perp, \parallel}^{\rm L, R} = \tfrac{1}{\sqrt{2}}(H_{+}^{\rm L,R} \mp H_{-}^{\rm L,R})\quad,\quad A_0^
{\rm L,R} = H_0^{\rm L,R}\quad,\quad A_{t} = H_{t}~.
\end{align}
The indices L and R refer to left- and right-handed chiralities of the dilepton system, $(\bar{\ell} \gamma^{\mu} P_{\rm L, R} \ell)$.
The left-handed chirality depends on the Wilson coefficient combination $C_{9} - C_{10}$ and the right-handed on $C_9 + C_{10}$.
Beyond the SM operator basis additional amplitudes can be present, for example a ``scalar'' amplitude in models with non-zero $C_S$~\cite{Altmannshofer:2008dz} or tensor amplitudes~\cite{Beaujean:2015gba}.
The relation between the angular coefficients, which are observables, and the
transversity amplitudes, which are not, is governed by a set of symmetries that
have been systematically worked out in \cite{Egede:2010zc,Matias:2012xw}.

As introduced in Sec.~\ref{sec:framework}, the amplitudes also depend on form-factors for the $B \to V$ transition.
Assuming naive factorisation, the amplitudes can be schematically written as
\begin{align}
A_{\parallel,\perp,0}^{\rm L, R} = \kappa_{\parallel,\perp,0}^{\rm L, R}  \cdot F_{\parallel,\perp,0}^{}(q^2) + \text{non-factorisable~corrections}
\end{align}
where the short distance contribution to the decay is in $\kappa_{\parallel,\perp,0}^{\rm L, R}$.
At large hadronic recoil (low $q^2$) there are relationships between the different form-factors such that $F_{\parallel} \approx F_{\perp}$.
In the limit that $C'_{7,9,10} \to 0$, $A_{\parallel}^{\rm L,R} \approx -A_{\perp}^{\rm L,R}$.
On the other hand at zero recoil, $q^2 = (m_B - m_V)^2$, $A_{\perp}^{\rm L,R} = 0$ and there is an exact relationship between $A_{0}^{\rm L,R}$ and $A_{\parallel}^{L,R}$ such that $A_{0}^{\rm L,R} = A_{\parallel}^{L,R}$, i.e. the three helicity amplitudes become equal \cite{Hiller:2013cza}.
At large $q^2$ the OPE also predicts that $A_{\parallel}^{L,R}$ and $A_{0}^{L,R}$ have the same short distance dependence.

It is possible to exploit these symmetry relations to construct observables that are free from form-factor uncertainties at leading order in a $1/m_b$ expansion.
For example, at low-$q^2$, the transverse asymmetry \cite{Kruger:2005ep}
\begin{align}
P_1\equiv A_{\rm T}^{(2)}  = \frac{S_3}{2 S_2^s}
\end{align}
is sensitive to new right-handed currents.
Here the form-factors are expected to cancel up to corrections of order $\Lambda_\text{QCD}/m_b$.
It is also possible to build other ``clean''  observables at low-$q^2$ exploiting the form-factor cancellation. This includes the so-called $P'$ series of observables
\cite{Descotes-Genon:2013vna}, defined as\footnote{Note that $P'_6 \propto S_7$ is
not a misprint but an unfortunate convention.}
\begin{align}
P'_4 &= \frac{S_4}{2\sqrt{-S_2^s S_2^c}} \,,
&
P'_5 &= \frac{S_5}{2\sqrt{-S_2^s S_2^c}} \,,
&
P'_6 &= \frac{S_7}{2\sqrt{-S_2^s S_2^c}} \,,
&
P'_8 &= \frac{S_8}{2\sqrt{-S_2^s S_2^c}} \,,
\end{align}
and also
\begin{align}
P_2 &= \frac{S_6^s}{8 S_2^s} \,,
&
P_3 &= \frac{S_9}{4 S_2^s} \,,
\end{align}
in addition to $P_1$ defined above.
Similar observables suited for high $q^2$ have also been constructed \cite{Bobeth:2010wg,Bobeth:2012vn}.

For the reader's convenience, in Table~\ref{tab:dictionary} we provide a
dictionary between the most commonly used conventions for the
$B\to V\ell^+\ell^-$ angular observables in the literature.
In this review, we adopt the LHCb conventions.

\begin{table}[tbp]
\centering
\renewcommand{\arraystretch}{1.2}
\begin{tabular}{cccc}
\hline
 LHCb \cite{Aaij:2015dea,Aaij:2015oid} &  ABBBSW \cite{Altmannshofer:2008dz}  & DHMV \cite{Descotes-Genon:2013vna} & \\
\hline
 \boldmath $A_\text{\textbf{FB}}$ &  $-A_\text{FB}$ &  $A_\text{FB}$ \\
 \boldmath $F_L$ &  $F_L$ &  $F_L$ \\
 \boldmath $S_3$ & $S_3$ &  \\
 \boldmath $S_4$ & $-S_4$ &  \\
 \boldmath $S_5$ & $S_5$ &  \\
 \boldmath $S_7$ & $-S_7$ &  \\
 \boldmath $S_8$ & $S_8$ &  \\
 \boldmath $S_9$ & $-S_9$ &  \\
 \boldmath $A_3$ & $A_3$ &  \\
 \boldmath $A_4$ & $-A_4$ &  \\
 \boldmath $A_5$ & $A_5$ &  \\
 \boldmath $A_{6s}$ & $A_6^s$ &  \\
 \boldmath $A_7$ & $-A_7$ &  \\
 \boldmath $A_8$ & $A_8$ &  \\
 \boldmath $A_9$ & $-A_9$ &  \\
 \boldmath $P_1=A_T^{(2)}$ &  & $P_1=A_T^{(2)}$ &\\
 \boldmath $P_2=\frac{1}{2}A_T^\text{\textbf{Re}}$ &  & $-P_2=-\frac{1}{2}A_T^\text{Re}$ \\
 \boldmath $P_3$ &  & $-P_3$ \\
 \boldmath $P_4'$ &  & $-\frac{1}{2}P_4'$ \\
 \boldmath $P_5'$ &  & $P_5'$ \\
 \boldmath $P_6'$ &  & $P_6'$ \\
 \boldmath $P_8'$ &  & $-\frac{1}{2}P_8'$ \\
 \boldmath $A_T^\text{\textbf{Im}}$ & & $-2P_3^\text{CP}$ & \\
\hline
\end{tabular}
\caption{Conventions for $B\to V\ell^+\ell^-$ angular observables
used in the literature, relative to the conventions used by the LHCb
collaboration.
The table should be interpreted such that, for instance,
$P_4'|_\text{LHCb} = -\frac{1}{2}P_4'|_\text{DHMV}$.
Note that the sign of the relation
between the LHCb and theory conventions of $S_{7,8,9}$, $P_{6,8}'$ and $A_{7,8,9}$ was
given correctly in \cite{Gratrex:2015hna} for the first time.
The LHCb convention is also used e.g.\ in \cite{Altmannshofer:2014rta,Hurth:2016fbr}.
The ABBBSW convention is also used e.g.\ in \cite{Altmannshofer:2011gn,Altmannshofer:2012az,Altmannshofer:2013foa}.
The DHMV convention is also used e.g.\ in \cite{Descotes-Genon:2013wba,Hurth:2013ssa,Descotes-Genon:2014uoa,Hurth:2014vma,Descotes-Genon:2015uva}.
}
\label{tab:dictionary}
\end{table}

\subsubsection{Angular distribution of \texorpdfstring{$B \to V\ell^+ \ell^-$}{B->Vll} decays at small values of \texorpdfstring{$q^2$}{q²}}

At small dilepton masses the angle, $\phi$, between the decay plane containing the dilepton pair and the decay plane containing the daughters of the vector meson is sensitive to the photon polarisation.
If the photon is not purely left-hand polarised the angular distribution of the $\phi$ angle can get a modulation proportional to
the transverse asymmetry
\begin{align}
P_1 \equiv A_{\rm T}^{(2)} \approx \frac{2{\rm Re}( C_7 C'_7 )}{|C_7|^2 + |C'_7|^2}\,.
\end{align}
In order to maximise sensitivity to the virtual photon, it is necessary to go to very small dilepton masses.
The most sensitive measurement comes from the LHCb experiment using $B^0 \to K^{*0} e^+ e^-$ decays~\cite{Aaij:2015dea}.
In the Run\,1 dataset LHCb measures $A_{\rm T}^{(2)} = -0.23 \pm 0.23_{\rm stat} \pm 0.05_{\rm syst}$, which is consistent with zero and pure left-hand polarisation.

\subsubsection{Measurements of angular observables}

The two most well measured angular observables are
$
F_{\rm L} \equiv -S_2^c
$
and
$
A_{\rm FB} \equiv  \frac{3}{4} S_6^s+ \frac{3}{8}S_6^c
$.
These can be extracted by experiments by simply fitting the distribution of events in $\cos\theta_{\ell}$ and $\cos\theta_K$.
The observable $S_3$ and $S_9$ ($A_3$ and $A_9$) can be determined from the $\phi$ angle.
The remaining observables cancel when integrating over one or more angles and can only be determined from a full angular analysis of the decay.
Such an analysis is only  possible with the large dataset.
The first full angular analysis of the decay has been performed by the LHCb experiment~\cite{Aaij:2015oid}.
The Belle collaboration has recently also studied the set of optimised observables that correspond to $S_3$--$S_9$~\cite{Abdesselam:2016llu}.
In order to measure these observables with a modest number of events, Belle exploits the folding technique introduced in Ref.~\cite{Aaij:2013qta}.
This reduces the amount of information that is extracted from the angular distribution in a single fit to the data.

\begin{figure}[!htb]
\centering
\includegraphics[width=0.45\linewidth]{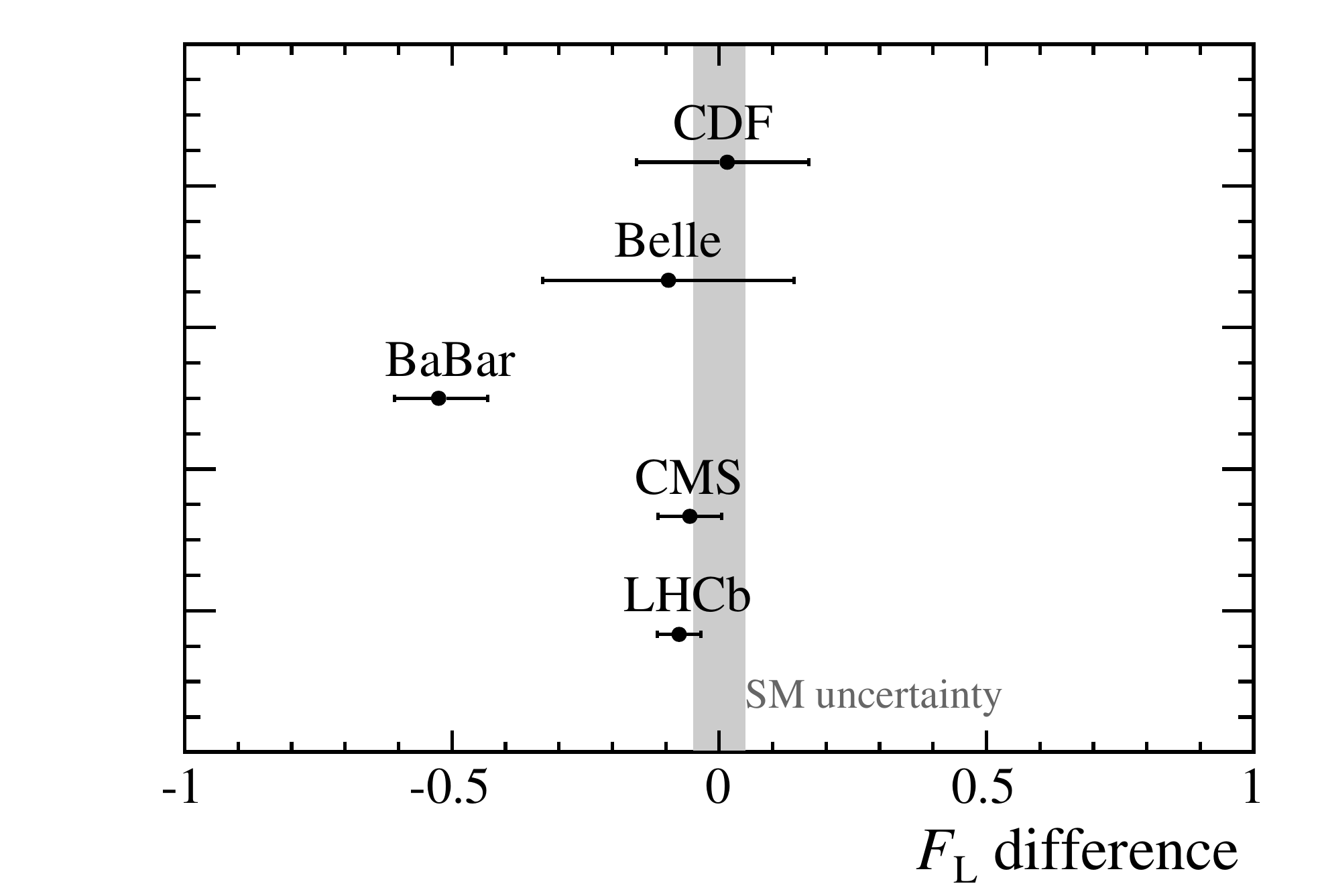}
\includegraphics[width=0.45\linewidth]{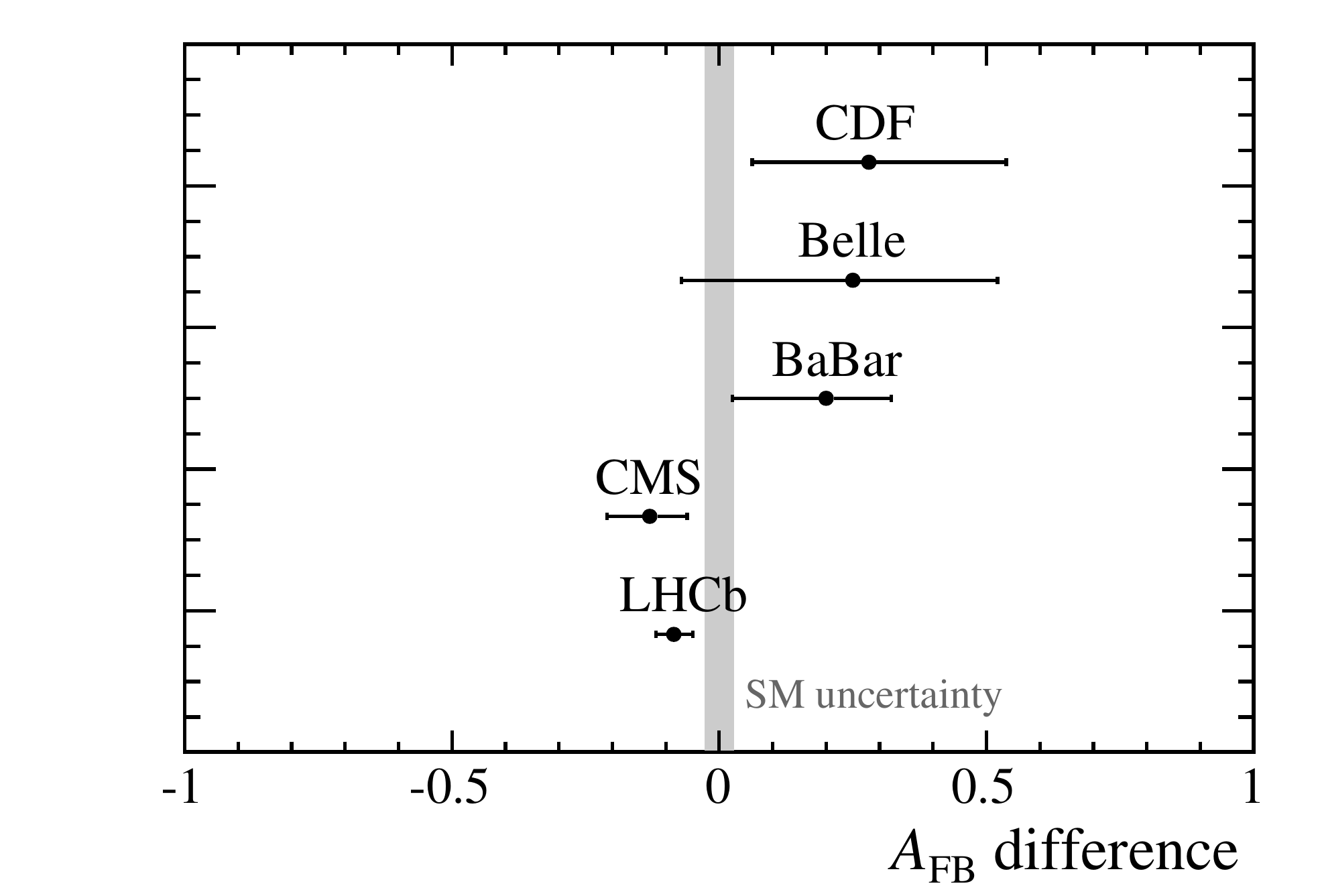} \\
\caption{
Fraction of longitudinal polarisation $F_{\rm L}$ of the $K^*$ system, $F_{\rm L}$, and dilepton system forward-backward asymmetry $A_{\rm FB}$ measured by the BaBar~\cite{Lees:2015ymt}, Belle~\cite{Wei:2009zv}, CDF~\cite{Aaltonen:2011ja}, CMS~\cite{Khachatryan:2015isa} and LHCb~\cite{Aaij:2015oid} collaborations in the dimuon mass squared range $1 < q^2 < 6{\rm GeV}^{2}$. The SM central values for the observables has been subtracted using the SM predictions from Ref.~\cite{Straub:2015ica}.
\label{fig:semileptonic:AFB}
}
\end{figure}

The latest measurements of $F_{\rm L}$ and $A_{\rm FB}$ from BaBar~\cite{Lees:2015ymt}, Belle~\cite{Wei:2009zv}, CDF~\cite{Aaltonen:2011ja}, CMS~\cite{Khachatryan:2015isa} and LHCb~\cite{Aaij:2015oid} in the range $1 < q^2 < 6 \,{\rm GeV}^2$ are shown in Fig.~\ref{fig:semileptonic:AFB}.
At the B-factories angular measurements combine isospin modes and electron and muon final states.
At the LHC, the experiments have so far only considered the $B \to K^{*0} \mu^+\mu^-$ final-state.
The measurements are generally in good agreement with SM predictions, with the exception of the BaBar measurement of $F_{\rm L}$.
This measurement is about $3\sigma$ below the SM expectation and appears to exhibit a sizeable isospin asymmetry between $B^0$ and $B^+$ decays~\cite{Lees:2015ymt}.
In the SM $A_{\rm FB}$ has a characteristic behaviour, where it starts with one sign and changes sign at $q^2 \approx 4\,{\rm GeV}^2$~\cite{Ali:1999mm}.
The change in sign comes from the interplay between the contributions from the ${\cal C}_7$ and ${\cal C}_9$ Wilson coefficients to the decay amplitudes.
This behaviour is reproduced by both the LHCb~\cite{Aaij:2015oid}  and CMS~\cite{Khachatryan:2015isa} measurements.
The LHCb collaboration measures the zero-crossing-point of $A_{\rm FB}$ to be $q^2 \in [ 3.7, 4.8 ] {\rm GeV}^2$ at 68\% CL~\cite{Aaij:2015oid}.

\begin{figure}[!htb]
\centering
\includegraphics[width=0.6\linewidth]{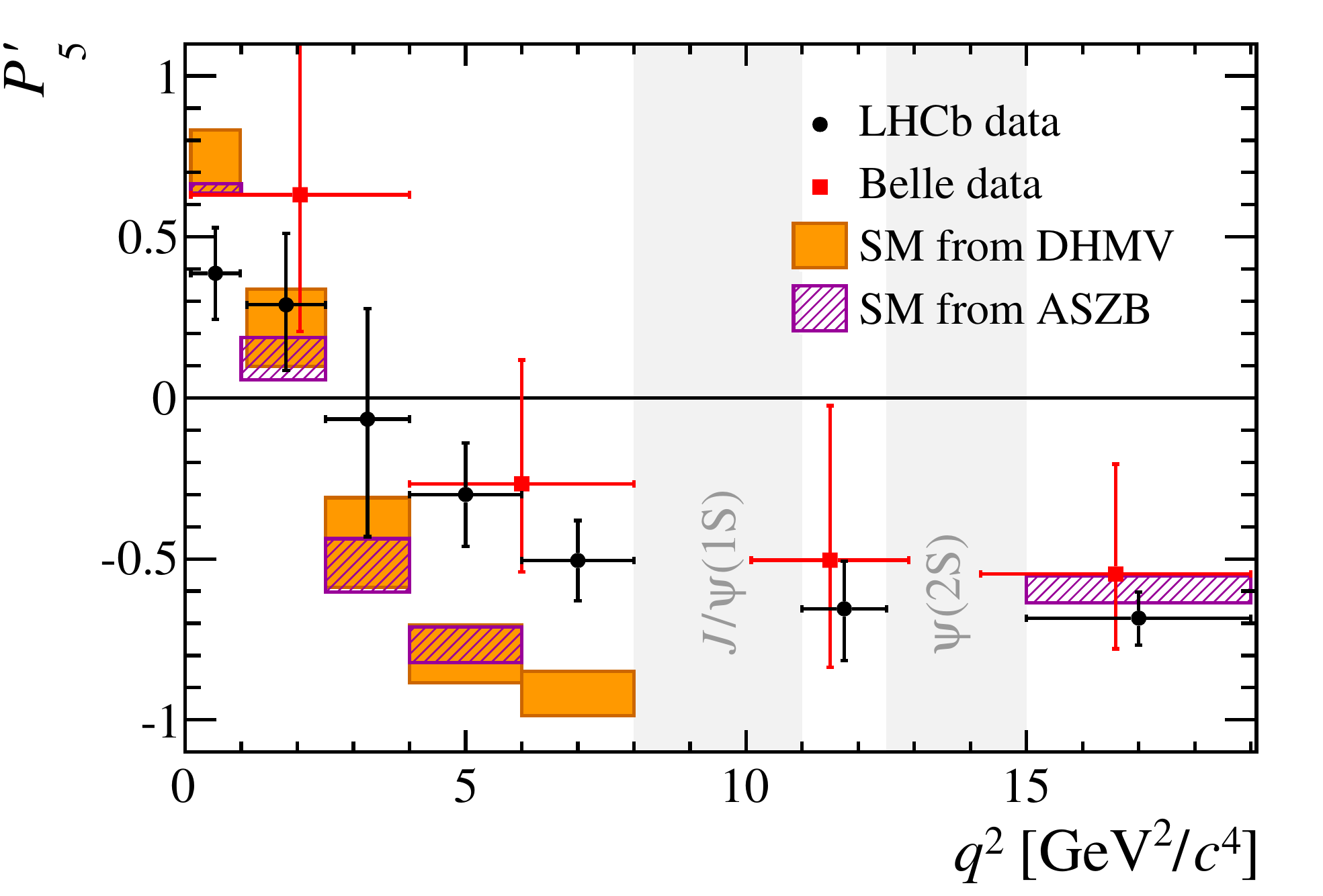}
\caption{
Optimised observable $P'_5$ measured by the Belle~\cite{Abdesselam:2016llu} and LHCb~\cite{Aaij:2015oid} collaborations as a function of dimuon invariant squared, $q^2$. The data overlay SM predictions from Refs.~\cite{Descotes-Genon:2014uoa} (DHMV) and Refs.~\cite{Straub:2015ica,Altmannshofer:2014rta} (ASZB).  No predictions are included close to the narrow charmonium resonances where the SM calculations are thought to break down.
\label{fig:semileptonic:P5p}
}
\end{figure}

The majority of the observables measured by the LHCb experiment in its full angular analysis are also in good agreement with SM predictions.
However, in the large recoil (low $q^2$) region two of the LHCb measurements of the $P'_5$ observable are about $3\,\sigma$ from the SM predictions.
The LHCb measurement and two different theoretical predictions are shown in Fig.~\ref{fig:semileptonic:P5p}.
The discrepancy seen by LHCb is also seen by Belle~\cite{Abdesselam:2016llu} in their analysis of the $P'_5$ angular observable.
There is no evidence for any non-zero $C\!P$ asymmetry in any of the angular observables measured by LHCb.

The LHCb experiment has also performed a separate analysis of the $B \to K^* \mu^+ \mu^-$ decay using angular moments~\cite{Aaij:2015oid,Beaujean:2015xea}.
The results of the moment analysis are consistent with those of the maximum-likelihood fits that are typically used to measure the angular observables.
The moment technique ultimately provides a less precise determination of the angular observables but has allowed the LHCb experiment to determine the observables in finer $q^2$ ranges than would be possible using a maximum-likelihood fit.
The moment analysis also makes fewer assumptions about the structure of the angular distribution \emph{e.g.} it makes no assumption about the presence of scalar and tensor operators.

Unfortunately it is not possible to measure the $P'_5$ observable in $B_s \to \phi \ell^+\ell^-$ decays without flavour tagging and a flavour tagged analysis of the $B_s \to \phi \mu^+\mu^-$ decay is not possible with the current LHCb dataset; the effective tagging power of the LHCb experiment is around 4\%~\cite{Aaij:2014zsa}. The time-integrated observables measured by LHCb in this analysis are fully consistent with SM expectations~\cite{Aaij:2015esa}.

In $\Lambda_b \to \Lambda \ell^+\ell^-$ decays, there is a new forward-backward asymmetry, $A_{\rm FB}^h$ that can be measured in the $\Lambda$ decay.
This asymmetry is only present because the $\Lambda$ baryon decays weakly.
The angular observables $A_{\rm FB}$ and $A_{\rm FB}^{h}$ in $\Lambda_b \to \Lambda \mu^+\mu^-$ decays have been measured by the LHCb experiment by fitting the projections of $\cos\theta_{\ell}$ and the $\Lambda$ decay angle (the equivalent to $\cos\theta_K$).
In the  $15 < q^2 < 20\,{\rm GeV}^2$ region LHCb measures~\cite{Aaij:2015xza}
\begin{align}
A_{\rm FB} &= -0.05\pm0.09_{\rm stat} \pm 0.03_{\rm syst}  \\
A_{\rm FB}^h & = -0.29 \pm 0.07_{\rm stat} \pm 0.03_{\rm syst}
\end{align}
These are reasonably consistent with SM expectation of
\begin{align}
A_{\rm FB}^{\rm SM} &=   -0.350 \pm 0.013  \\
A_{\rm FB}^{h, {\rm SM}} & = -0.271 \pm 0.009
\end{align}
from Ref.~\cite{Detmold:2016pkz} that uses form-factors from Lattice QCD.
However, it should be noted that at present no dilepton system forward-backward $A_{\rm FB}$ is seen.

\subsection{Lepton universality tests}
\label{sec:semileptonic:universality}

\begin{figure}[!htb]
\centering
\includegraphics[width=0.6\linewidth]{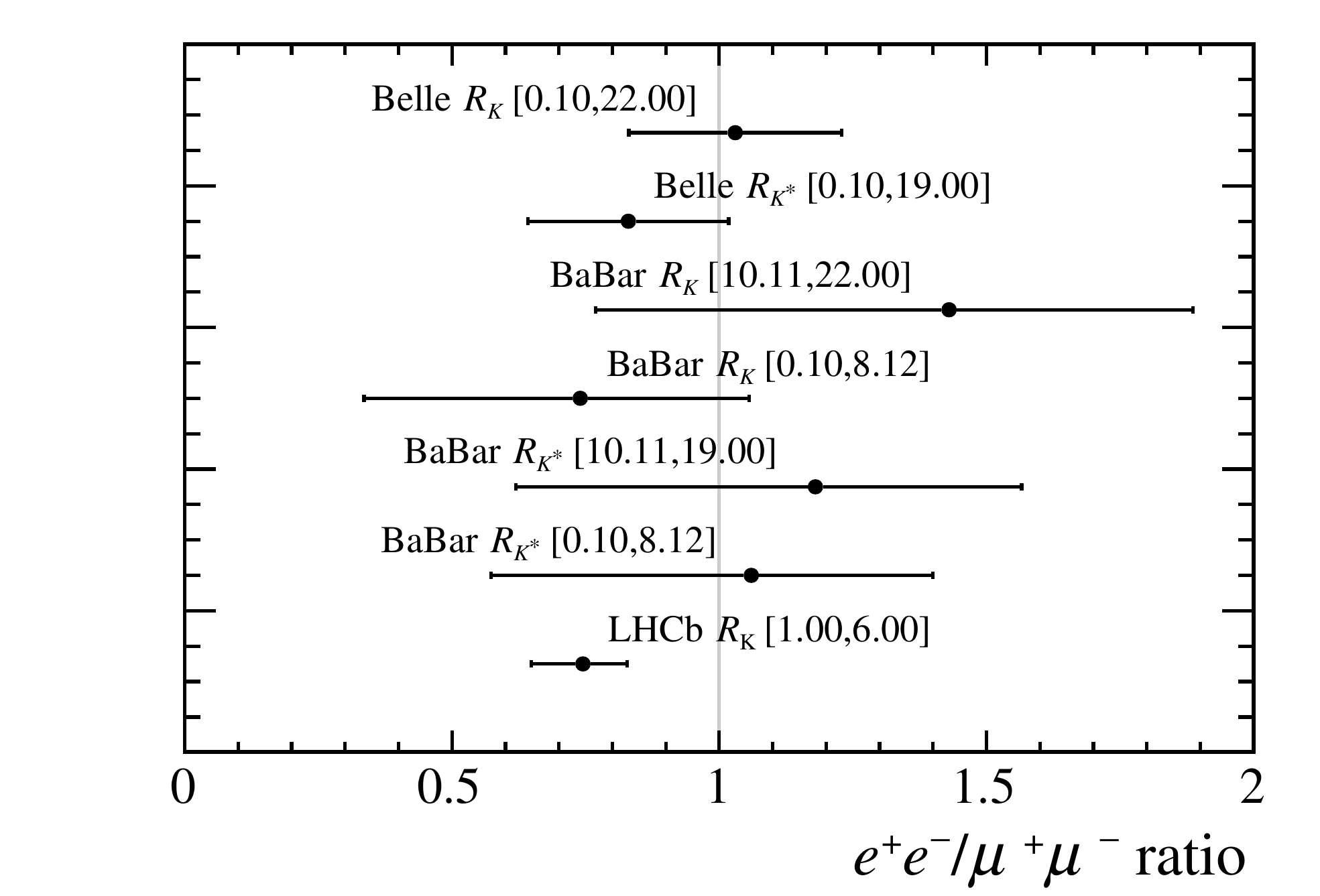}
\caption{
The ratios $R_{K^{(*)}} = \Gamma[B\to K^{(*)}\mu^+\mu^-] / \Gamma[B\to K^{(*)}\mu^+\mu^-]$ measured by the B-factories~\cite{Lees:2012tva,Wei:2009zv} and LHCb experiment~\cite{Aaij:2014ora}.  The SM expectation is $R_{K^{(*)}} \simeq 1$.
\label{fig:semileptonic:rratio}
}
\end{figure}

Whilst individual branching fractions can be affected by hadronic uncertainties, the ratio of partial widths of semileptonic decays with different flavours of leptons in the final state constitute precise tests of the SM.
In the SM, the gauge bosons couple equally to the different flavours of lepton.
The Higgs boson's coupling to mass is the only non-universal coupling and this has a negligible effect on the partial widths of the decays.
Ratios of the partial widths, referred to as $R$-ratios are therefore expected to be unity up to corrections from phase-space differences due to the different masses of the leptons.
For example the ratio
\begin{align}
R_{K^{(*)}} = \Gamma[B\to K^{(*)}\mu^+\mu^-] \Big/ \Gamma[B\to K^{(*)}e^+e^-]
\end{align}
evaluated in the $q^2$ range $1 < q^2 < 6\,{\rm GeV}^{2}/c^4$ is expected to be $R_{\rm K^{(*)}}^{\rm SM}[1,6] = 1.000\pm0.001$~\cite{Bobeth:2007dw}.
Similarly precise predictions can be made for other hadronic systems.
From the experimental perspective the main challenge in performing such a measurement is in understanding the difference in performance to reconstruct electrons and muons.
The most important difference arises from final state radiation from the electrons, which causes the electrons to radiate a significant amount of energy in the detector. The known QED corrections to the decay process are typically smaller than the energy loss due to Bremsstrahlung in the LHCb detector.
An overview of the available experimental measurements of the $R$-ratios is provided in Fig.~\ref{fig:semileptonic:rratio}.
The single most precise measurement comes from the LHCb experiment.
Using its full Run\,1 dataset the LHCb experiment measures~\cite{Aaij:2014ora}
\begin{align}
R_{K}[1,6] = 0.745 \,^{+0.090}_{-0.074} ({\rm stat}) \,^{+0.035}_{-0.035}({\rm syst})\,~,
\end{align}
which is 2.6 standard deviations from the SM expectation of unity.

\begin{figure}[!htb]
\begin{center}
\includegraphics[width=0.45\linewidth]{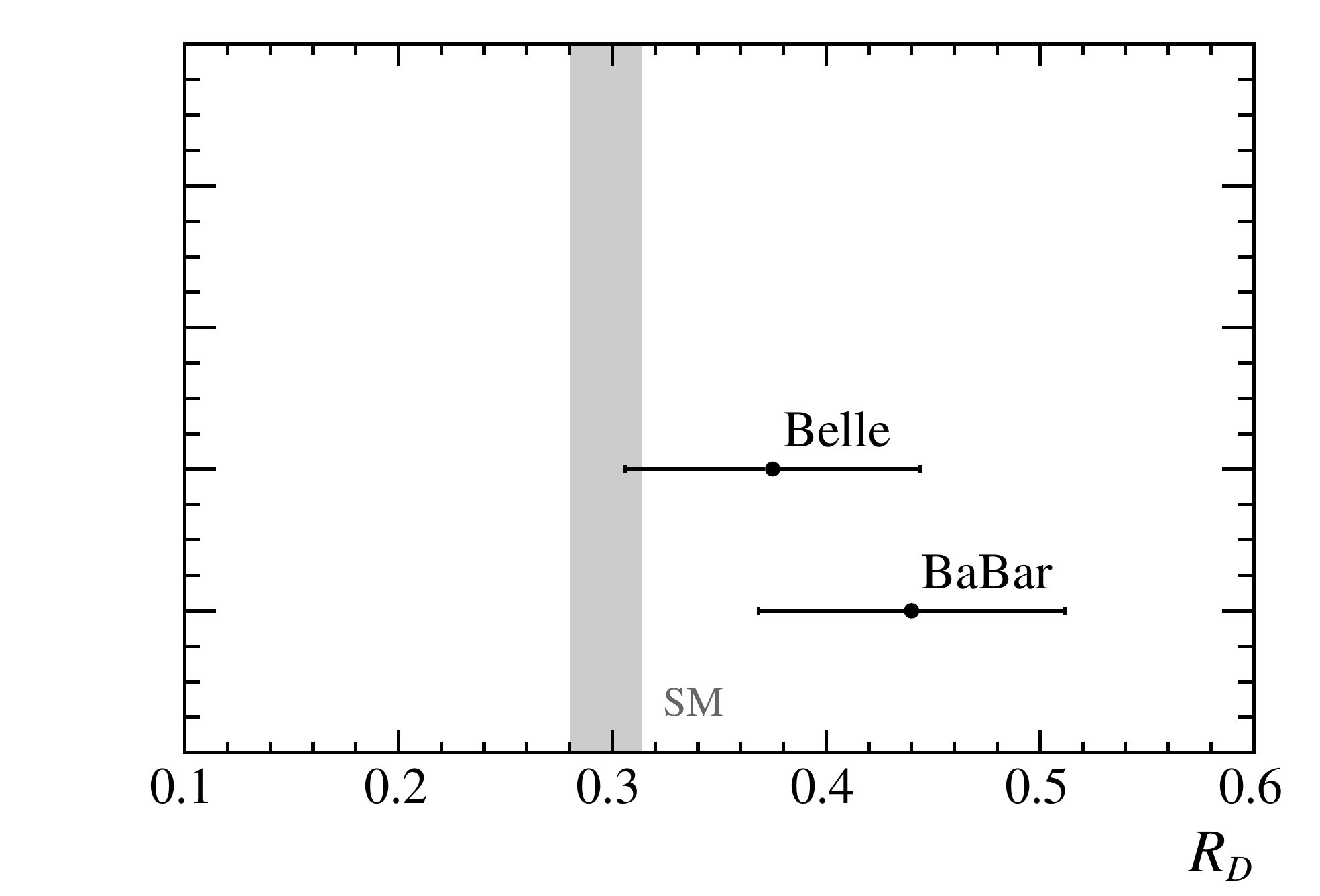}
\includegraphics[width=0.45\linewidth]{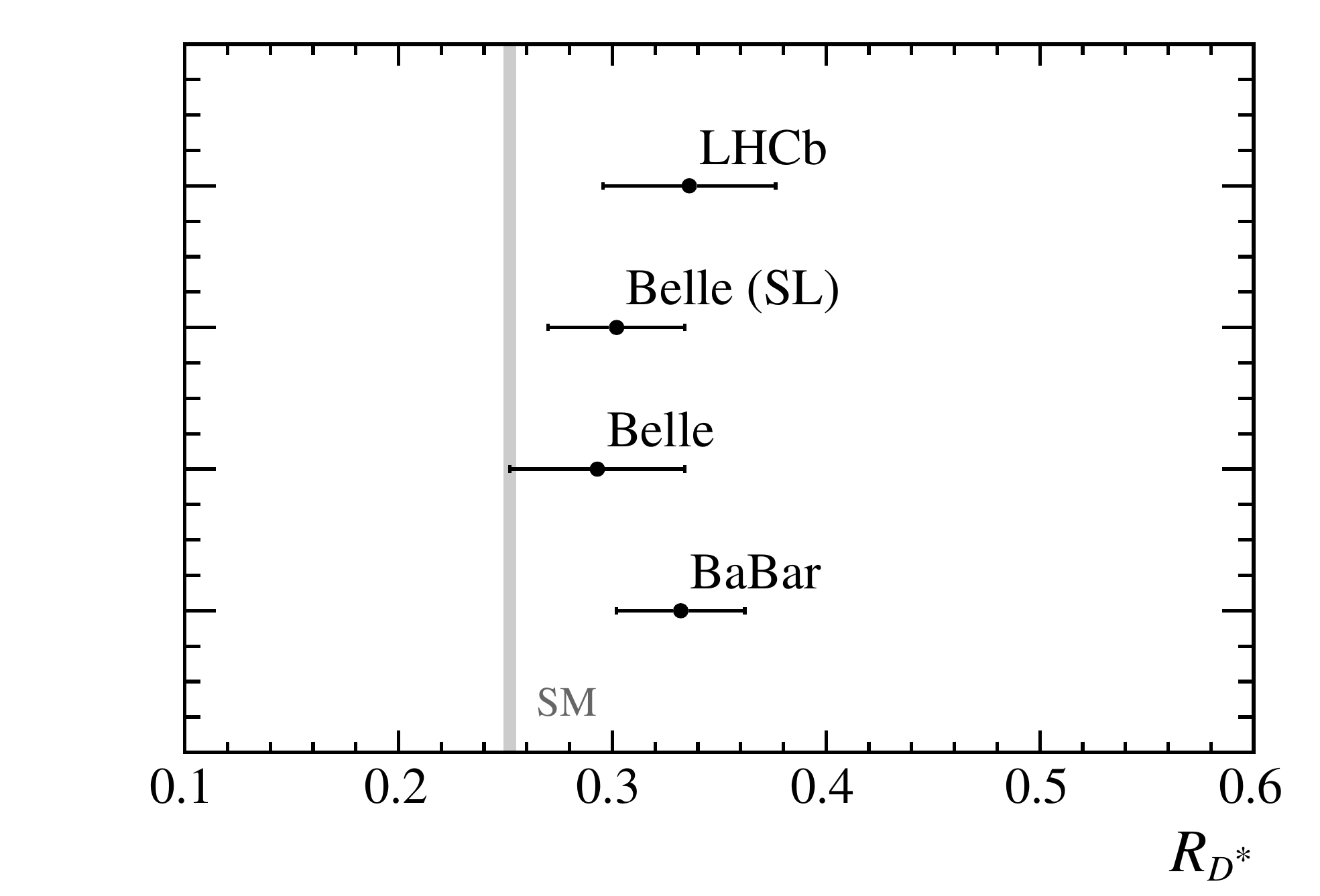}
\end{center}
\caption{
Measurement of the ratios $R_{D}$ and $R_{D^{*}}$ by the BaBar~\cite{Lees:2012xj}, Belle~\cite{Huschle:2015rga, Abdesselam:2016cgx} and LHCb~\cite{Aaij:2015yra} collaborations. Two results are shown from Belle,  one uses a hadronic reconstruction of the other $B$ meson in the event and the other reconstructs it through its dominant semileptonic decays.
The SM predictions from Refs.~\cite{Kamenik:2008tj} and \cite{Fajfer:2012vx} are indicated by the shaded regions.
\label{fig:semileptonic:RD}
}
\end{figure}

The LHCb measurement of $R_{K}$ is not the only hint of lepton non-universality that has appeared in recent years.
In semileptonic decays the ratios
\begin{align}
R_{D^{(*)}} = \Gamma[B \to D^{(*)} \tau\nu] \Big/ \Gamma[B \to D^{(*)} \mu \nu]
\end{align}
are both larger than the corresponding SM expectation.
A summary of  $R_D$ and $R_{D^*}$ measurements\footnote{This summary does not include the recently published Belle measurement of the $\tau$ lepton polarization in the  decay ${\bar B} \rightarrow D^* \tau^- {\bar \nu_{\tau}}$~\cite{Abdesselam:2016xqt}.} is provided in Fig.~\ref{fig:semileptonic:RD}.
The SM predictions from Refs.~\cite{Kamenik:2008tj} and \cite{Fajfer:2012vx}, based on HQET form factors, are also indicated in the figure.
A recent combination by HFAG indicates a tension with the SM predictions at 4.0 standard deviations~\cite{Amhis:2014hma}.
Note, in this case this is a large effect in a tree-level, Cabibbo favoured, decay mode.

\subsubsection{Experimental prospects}

The LHC experiments will collect very large samples of exclusive $b \to s\mu^+\mu^-$ decays during run 2 of the LHC, particularly modes without $K_S$ or $\pi^0$ in the final-state. With the planned upgrade of the LHCb experiment, the precision on angular observables in the $B \to K^* \mu^+ \mu^-$ decay will reach a level of $\lesssim 0.01$ in the current $q^2$ binning. In many cases the experimental measurements of the angular observables will be much more precise than their corresponding SM predictions. The large datasets will also enable the experiments to further sub-divide the data to better test the $q^2$ dependence of the SM predictions.
By the end of data taking at the LHC, the LHC experiments will also record sizeable samples of exclusive $b \to d \mu^+\mu^-$ decays. With $50\,{\rm fb}^{-1}$ the LHCb experiment expects to reach a precision of $2.5\%$.

The ultimate precision that can be achieved on exclusive  $b \to s\mu^+\mu^-$ branching fraction measurements is already limited by our knowledge of the $B \to J/\!\psi [K,K^*,\ldots]$ branching fractions. These branching fractions are needed as an input to normalise the rates measured by the LHC experiments (where the $b\bar{b}$ production is not precisely known). It should be possible to significantly improve the branching fractions of these normalisation channels using the Belle II dataset. The LHCb and Belle II experiments will also provide excellent precision on the branching fraction of exclusive $b \to s e^+ e^-$ decays. Belle II is expected to measure $R_{K^{(*)}}$ to a precision of $0.02$~\cite{Aushev:2010bq}. The upgraded LHCb experiment, with $50\,{\rm fb}^{-1}$ of integrated luminosity, should reach a similar precision assuming some improvement of the 0.035 systematic uncertainty on the current measurement can be achieved. The large samples of exclusive $b \to s e^+ e^-$ decays available at LHCb and Belle II will also allow for comparisons to be made between the angular distributions of $b \to s\mu^+\mu^-$ and $b \to s e^+ e^-$ decays.

\subsection{Decays with neutrinos in the final state}
\label{sec:semileptonic:neutrinos}

Decays based on the FCNC transition $b\to q\nu\bar\nu$ lead to final states
with missing energy in experiments. Beyond the SM, the same signature could
also arise from the presence of light feebly interacting particles in the
final state, as will be discussed in Sec.~\ref{sec:hidden}.
The exclusive decays $B\to K \nu\bar\nu$ and $B\to K^* \nu\bar\nu$ are the most promising probes of the
$b\to s\nu\bar\nu$ transition while the  $b\to d\nu\bar\nu$ transition is
probed by the $B\to (\pi,\rho) \nu\bar\nu$ decays.
The exclusive decays $B^+\to \pi^+\nu\bar\nu$
and $B^+\to \rho^+\nu\bar\nu$ unfortunately receive large backgrounds from $B^+ \to \tau^+  (\to \pi^+ \bar{\nu}) \nu$ and $B^+ \to \tau^+  (\to \rho^+ \bar{\nu}) \nu$
decays, respectively;
a similar contribution is present for $B^+\to K^{(*)+}\nu\bar\nu$ decays but
in this case only amounts to a 5--10\% correction in the SM
\cite{Kamenik:2009kc}.

\subsubsection{Standard Model predictions}

Compared to exclusive semi-leptonic decays with charged leptons, $b \to q \nu\bar{\nu}$ decays  are
theoretically cleaner as there are no hadronic uncertainties beyond the QCD
form factors. Photon-mediated contributions to these decays are absent.
Using $B\to K^*$ form factors from a combined fit to LCSR and
lattice results \cite{Straub:2015ica},
$B\to \rho$ form factors from LCSR \cite{Straub:2015ica},
and $B\to K$ and $B\to \pi$ form factors from lattice QCD
\cite{Bailey:2015dka,Lattice:2015tia,Bailey:2015nbd},
one obtains the SM predictions
\begin{align}
\begin{split}
\text{BR}(B^+\to K^{+}\nu\bar\nu)_\text{SM}
&= ( 4.6 \pm 0.5 ) \times 10^{-6}
\,,\\
\text{BR}(B^+\to K^{*+}\nu\bar\nu)_\text{SM}
&= ( 8.4 \pm 1.5 ) \times 10^{-6}
\,,\\
\text{BR}(B^0\to \pi^0\nu\bar\nu)_\text{SM}
&= ( 0.6 \pm 0.3 ) \times 10^{-7}
\,,\\
\text{BR}(B^0\to \rho^0\nu\bar\nu)_\text{SM}
&= ( 2.0 \pm 0.4 ) \times 10^{-7}
\,,
\end{split}
\end{align}
where the tree-level contributions involving an intermidiate $\tau$ have not been included in the $B^+$ modes.
A sum over all three neutrino flavours is implied.
For the $B^0\to \pi^0\nu\bar\nu$ decay, a more precise prediction can be obtained
by extracting the $B\to \pi$ form factors from $B\to \pi\ell\nu$ decays
\cite{Du:2015tda}.
The partial widths of the charged and neutral modes are equal for the
$B\to K^{(*)}\nu\bar\nu$ modes.
For $B^+\to \rho^+\nu\bar\nu$
and $B^+\to \pi^+\nu\bar\nu$, the FCNC contributions are both a factor 2 larger than for the
corresponding neutral modes, but they are affected by the tree-level background
mentioned above. The direct $C\!P$ asymmetries of all neutral modes vanish
in the SM and beyond as there is no strong phase. An angular analysis
of the $B\to K^*(\to K\pi)\nu\bar\nu$ decay would allow to access the $K^*$
longitudinal polarisation fraction $F_L$ that is sensitive to right-handed
currents beyond the SM \cite{Altmannshofer:2009ma}.

\subsubsection{New physics sensitivity}

While $b\to q\nu\bar\nu$ transitions are governed by the operator $Q_\nu$ in
the SM, beyond the SM there can be six different operators
$(Q_\nu^{(\prime)})_s^{e,\mu,\tau}$
for $b\to s\nu\bar\nu$ and the same number for $b\to d\nu\bar\nu$.
The modification of the branching fractions to pseudoscalar or vector mesons
can be written as \cite{Buras:2014fpa}
\begin{align}
\frac{\text{BR}(B\to P\nu\bar\nu)}{\text{BR}(B\to P\nu\bar\nu)_\text{SM}}   & =
  \frac{1}{3}\sum_\ell (1 - 2\,\eta_\ell)\epsilon_\ell^2
 \,, &
\frac{\text{BR}(B\to V\nu\bar\nu)}{\text{BR}(B\to V\nu\bar\nu)_\text{SM}}
& =
  \frac{1}{3}\sum_\ell (1 +  \kappa_V \eta_\ell)\epsilon_\ell^2
  \,,
\label{eq:bqnunu-np}
\end{align}
where
\begin{equation}  \label{eq:epsetadef2}
 \epsilon_\ell = \frac{\sqrt{ |C_\nu^\ell|^2 + |C_\nu^{\prime\ell}|^2}}{|C_\nu^\text{SM}|}~,
\qquad
 \eta_\ell = \frac{-\text{Re}\left(C_\nu^\ell C_\nu^{\prime\ell *}\right)}{|C_\nu^\ell|^2 +
|C_\nu^{\prime\ell}|^2}
\end{equation}
and $\kappa_V$ is a process-dependent ratio of integrals over hadronic form factors \cite{Buras:2014fpa}.
This dependence shows that the decays with pseudoscalar mesons and with vector
mesons are complementary and combining them would allow to disentangle
new physics in left-handed vs. right-handed currents.

An important point in any new physics model contributing to $b\to q\nu\bar\nu$ is
the possible correlation with the $b\to q\ell^+\ell^-$ decays, that are strongly
constrained experimentally, especially for $q=s$. It is instructive to consider several
benchmark cases.
\begin{itemize}
 \item Models with left-handed flavour-changing $Z$ couplings predict
\begin{equation}
 [(C_{10})^\ell_q]^\text{NP} = [(C_\nu)^\ell_q]^\text{NP} \,,\qquad
 [(C_{9})^\ell_q]^\text{NP} = -0.08 \, [(C_\nu)^\ell_q]^\text{NP} \,,
\label{eq:bsnunu-lhz}
\end{equation}
 and are lepton flavour universal. This implies that an enhancement
 or suppression of $B\to K^{(*)}\nu\bar\nu$ requires a similar
 enhancement  or suppression of $B_s\to \ell^+\ell^-$ and an enhancement
 or suppression of $B\to (\pi,\rho)\nu\bar\nu$ requires a similar
 enhancement  or suppression of $B_d\to \ell^+\ell^-$.
\item In models where the tree-level exchange of a leptoquark contributes
to $b\to q\nu\bar\nu$, the correlation with $b\to d\ell^+\ell^-$ can be used
to determine
the spin and gauge quantum numbers of the leptoquark \cite{Buras:2014fpa}
(see also \cite{Dorsner:2016wpm}).
\item In models with new physics coupling dominantly to third-generation leptons,
the experimental limits on $B\to K^{(*)}\nu\bar\nu$ are already relevant,
as there are few bounds on $b\to s\tau^+\tau^-$ processes
(cf. Sec.~\ref{sec:semileptonic:universality}).
\end{itemize}
Finally, we note that in models with lepton flavour violation, also
operators of the type $(Q_\nu^{(\prime)})_q^{\ell_i\ell_j}$ with
$\ell_i\neq\ell_j$ contribute to the same experimental final state, as the
flavour of the neutrinos is not detected.

%------------------------------------------------------------------
\subsubsection{Searches for \texorpdfstring{$B \to (K, K^{*}) +  X_{\rm invisible}$}{B ->(K, K*) +  X(invisible)}}
\label{sec:bqnunu-exp}
%------------------------------------------------------------------

The experimental challenge is to distinguish a $B$ decay to
an $X_s$ system and two missing neutrinos, as it is difficult to suppress the background at the level of the SM predictions.
%The smallness of the SM branching fractions
%makes suppression of the background difficult.

\begin{table}[tbp]
\centering
\renewcommand{\arraystretch}{1.2}
\begin{tabular}{lcc}
\hline
Mode & BR upper limit & Ref. \\
\hline
$\text{BR}(B^+ \to K^+ \nu \overline{\nu})$ & $< 1.6 \times 10^{-5}$  & \cite{delAmoSanchez:2010bk} \\
$\text{BR}(B^0 \to K^0 \nu \overline{\nu})$ & $< 4.9 \times 10^{-5}$   & \cite{delAmoSanchez:2010bk} \\
$\text{BR}(B^+ \to K^{*+} \nu \overline{\nu})$ & $< 4.0 \times 10^{-5}$  & \cite{Lutz:2013ftz} \\
$\text{BR}(B^0 \to K^{*0} \nu \overline{\nu})$ & $< 5.5 \times 10^{-5}$   & \cite{Lutz:2013ftz} \\
\hline
$\text{BR}(B^+ \to \pi^+ \nu \overline{\nu})$ & $< 9.8 \times 10^{-5}$    & \cite{Lutz:2013ftz} \\
$\text{BR}(B^0 \to \pi^0 \nu \overline{\nu})$ &  $< 6.9 \times 10^{-5}$   &  \cite{Lutz:2013ftz}\\
$\text{BR}(B^+ \to \rho^+ \nu \overline{\nu})$ & $< 2.1 \times 10^{-4}$  & \cite{Lutz:2013ftz} \\
$\text{BR}(B^0 \to \rho^0 \nu \overline{\nu})$ & $< 2.1 \times 10^{-4}$  &  \cite{Lutz:2013ftz}\\
\hline
\end{tabular}
\caption{90\% CL\ upper bounds on $b\to q\nu\bar\nu$ branching fractions
from BaBar \cite{Aubert:2008am,delAmoSanchez:2010bk,Lees:2013kla} and Belle \cite{Lutz:2013ftz}.
Note that the modes $B^+ \to \pi^+ \nu \overline{\nu}$ and
$B^+ \to \rho^+ \nu \overline{\nu}$ are affected by large tree-level backgrounds
as discussed in the text.}
\label{tab:bsnunu-exp}
\end{table}

A summary of the strongest experimental limits from the BaBar and Belle experiments
is given in Table~\ref{tab:bsnunu-exp}.
The limits are already about 3 and 5
times the SM predictions for $K \nu \overline{\nu}$  and $K^* \nu \overline{\nu}$, respectively,
but the backgrounds are severe. Belle II with a data sample of about  $50$ ab$^{-1}$ could possibly observe
either of these decays.
The limits on the $b\to d\nu\bar\nu$ decays are instead almost three orders of
magnitude above the SM expectations.

%-----------------------------------------------------------------------------------
\section{Searches for light particles in rare \texorpdfstring{$B$}{B} decays}
\label{sec:hidden}
%-----------------------------------------------------------------------------------

To solve the hierarchy problem, most extensions of the SM introduce new particles at the TeV mass-scale with sizeable couplings to the SM particles.
These models can also provide dark matter candidates, for example the lightest supersymmetric partner in SUSY models.
The lack of any evidence for new particles in LHC collisions has rekindled the interest in hidden sector theories  (see for example Ref.~\cite{Essig:2013lka}).
In contrast to most SM extensions, these theories postulate that dark matter particles do not carry SM charges and couple only feebly to the SM particles.
Hidden sector particles are singlet states under the SM gauge interactions.
Couplings between the SM and hidden-sector particles arise via mixing of the hidden-sector field with a SM ``portal'' operator.
As an example, consider a new spontaneously broken $U(1)$ gauge symmetry.
The mediator of this symmetry can be a new vector boson, a dark sector photon.
The only (renormalisable) interaction of the dark photon with the SM is mixing with the SM hypercharge gauge boson (the photon and the $Z^0$).
If the mediator is a scalar it can couple to the SM-like Higgs via mixing.

More generally, lower dimensional renormalisable portals in the SM can be classified
into the following  types:
\begin{center}
\begin{tabular}{rl}
Portal & Coupling \\
\hline
\smallskip
Dark Photon, $A_{\mu}$ &  $-\tfrac{\epsilon}{2 \cos\theta_W} F'_{\mu\nu} B^{\mu\nu}$ \\ \smallskip
Dark Higgs, $S$  & $(\mu S + \lambda S^2) H^{\dagger} H$  \\ \smallskip
Axion, $a$ &  $\tfrac{a}{f_a} F_{\mu\nu}  \tilde{F}^{\mu\nu}$, $\tfrac{a}{f_a} G_{i, \mu\nu}  \tilde{G}^{\mu\nu}_{i}$ \\ \smallskip
Sterile Neutrino, $N$ & $y_N L H N$ \\
\end{tabular}
\end{center}
Here, $F'_{\mu \nu}$ is the field strength for the dark photon, which couples to the hypercharge field, $B^{\mu\nu}$; $S$ is a new scalar singlet that couples to the Higgs doublet, $H$, with dimensionless and dimensional couplings, $\lambda$ and $\mu$; $a$ is a pseudoscalar axion that couples to a dimension-4 diphoton or digluon operator; and $N$ is a new neutral fermion that couples to one of the left-handed doublets of the SM and the Higgs field with a Yukawa coupling $y_N$.

The axion is an important example of a dark sector theory.
The Peccei-Quinn axion~\cite{Peccei:1977hh} was originally introduced as a solution to the strong $C\!P$ problem where the complex nature of the QCD vacuum leads to a term in the QCD lagrangian,
\begin{align}
\frac{\theta}{32\pi} G_{i, \mu\nu}  \tilde{G}^{\mu\nu}_{i}.
\end{align}
This term is $C\!P$ violating and its absence in nature leads to a large fine tuning problem.
The mixing angle, $\theta$ associated to this term is constrained to be $\theta < 10^{-10}$ by existing limits on the neutron electric-dipole-moment.
The axion provides a natural explanation of why this term is small.

Decays of $b$ hadrons offer a unique window to probe dark-sector models with particles at the GeV mass-scale.
These states are either invisible, and  hence do not interact with the detector, or decay back into SM particles via mixing with the SM gauge bosons.
If the new states decay back to SM particles, their lifetime can be large leading to a striking experimental signature.
The decays of these new states may also violate quantum numbers that are accidentally conserved in the SM, for example they may violate charged lepton flavour
and number, or baryon number.

%-------------------------------------------------------------------------------------
\vskip 2mm \noindent
\subsection{Search for light particles decaying into invisible final states}
\label{ssec:light-invisible}
%-------------------------------------------------------------------------------------

The most promising processes for the invisible final states are:
\begin{equation}
 B \to X_{\rm invisible} (\gamma),  \;\;\; B\to (\pi, \rho) + X_{\rm invisible}, \;\;\; B\to (K,K^*) + X_{\rm invisible}, \;\;\; B^+ \to X_{\rm invisible} \ell^+ ,
\end{equation}
with $X_{\rm invisible}$ made of at least two particles for the first mode, but possibly only one for the others.
This also includes situations in which the invisible particle is not stable but performs cascade decays in the hidden
sector, e.g. $X_{\rm invisible} \to  Y_{\rm invisible}Y_{\rm invisible}$.
If the width of $X_{\rm invisible}$ is not too large, then it is produced mostly
on-shell with the kinematics of a two-body decay.
Searches for the final states  $B\to (\pi, \rho) + X_{\rm invisible}$
and $B\to (K,K^*) + X_{\rm invisible}$ have already been discussed in
Sec.~\ref{sec:bqnunu-exp}.

%-------------------------------------------------------
\vskip 2mm
\subsubsection{\texorpdfstring{$B \to X_{\rm invisible} (\gamma)$}{B->X(invisible)γ}}
%-------------------------------------------------------

The SM process that fits into this category is the $B \to \nu \overline{\nu} (\gamma)$.
The  decay $B \to \nu \overline{\nu} (\gamma)$ is similar from a theoretical point of view to the leptonic
decays $B \to \ell^+ \ell^- (\gamma) $ (see Sec.~\ref{sec:leptonic}).
Without a photon, this decay is extremely suppressed in the SM due to helicity conservation, and it can only occur
due to the tiny but non-zero neutrino mass.
The radiation of a photon from an initial-state quark can remove the helicity suppression, resulting
in a larger branching fraction value than the non-radiative process.
Standard Model branching fractions for $B^0 \to \nu \overline{\nu}$
and $B^0 \to \nu \overline{\nu} \gamma$ have been computed to be
$\sim 1\times 10^{-25}$ and $2 \times 10^{-9}$, respectively~\cite{Badin:2010uh}.

Beyond the SM, there are several models that predict the existence of {\it light hidden particles} that do not interact in the detector.
For example, decays to pairs of scalar particles, $B^0 \to \chi_0 \chi_0 (\gamma)$ are not helicity suppressed and
hence can occur at rates much larger than the SM rate of $B \to \nu \overline{\nu} (\gamma)$ modes.
A branching fraction of $10^{-7}-10^{-6}$ is computed for the decay $B^0 \to \overline{\nu} \chi^0$,
in a phenomenological model where $\chi^0$ is a neutralino~\cite{Dedes:2001zia}.

Light hidden particles can also arise in models with large extra dimensions, which aim to provide
a possible solution to the hierarchy problem. In such models, a small rate of invisible $B^0$ decays can arise \cite{Agashe:2000rwa, Agashe:2000rw, Davoudiasl:2002fq}.

Because the signature for $B^0 \to  X_{\rm invisible} (\gamma)$ decays is
the absence of any detector activity, apart possibly from a neutral cluster in the calorimeter
associated to a photon, these decays can be searched for only at  B-factory experiments where
the decay of interest can be {\it tagged} from the other $B$ in the event.
The missing energy represents missing energy from any source including SM decays with neutrinos and non-SM decays with {\it light hidden particles}.

To date, the best limit on the $B^0 \to X_{\rm invisible}$ mode comes from the BaBar collaboration and is based on the full data sample
\cite{Lees:2012wv}, where an upper limit of $2.4 \times 10^{-5}$ at 90\% CL is set.
BaBar also reports limits on the $B^0 \to X_{\rm invisible}  \gamma$ branching fraction for $E_{\gamma} > 1.2$ GeV,
assuming decay kinematics as in the $B \to \nu \overline{\nu} (\gamma)$ analysis, based on a constituent quark model \cite{Lu:1996et},
of ${\rm BR}(B^0 \to X_{\rm invisible} + \gamma) < 1.7 \times 10^{-5}$ at 90\% CL.

%-------------------------------------------------------
\vskip 2mm
\subsubsection{\texorpdfstring{$B^+ \to X_{\rm invisible} \ell^+$}{B->X(invisible)l}}
%-------------------------------------------------------
Belle has performed a search for light non-SM particles with a mass in the range $0.1$--$1.8$ GeV/c$^2$ in the decays
$B^+ \to \mu^+ X_{\rm invisible}$ and $B^+ \to e^+ X_{\rm invisible}$~\cite{Park:2016gek}.
In this case there are further possibilities for the $X_{\rm invisible}$ candidate
hypotheses of new physics beyond the SM. One is sterile neutrinos in large extra dimensions~\cite{Agashe:2000rwa}
and in the neutrino minimal standard model ($\nu$MSM) that incorporates three light
singlet right-handed fermions~\cite{Gorbunov:2007ak}.
Another option is the lightest supersymmetric particle (LSP) in the minimal supersymmetric standard
model (MSSM)~\cite{Dedes:2001zia} with $R-$parity violation.
Belle finds no evidence of a signal and sets upper limits on the branching fraction of the order of $10^{-6}$ at 90\% CL.

%------------------------------------------------------------------------------------
\subsection{Search for light particles decaying into visible final states}
\label{ssec:light-displaced}
%------------------------------------------------------------------------------------

It is also possible for $B$ mesons to decay into new states with sizeable lifetime, which then decay into SM particles.
These states can be detected by the presence of
a displaced vertex with respect to the $B$ decay point.
Several models predict the presence of weakly-interacting long-lived particles.

Many theories predict that TeV-scale dark matter particles interact via GeV-scale bosons
\cite{ArkaniHamed:2008qn, Pospelov:2008jd, Cheung:2009qd}.
Previous searches for such GeV-scale particles have been performed using large data
samples from many types of experiments (see Ref.~\cite{Alekhin:2015byh} for a summary).
One class of models involves the scalar portal hypothesises that such a field was responsible for an inflationary period
in the early universe~\cite{Bezrukov:2009yw}, and may have generated the
baryon asymmetry observed today \cite{Hertzberg:2013jba,Hertzberg:2013mba}.
The associated inflaton particle is expected to have a mass in the range $270$--$1800$ MeV \cite{Bezrukov:2009yw}.
Another class of models invokes the axial-vector portal in theories of dark matter that seek
to address the cosmic ray anomalies~\cite{Chang:2008aa,Adriani:2008zr,Adriani:2011xv,Adriani:2013uda,FermiLAT:2011ab,Aguilar:2014mma},
and to explain the suppression of $C\!P$ violation in strong interactions~\cite{Peccei:2006as}.
To couple the axion portal to a hidden sector containing a TeV-scale dark matter particle, while also explaining the suppression of $C\!P$
violation in strong interactions, Ref.~\cite{Nomura:2008ru} proposes an axion with $360$--$800\,{\rm MeV}$ and an energy
scale, $f$, at which the symmetry is broken in the range 1--3 TeV. A broader range of mass and
energy scale values is allowed in other dark matter scenarios involving axions or axion-like states~\cite{Mardon:2009gw, Freytsis:2009ct}.
In both cases the hidden particle can be long-lived, hence showing a detached vertex with respect to the production point.

The LHCb experiment has published a search for a hidden-sector boson $\chi$ produced in the decay $B^0 \to K^{*0} \chi$
with $K^{*0} \to K^+ \pi^-$ and $\chi \to \mu^+ \mu^-$ based on $3\,{\rm fb}^{-1}$~\cite{Aaij:2015tna}.
No evidence for a signal is observed, and upper limits are placed on $\text{BR}(B^0 \to K^{*0} \chi)\times \text{BR}(\chi \to \mu^+\mu^-$)
as a function of mass and lifetime of the $\chi$ boson. These limits are of the order of $10^{-9}$ for $\chi$ lifetimes less than 100 ps and for
$m_{\mu\mu} < 1\,{\rm GeV}$.

%----------------------------------------------------------------------------------------
\section{Lepton flavour and lepton number violating \texorpdfstring{$b$}{b} hadron decays}
\label{sec:lfv}
%----------------------------------------------------------------------------------------

\subsection{Lepton flavour violating \texorpdfstring{$b$}{b} hadron decays}
\label{ssec:ssec_lfv}

Lepton flavour violating decays are forbidden  modes in the SM
as they violate charged lepton family numbers that are accidentally
conserved in the SM in the absence of neutrino masses.
However, lepton flavour is not protected by any fundamental symmetry in the SM
and in fact the existence of
neutrino mixing explicitly requires that lepton flavour is not conserved in the neutrino sector.
This in turn implies lepton flavour violation (LFV) in the charged lepton sector as well, via loop processes which contain neutrinos.
However, the expected rate for such processes is many orders of magnitude below the current or
foreseen experimental sensitivity to these decay modes.

Observation of LFV in $B$ decays would therefore be an unambiguous evidence for physics beyond the SM.
These decays are allowed in some scenarios beyond the SM that
include models with heavy singlet Dirac neutrinos~\cite{Ilakovac:1999md},
supersymmetric models~\cite{Diaz:2004mk}, and the Pati-Salam model~\cite{Pati:1974yy}.
In models with Higgs-mediated LFV, modes with heavier leptons generally are expected to exhibit larger
LFV than modes with lighter leptons.
For example, in the general flavour-universal MSSM, the branching fractions allowed for
$B^0 \to \ell^{\pm} \tau^{\mp}$ are $\sim 2\times 10^{-10}$~\cite{Dedes:2002rh}.

As described earlier, experimental searches for modes containing $\tau$ leptons
in the final state tend to be more difficult due to the multiple decay modes of the $\tau$ and missing energy
resulting from the presence of one or more neutrinos.
Consequently, experimental limits on $\mu-e$ LFV modes tend to be more stringent than $\tau-e$ or $\tau-\mu$.
The current best experimental limits on the LFV modes $B^0_{(s)} \to e^{\pm} \mu^{\mp}$ come from the LHCb collaboration~\cite{Aaij:2013cby},
$\text{BR}(B^0_s \to e^{\pm} \mu^{\mp}) < 1.1 \times 10^{-8}$ and
$\text{BR}(B^0 \to e^{\pm} \mu^{\mp}) < 2.8 \times 10^{-9}$ at 90\% CL, and supersede the results from CDF from Run 2 at the Tevatron~\cite{Aaltonen:2009vr}.
% overlap:
%The BaBar collaboration has performed a search for $B^0 \to \tau^{\pm} \ell^{\mp}$ (with $\ell = \mu$ or $e$)
%based on a data sample of 378 million $B^0 \overline{B}^0$ pairs~\cite{Aubert:2008cu}.
The BaBar collaboration has searched for the decay $B^0 \to \tau^{\pm} \ell^{\mp}$ (with $\ell = \mu$ or $e$)
using a sample of 378 million $B^0 \overline{B}^0$ pairs~\cite{Aubert:2008cu}.
No significant  signal  is  seen  in  either mode  and  limits  of ${\rm BR}(B^0 \to \tau^{\pm} e^{\mp}) <2.8 \times 10^{-5}$  and
${\rm BR}(B^0 \to \tau^{\pm} \mu^{\mp})< 2.2 \times 10^{-5}$ at 90 \% CL are obtained.
A summary of the limits for lepton flavour violating modes is reported in Table~\ref{tab:LFV}.
The best upper limits on semileptonic modes come from BaBar~\cite{Aubert:2006vb, Lees:2012zz, Aubert:2007mm}.

\begin{table}
\centering
\begin{tabular}{lcc}
\hline
Mode & BR upper limit &  Ref. \\
\hline
$B^0 \to \mu^{\mp} e^{\pm}$ & $< 2.8\times 10^{-9}$  & \cite{Aaij:2013cby} \\
$B^0 \to \tau^{\mp} e^{\pm}$ & $< 2.8\times 10^{-5}$  & \cite{Aubert:2008cu} \\
$B^0 \to \tau^{\mp} \mu^{\pm}$ & $< 2.2\times 10^{-5}$ & \cite{Aubert:2008cu} \\
$B_s \to \mu^{\mp} e^{\pm}$ & $< 1.1\times 10^{-8}$  & \cite{Aaij:2013cby}  \\
\hline
$B^+ \to K^+ \mu^{\mp} e^{\pm}$ & $< 9.1 \times 10^{-8}$  & \cite{Aubert:2006vb} \\
$B^+ \to K^{*+} \mu^{\mp} e^{\pm}$ & $< 1.4\times 10^{-6}$  & \cite{Aubert:2006vb} \\
$B^+ \to K^{+} \tau^{\mp} e^{\pm}$ & $< 3.0\times 10^{-5}$  & \cite{Lees:2012zz} \\
$B^+ \to K^{+} \tau^{\mp} \mu^{\pm}$ & $< 4.8\times 10^{-5}$  & \cite{Lees:2012zz} \\
$B^+ \to \pi^+ \mu^{\mp} e^{\pm}$ & $< 1.7 \times 10^{-7}$  & \cite{Aubert:2007mm} \\
$B^+ \to\pi^{+} \tau^{\mp} e^{\pm}$ & $< 7.5\times 10^{-5}$  & \cite{Lees:2012zz} \\
$B^+ \to \pi^{+} \tau^{\mp} \mu^{\pm}$ & $< 7.2\times 10^{-5}$  & \cite{Lees:2012zz} \\
$B^0 \to K^0 \mu^{\mp} e^{\pm}$ & $< 2.7 \times 10^{-7}$  & \cite{Aubert:2006vb} \\
$B^0 \to \pi^0 \mu^{\mp} e^{\pm}$ & $< 1.4 \times 10^{-7}$  & \cite{Aubert:2007mm} \\
$B^0 \to K^{*0} \mu^{\mp} e^{\pm}$ & $< 5.8\times 10^{-7}$  & \cite{Aubert:2006vb} \\
\hline
\end{tabular}
\caption{
Summary of experimental searches for lepton flavour violating decay modes (upper limits are at 90 \% CL).
\label{tab:LFV}
}
\end{table}

\subsection{Lepton number violating \texorpdfstring{$b$}{b} hadron decays}
\label{ssec:ssec_lnv}

In a similar fashion to lepton flavour, lepton number is not conserved in the SM.
It can be explicitly violated if neutrinos are of Majorana type, hence if they are their own antiparticles.
In fact, neutrino oscillations as a result of non-zero neutrino masses hint at the existence of
new degrees of freedom such as right-handed Majorana neutrinos which can provide an elegant
way to incorporate non-zero neutrino masses through the seesaw
mechanism~\cite{Minkowski:1977sc, Glashow:1979nm, Mohapatra:1979ia, Mohapatra:1980yp, GellMann:1980vs}.
The smallness of the neutrino masses could be driven by either the existence of super-heavy
Majorana neutrinos, which have $O(1)$ Yukawa couplings, or by the existence of Majorana
neutrinos with masses at the Fermi scale but with Yukawa couplings smaller than that of the electron.
Consequently, searches for lepton number violation (LNV) can provide insight into the nature of neutrinos.

\begin{figure}[htb]
\begin{center}
\includegraphics[width=0.45\linewidth]{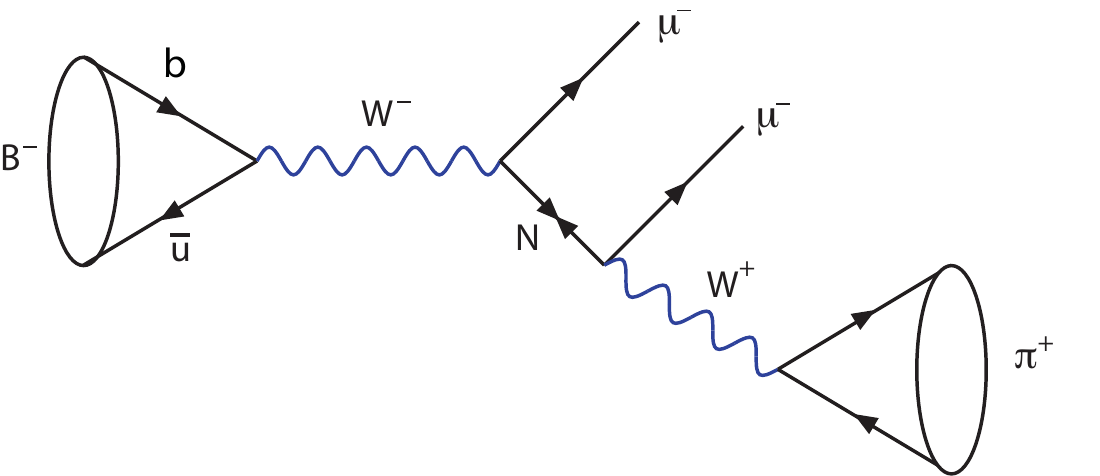}
\end{center}
\caption{Diagram for the lepton number violating decay $B^+ \to \pi^- \ell^+ \ell^+$ mediated by a Majorana neutrino in the s-channel.}
\label{fig:HNL}
\end{figure}

Lepton number violating processes can occur in  meson decays into two like-sign leptons
and another meson:
\begin{equation}
M_1^+  \to \ell^+  \ell^+ M^-_2
\label{eq:HNL}
\end{equation}
where the process occurs mostly via the diagram shown in Fig.~\ref{fig:HNL}
where $N$ is the Majorana neutrino.
For a heavy Majorana neutrino with a mass of a
few GeV, the $s$-channel process  is expected to give the dominant contribution.
If this decay is the result of the exchange of a
Majorana neutrino, then the reconstructed invariant mass
of the hadron $h$ with the opposite-sign lepton, $m_{h^+ \ell^-}$, can
be related to the Majorana neutrino mass \cite{Atre:2009rg, Han:2006ip}.
Note that it is possible for virtual Majorana neutrinos of any
mass to contribute to this decay.

Searches for lepton number violating modes have been performed in numerous experiments.
Table~\ref{tab:LNV} summarises the world best upper limits on lepton-number violating $B$ decays.
BaBar reports a search for the LNV decays
$B^+ \to  h^- \ell^+ \ell^+$ (where $h = K;\pi$) based on 471 million $B\overline{B}$
pairs~\cite{BABAR:2012aa}.
No significant signals are observed, and branching fraction upper
limits are determined in the range $[2, 11] \times 10^{-8}$ at the
90\% CL.

\begin{table}
\centering
\begin{tabular}{lcc}
\hline
Mode & BR upper limit &  Ref. \\
\hline
$B^+ \to \pi^- e^+ e^+$ & $<2.3 \times 10^{-8}$ at 90\% CL             & \cite{BABAR:2012aa} \\
$B^+ \to K^- e^+e^+$  &  $<3.0 \times 10^{-8}$ at 90\% CL             & \cite{BABAR:2012aa} \\
$B^+ \to \pi^- \mu^+ \mu^+$ &  $<4.0 \times 10^{-9} $ at 95\% CL  & \cite{Aaij:2014aba}\\
$B^+ \to K^- \mu^+ \mu^+$ &  $<4.4 \times 10^{-8}$ at 90\% CL     & \cite{Aaij:2011ex} \\
$B^+ \to D^- e^+e^+$ & $<2.6 \times 10^{-6} $ at 90\% CL             & \cite{Seon:2011ni}\\
$B^+ \to D^- \mu^+ e^+$ & $<1.8 \times 10^{-6}$ at 90\% CL         & \cite{Seon:2011ni} \\
$B^+ \to D^- \mu^+ \mu^+$ & $<6.9 \times 10^{-7}$ at 95\% CL     & \cite{Aaij:2012zr}\\
$B^+ \to D^{*-} \mu^+ \mu^+$ & $<2.4 \times 10^{-6}$ at 95\% CL & \cite{Aaij:2012zr}\\
$B^+ \to D_s^- \mu^+ \mu^+$ & $<5.8 \times 10^{-7}$ at 95\% CL  & \cite{Aaij:2012zr}\\
\hline
\end{tabular}
\caption{
Summary of experimental searches for lepton number violating decay modes.
\label{tab:LNV}
}
\end{table}

Since $b \to c$ decays are in general favoured over charmless
$B$ decays, it is interesting to extend the search for
lepton number violating processes to $B^+ \to X_c^- \ell^+ \ell^+$ decays, where $X_c^-$ is
any charmed hadron with opposite charge to the
leptons. Using a sample of 772 million $B\overline{B}$ pairs, Belle reports a measurement of the
$B^+ \to D^-\ell^+ \ell'^-$ decays~\cite{Seon:2011ni}, where $\ell, \ell' = e$ or $\mu$  in any combination.
There was no event observed in the signal region of any mode.
The results are summarised in Table~\ref{tab:LNV}  and are in the range $[1.8, 2.6] \times 10^{-6}$.

The LHCb experiment also searched for on-shell Majorana
neutrinos coupling to muons in the $B^- \to N  \mu^-$, $N \to \pi^+ \mu^-$ decay
channel~\cite{Aaij:2014aba}. The search was performed
as a function of the Majorana neutrino mass between 250 and $5000\,{\rm MeV}/c^2$
and for neutrino lifetimes up to $\sim1\,{\rm ns}$.
No signal is seen by LHCb and upper limits on the $B^- \to \pi^+ \mu^- \mu^-$ branching fraction
have been set. In Table~\ref{tab:LNV} the upper limit on the ${\rm BR}(B^+ \to \pi^- \mu^+ \mu^+)$ is reported for
a Majorana neutrino lifetime $\tau_{\rm N} < 1\,{\rm ps}$.

From the non-observation of these LNV rare meson decay modes one can determine
constraints on mixing parameters of the Majorana neutrinos with the active neutrinos
as a function of the heavy neutrino mass (see for example Ref.~\cite{Atre:2009rg}).

\section{Global analyses and determination of Wilson coefficients}
\label{sec:globalfits}

\noindent
Barring the possibility of feebly interacting light new particles (as discussed
in Sec.~\ref{sec:hidden}), new physics contributions to rare $b$ hadron decays
are described by the modification of the SM Wilson coefficients and generation
of Wilson coefficients that vanish in the SM. There are a  large number of
observables in leptonic, radiative, and semi-leptonic inclusive and exclusive
decays that depend in different ways on combinations of Wilson coefficients.
A \textit{global} analysis of the measurements is needed in order to determine the numerical values of the Wilson coefficients.
In the recent past, several such analyses have been performed.
\begin{itemize}
 \item
 Altmannshofer et al.\ \cite{Altmannshofer:2011gn,Altmannshofer:2012az,Altmannshofer:2013foa}
 considered global fits of $(C_7^{(\prime)})_s$ and $(C_{9,10}^{(\prime)})_s^\mu$,
 extended to include $(C_{9,10}^{(\prime)})_s^e$ in Ref.~\cite{Altmannshofer:2014rta},

  \item
 Bobeth et al.\ \cite{Bobeth:2011nj,Beaujean:2012uj,Beaujean:2013soa}
  considered global fits of the SM operator basis $(C_7)_s$ and $(C_{9,10})_s^\mu$ ,

  \item
  Descotes-Genon et al.\
 \cite{Descotes-Genon:2013wba,Descotes-Genon:2015uva}
  considered global fits of $(C_7^{(\prime)})_s$ and $(C_{9,10}^{(\prime)})_s^\mu$
 extended to include $(C_{9,10}^{(\prime)})_s^e$ in Ref.~\cite{Descotes-Genon:2015uva},

   \item
   Hurth et al.\
 \cite{Hurth:2013ssa}
  considered global fits of $(C_7)_s$ and $(C_{9,10})_s^\mu$ ,
 extended to include $(C_7')_s$, $(C_{9,10}')_s^\mu$ and
 $(C_{9,10}^{(\prime)})_s^e$ in Ref.~\cite{Hurth:2014vma,Hurth:2016fbr}.
\end{itemize}
The impact of scalar and pseudoscalar operators has been studied in
Refs.~\cite{Becirevic:2012fy,Beaujean:2015gba}.
In all of the above analysis, the four-quark operators $Q_{1}$--$Q_{6}$ were
assumed to be free from new physics. A global analysis of \textit{non-leptonic}
$B$ decays relaxing this assumption~\cite{Brod:2014bfa} was performed in Ref.~\cite{Bobeth:2014rra}.
The baryonic decay $\Lambda_b\to \Lambda\ell^+\ell^-$ has recently been
included in a global fit for the first time Ref.~\cite{Meinel:2016grj}.

\subsection{Global analyses of \texorpdfstring{$b\to s(\gamma,\mu^+\mu^-)$}{b->s(γ,μμ)} transitions}

\begin{figure}[tbp]
\includegraphics[width=0.33\textwidth]{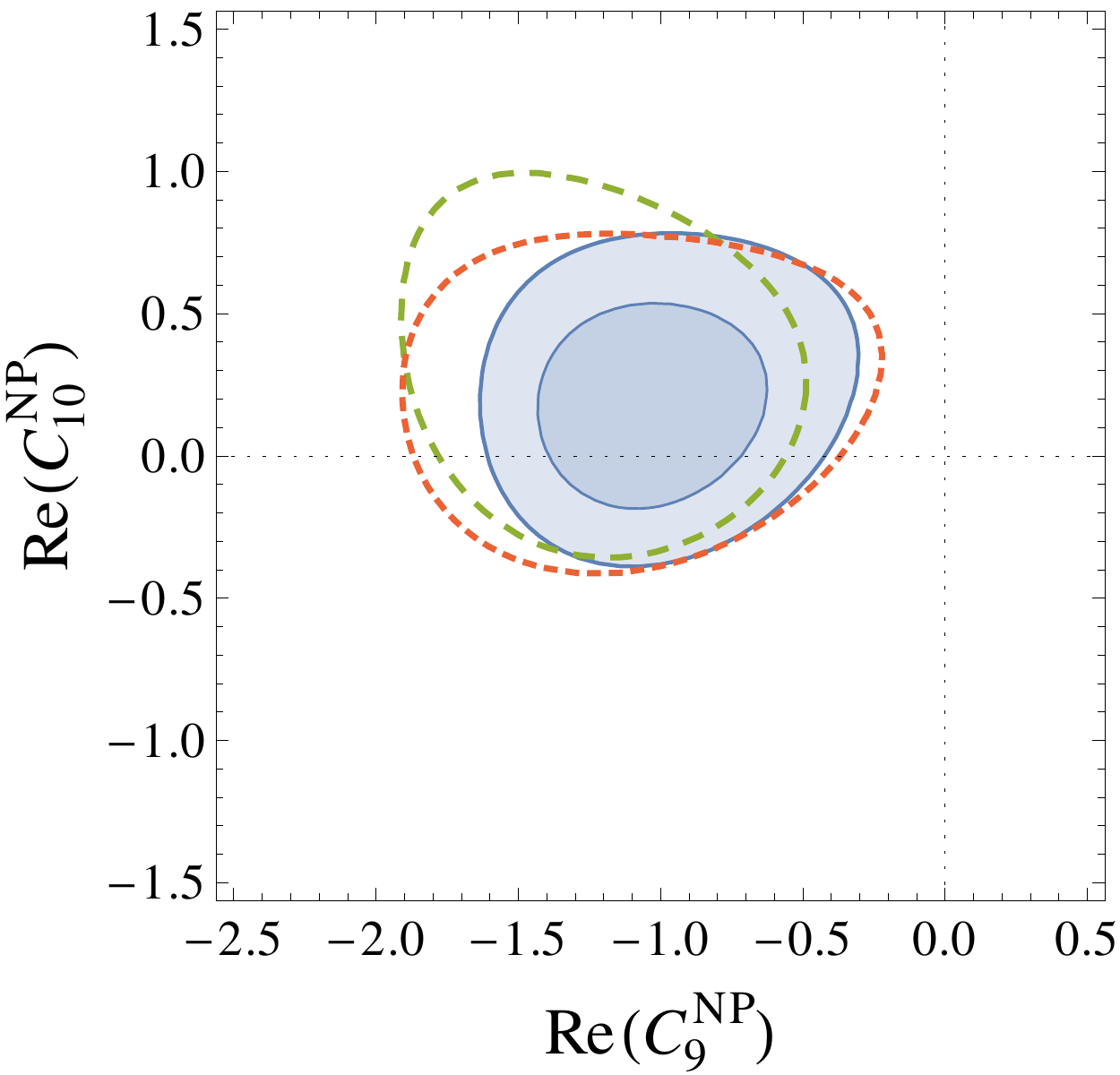}%
\includegraphics[width=0.31\textwidth]{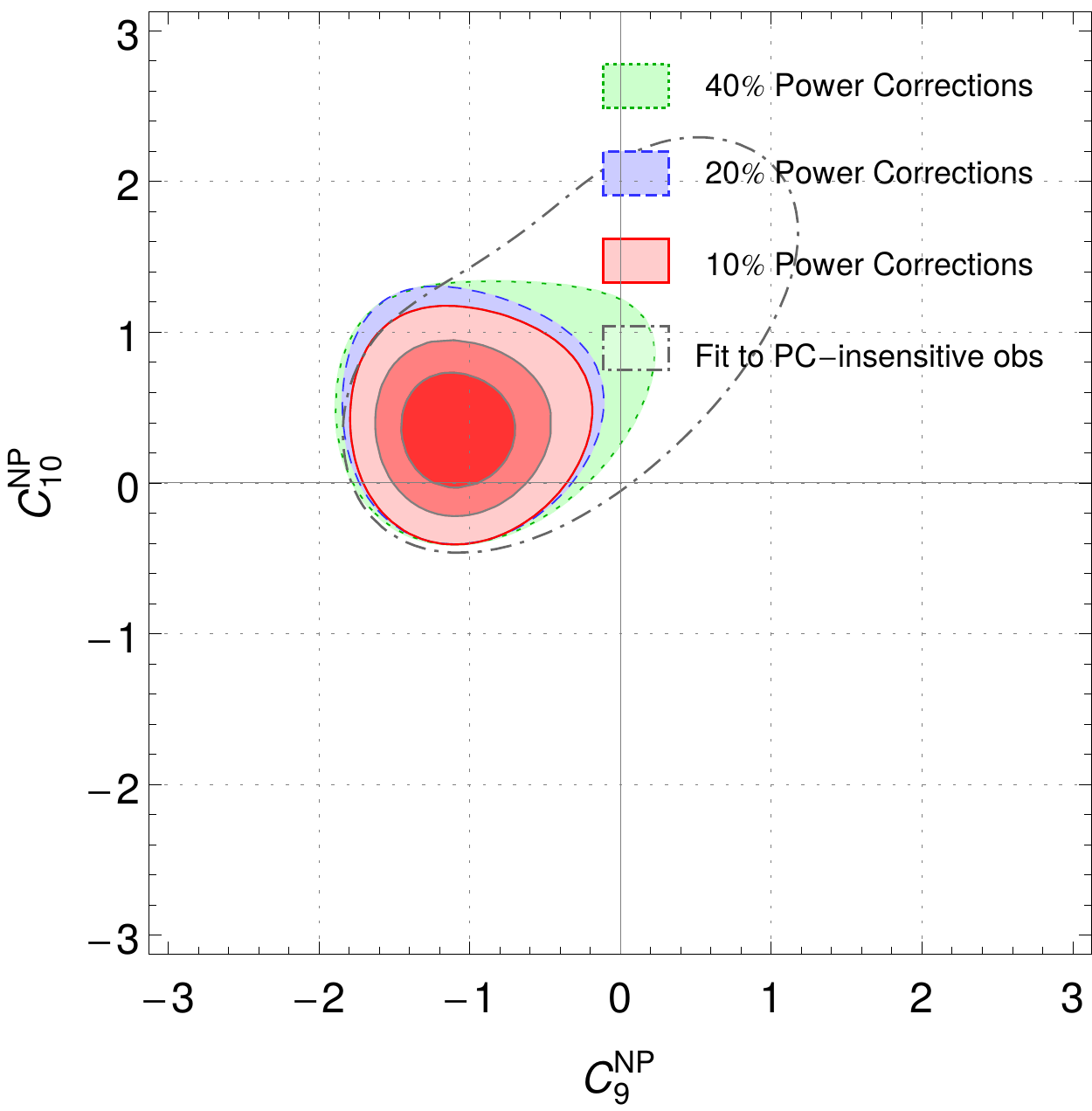}%
\raisebox{0.6cm}{\includegraphics[width=0.33\textwidth]{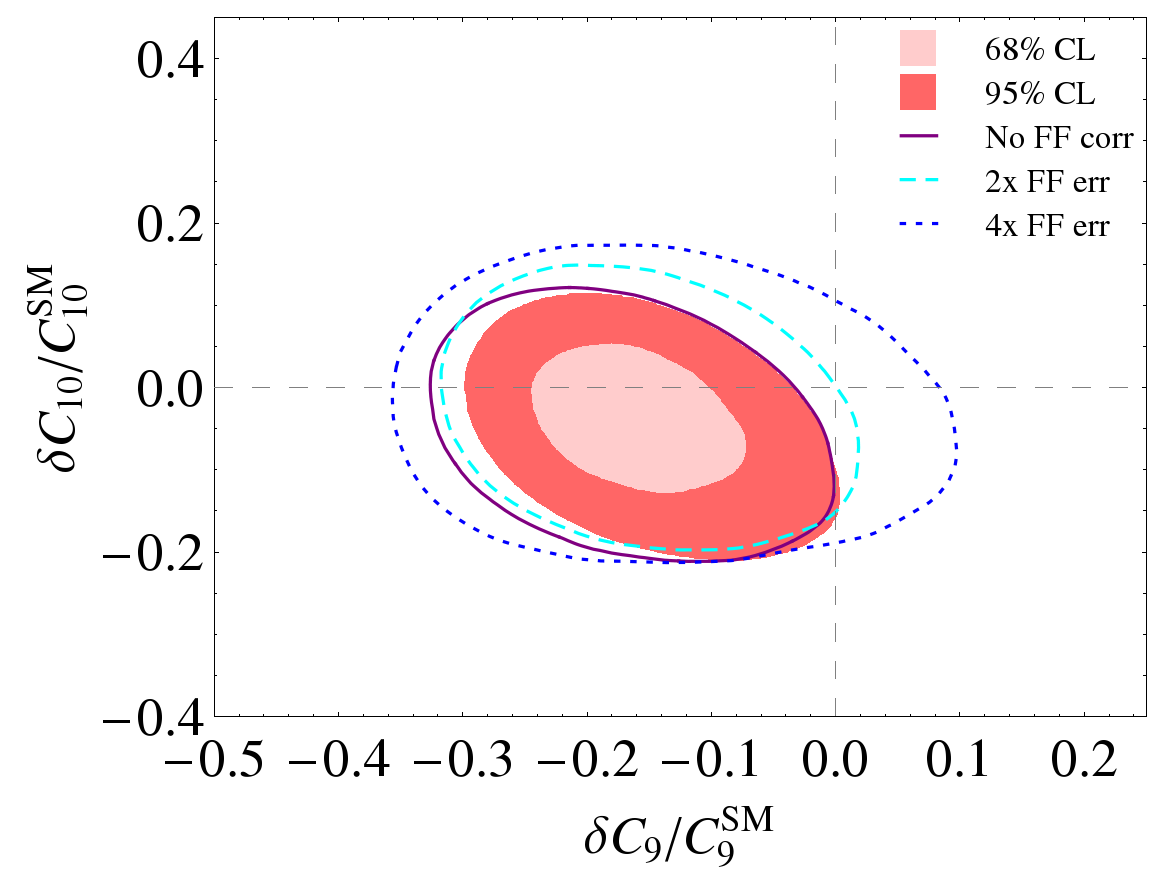}}%
\caption{Allowed regions in the plane of new physics contributions to $(C_9)_s^\mu$
and $(C_{10})_s^\mu$ from three different recent global fits of $b\to s(\gamma, \mu^+\mu^-)$
transitions (left: \cite{Altmannshofer:2014rta}, centre: \cite{Descotes-Genon:2015uva},
right: \cite{Hurth:2016fbr}).
In all three plots, the dashed lines refer to scenarios with more conservative assumptions
on the theoretical hadronic uncertainties.
Note that the axes on the rightmost plot are normalised to $C_{10}^\text{SM}\approx -4.2$
and $C_{9}^\text{SM}\approx 4.1$.}
\label{fig:globalfits}
\end{figure}

The most recent global analyses of new physics in $b\to s(\gamma,\mu^+\mu^-)$ transitions,
including the 3~fb$^{-1}$ $B\to K^*\mu^+\mu^-$ angular analysis by LHCb, are
Ref.~\cite{Altmannshofer:2014rta}, \cite{Descotes-Genon:2015uva} and
\cite{Hurth:2016fbr}. These analyses, while differing in the detailed treatment
of theoretical uncertainties and the inclusion of observables, agree on the
conclusion that
\begin{itemize}
 \item there is a tension with SM expectations driven both by
$B\to K^*\mu^+\mu^-$ angular observables and by several exclusive
$b\to s\mu^+\mu^-$ branching fraction measurements,
 \item the tension can be relieved by a new physics effect in $(C_9)_s^\mu$
 interfering destructively with the SM, with
 a good fit obtained also with a simultaneous contribution to $(C_{10})_s^\mu$,
 again interfering destructively with SM.
 The best-fit values are $[(C_9)_s^\mu]^\text{NP}\approx -1.1$ for new physics
 in $(C_9)_s^\mu$ only and
 $[(C_9)_s^\mu]^\text{NP}=-[(C_{10})_s^\mu]^\text{NP}\approx -0.5$ for a fit
 fixing the new physics contribution to $(C_9)_s^\mu$ and $(C_{10})_s^\mu$ to
 be equal with opposite sign (as is expected in models generating a
 four-fermion operator with left-handed quarks and leptons only).
 The best-fit regions in the plane $(C_9)_s^\mu$ vs. $(C_{10})_s^\mu$ are
 shown in Fig.~\ref{fig:globalfits} for the three of the global analyses~\cite{Altmannshofer:2014rta,Descotes-Genon:2015uva,Hurth:2016fbr}.
\end{itemize}
The fact that this tension mostly appears in an operator that couples
vectorially to leptons and involves left-handed quarks implies that it could
also be caused by an unexpectedly large non-factorisable hadronic effect
(see also \cite{Khodjamirian:2010vf,Lyon:2014hpa,Jager:2014rwa,Descotes-Genon:2014uoa}).
More precise measurements of the $q^2$ dependence of the deviation could allow
to disentangle a QCD effect from a possible new physics contribution
\cite{Altmannshofer:2015sma,Ciuchini:2015qxb}.
If the new physics contribution is not lepton flavour universal, measurements
of ratios of $b\to se^+e^-$ and $b\to s\mu^+\mu^-$ observables
\cite{Altmannshofer:2014rta,Hiller:2014ula} represent
theoretically clean ways to unambiguously identify it, since a possible hadronic
effect is photon-mediated and thus lepton flavour universal.

Other key results of the global $b\to s(\gamma,\mu^+\mu^-)$ analyses are:
\begin{itemize}
 \item The angular analysis of $B\to K^*\mu^+\mu^-$ and its overall good agreement
 with the SM has settled the
 question whether the Wilson coefficients $(C_{7,9,10})_s^{(\mu)}$ have the signs predicted in
 the SM\footnote{Note that a simultaneous sign flip of \textit{all} Wilson coefficients at the
 $b$ quark mass scale is not observable in $b$ hadron decays, as the observables  are always
 bilinear in the Wilson coefficients. We consider such a scenario physically implausible.
 } (cf. \cite{Haisch:2007ia} for a pre-LHC discussion);
 \item A non-standard phase in $(C_7)_s$ is still weakly constrained, given
 large theoretical uncertainties in the inclusive and exclusive
 $b\to s\gamma$ direct $C\!P$ asymmetries (cf. Sec.~\ref{sec:radiative});
 \item There is no evidence for right-handed FCNCs, i.e. non-zero values for
 the Wilson coefficients $(C_i')_s^{(\mu)}$;
 \item There is no evidence for non-standard $C\!P$ violation, i.e. non-zero imaginary
 parts in any of the Wilson coefficients.
\end{itemize}

\subsection{Testing lepton flavour universality in \texorpdfstring{$b\to s\ell^+\ell^-$}{b->sll} transitions}

While lepton flavour universality holds in the SM at the level of
fundamental interactions, up to the small Yukawa couplings and tiny neutrino masses, it could be
violated beyond the SM. Since the most precise and diverse constraints
on the $b\to s\ell^+\ell^-$ Wilson coefficients are obtained from processes
with muons in the final state, testing lepton flavour universality requires
improving the measurements of $b\to se^+e^-$ and $b\to s\tau^+\tau^-$ processes.
Since photon-mediated contributions are always lepton flavour universal,
the focus is on observables sensitive to semi-leptonic operators.

\subsubsection{\texorpdfstring{$\mu$}{μ}-\texorpdfstring{$e$}{e} universality}

Currently, the existing measurements of $b\to se^+e^-$ processes include
\begin{itemize}
 \item $B^+\to K^+e^+e^-$ branching fraction by LHCb as discussed in Sec.~\ref{sec:semileptonic:universality},
 \item $B\to X_se^+e^-$ branching fraction by BaBar \cite{Lees:2013nxa},
 \item $B\to K^*e^+e^-$ angular analysis at very low dilepton invariant mass by LHCb \cite{Aaij:2015dea}.
\end{itemize}
The latter are however dominated by the photon pole and thus provide little
sensitivity to (potentially lepton flavour non-universal) semi-leptonic
operators.
Recent global fits of the semi-leptonic operators $[C_{9,10}^{(\prime)}]_s^\mu$
vs. $[C_{9,10}^{(\prime)}]_s^e$ have been performed in
\cite{Altmannshofer:2014rta,Descotes-Genon:2015uva,Hurth:2016fbr}
(see also \cite{Ghosh:2014awa}). They agree on not finding evidence for
new physics in $b\to se^+e^-$, given the currently sizeable
experimental uncertainties. There is a tension with the SM prediction of
universality driven by the LHCb measurement of $R_K$
(see Sec.~\ref{sec:semileptonic:universality}) that could
be solved by new physics in either $b\to s\mu^+\mu^-$ or $b\to se^+e^-$
transitions~\cite{Hiller:2014yaa}.
The global fits to the $b\to s\mu^+\mu^-$ observables favour an effect in $b\to s\mu^+\mu^-$ rather than in $b\to s e^+e^-$ transitions.

\subsubsection{\texorpdfstring{$\mu$}{μ}-\texorpdfstring{$\tau$}{τ} universality}

Since no rare decay of the type $b\to s\tau^+\tau^-$ has been observed experimentally
yet, only limits can be set on the Wilson coefficients $(C_{9,10,S,P}^{(\prime)})_s^\tau$.
In Ref.~\cite{Bobeth:2011st}, it was shown that the Wilson coefficients
$(C_{9,10}^{(\prime)})_s^\tau$ can also
be constrained \textit{indirectly} from $b\to s\mu^+\mu^-$ processes
due to contributions analogous to
Fig.~\ref{fig:semileptonic:btoslldiagrams}(c) with $q\bar q$ replaced by
$\tau^+\tau^-$. These indirect bounds are comparable to the direct ones at present.
Models predicting large effects in $b\to s\tau^+\tau^-$ transitions with
\textit{left-handed} muons are constrained by $B\to K^{(*)}\nu\bar\nu$
processes as well due to $SU(2)_L$ symmetry \cite{Buras:2014fpa}.

\subsection{Testing the minimal flavour violation hypothesis using \texorpdfstring{$b \to d$}{b->d} transitions}

\begin{figure}[!htb]
\centering
\includegraphics[width=0.6\linewidth]{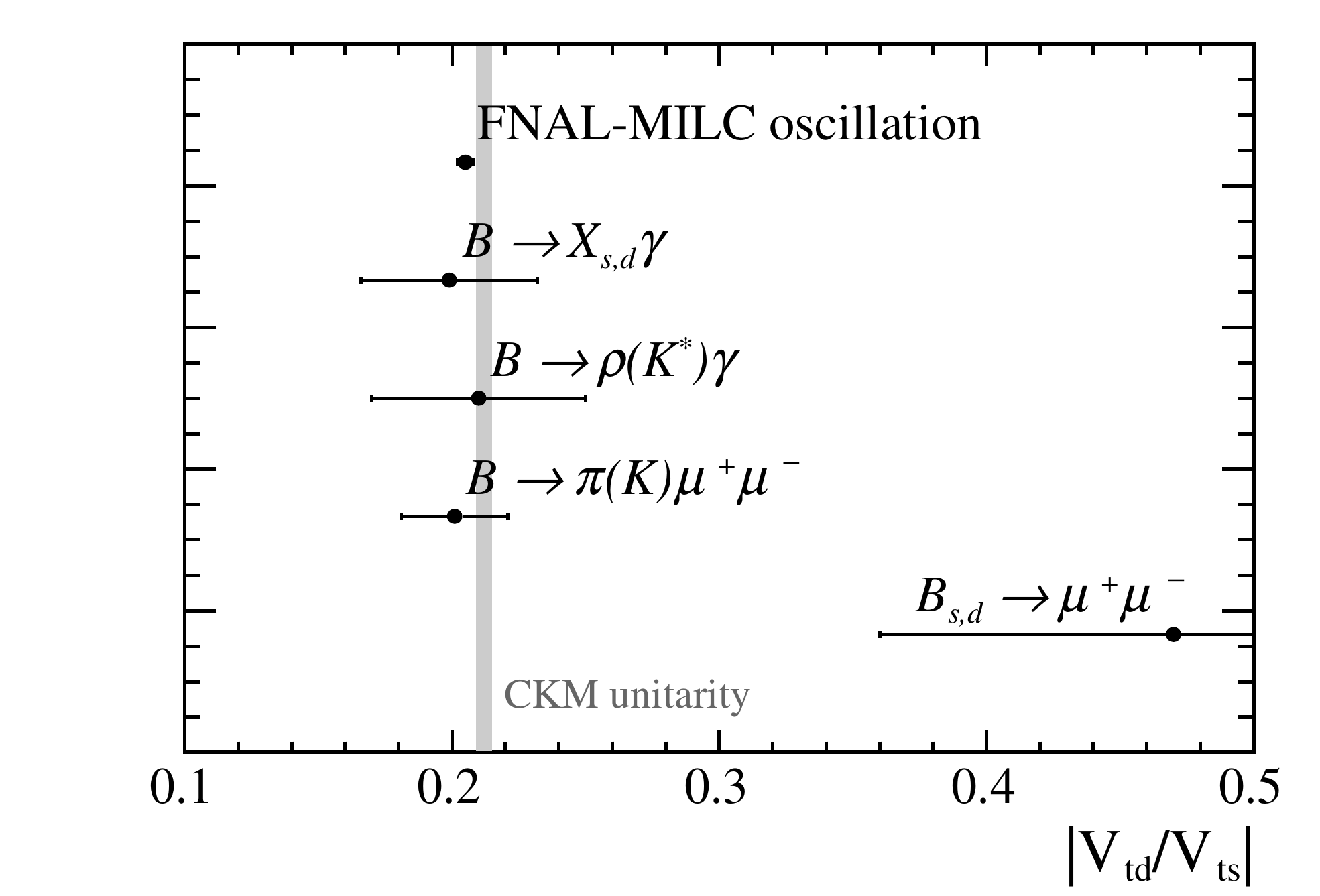}
\caption{
The ratio $|V_{td}/V_{ts}|$ determined using rare $b \to d$ and $b \to s$ processes.
The FNAL/MILC data point corresponds to the value of $|V_{td}/V_{ts}|$ obtained from $\Delta F = 2$ processes~\cite{Bazavov:2016nty}.
The SM expectation is given by the result of a global fit assuming CKM unitarity
\cite{Charles:2015gya}.
\label{fig:vtdvts}
}
\end{figure}

While \textit{global} analyses of new physics in $b\to d(\gamma,\ell^+\ell^-)$
transitions have not been performed so far, an indication of the agreement
with the SM can be given by comparing the extraction of the
ratio of CKM elements $|V_{td}/V_{ts}|^{2}$
from various rare $B$ decays discussed throughout the text to the extraction
from the oscillation frequencies of the neutral $B^0$ and $B_{s}$ systems.
These extractions should all agree, not only within the SM, but also in
all models beyond the SM satisfying the
 minimal flavour violation (MFV)
criterion \cite{Buras:2000dm,D'Ambrosio:2002ex} (cf. Sec.~\ref{sec:heffbsm}).
Figure~\ref{fig:vtdvts} summarises the measurements of $|V_{td}/V_{ts}|$ discussed in Sections~\ref{sec:leptonic}--\ref{sec:semileptonic}.
This figure includes constraints from the ratio of:
inclusive $\bar B \to X_s \gamma$ and $\bar B \to X_d \gamma$ decay rates measured by the B-factory experiments~\cite{delAmoSanchez:2010ae};
$B \to \rho \gamma$ to $B \to K^* \gamma$~\cite{Aubert:2008al};
$B^+ \to \pi^+\mu^+\mu^-$ to $B^+ \to K^+ \mu^+\mu^-$~\cite{Aaij:2015nea}
(see also \cite{Du:2015tda});
and $B^0 \to \mu^+\mu^-$ to $B_{s}\to \mu^+ \mu^-$.
The figure also includes the value of $|V_{td}/V_{ts}|$ expected from CKM unitarity and a recent determination of $|V_{td}/V_{ts}|$ from the oscillation frequencies of the neutral $B^0$ and $B_{s}$ systems by the FNAL/MILC lattice collaboration~\cite{Bazavov:2016nty}.
In general measurements are consistent with the SM (and the MFV hypothesis).
The largest deviation is currently seen in the ratio of branching fractions
of $B_{d,s}\to\mu^+\mu^-$ decays. As discussed in Sec.~\ref{sec:leptonic},
these modes allow the theoretically cleanest extraction of $|V_{td}/V_{ts}|$
among all rare $B$ decays, comparable in precision to the extraction
from meson oscillations. More precise experimental determinations of this ratio
are thus of utmost importance.

\section{Outlook}
\label{sec:outlook}

One of the most interesting puzzles in particle physics is that, on the one hand, new physics
is expected in the TeV energy range to solve the hierarchy problem and stabilise the Higgs mass; but
on the other hand, no sign of new physics has been detected through precision tests of the electroweak
theory or through flavour-changing (or $C\!P$-violating) processes in strange, charm or beauty hadron decays.

In the past decade, one of the major accomplishments of particle physics has been to gain an in-depth
understanding of the role of quark flavour. In this time frame, experimental measurements and
theoretical predictions have advanced tremendously allowing us to make very precise tests of the SM.
Rare  decays  of  $b$ hadrons  are playing  a  very  important  role  in  these tests.
The interplay of weak and Higgs interactions implies that FCNC  processes can occur only at higher orders in the electroweak
interactions and are strongly suppressed. This strong suppression makes FCNC
processes natural candidates to search for physics beyond the SM.
Indeed, while direct searches at the LHC are exploring particles with masses of up to few TeV,
FCNC processes (together with the searches for charged lepton flavour violation and low energy observables
like electric dipole moments) are probing mass scales of up to $10^3$\,TeV.
To avoid these constraints it is necessary to postulate a specific flavour structure for extensions of the SM.

A wealth of new experimental results have been produced by the ATLAS, CMS and LHCb collaborations during Run\,1 of the LHC.
This progress is expected to continue for the next two decades.
During Run\,2 of the LHC (2015-2018), the LHCb experiment expects to collect an additional 5\,fb$^{-1}$,
and ATLAS and CMS 100\,fb$^{-1}$ of integrated luminosity each.
This, together with an increased centre-of-mass energy of $\sqrt{s}= 13$\,TeV for the LHC $pp$ collisions,
will allow the experiments to collect datasets that are a factor of four (LHCb) and eight (ATLAS and CMS) larger than
those collected in Run\,1\footnote{This is valid only if the same trigger thresholds are kept by the experiments.}.
An additional gain is expected in Run\,2 for both for ATLAS and CMS due to the insertion of the Inner B-Layer (in ATLAS)  and the new pixel detector (in CMS).
This will improve the impact parameter resolution of the tracking systems of the experiments,
providing a better separation between displaced vertices and prompt background events.

On a longer term, an upgrade foreseen for the LHCb experiment in 2019--2020 will allow a
dataset of $50\,{\rm fb}^{-1}$ to be collected in about five years of operation~\cite{Bediaga:1443882}.
Major upgrades of the ATLAS and CMS detectors are also scheduled in 2023--2026.
The ultimate aim for ATLAS and CMS is to reach an integrated luminosity of about 3000\,fb$^{-1}$ by around 2035.
With a dataset corresponding to  3000\,fb$^{-1}$, ATLAS and CMS will be able to measure the branching fraction of the $B_s \to \mu^+ \mu^-$ decay with an accuracy
of about 10\% and the ratio of the branching fractions ${\rm BR}(B^0\to \mu^+\mu^-)/{\rm BR}(B_s \to \mu^+ \mu^-)$ with
an accuracy of about 20\%~\cite{CMS:2015iha}.
The increased yields will also allow to measure the $B_s \to \mu^+ \mu^-$ effective lifetime.
Discussions have started on a longer term upgrade of the LHCb experiment, with the aim of collecting a dataset of a few hundred fb$^{-1}$.

The huge datasets available at the LHC experiments will enable them to reach an unprecedented level of accuracy in the study of the branching fractions, $C\!P$-asymmetries and angular observables of rare $b \to s \ell^+\ell^-$ decays. This is particularly true for decay modes with dimuons in the final state where
an in-depth investigation of the anomaly observed in the angular distributions of the $B^0 \to K^{*0} \mu^+ \mu^-$
decay mode will become possible.
The LHCb experiment will also have a sizeable dataset of dielectron final states, allowing for precise comparisons of the rate and angular distribution of $b \to s \mu^+\mu^-$ and $b\to s e^+ e^-$ decays.

Apart from the LHC,
inclusive measurements, decay channels with neutrinos, $\tau$ leptons and $\pi^0$s in the final state will be mostly improved by Belle II~\cite{Aushev:2010bq}.
Belle II is expected to start data taking with its full detector in 2018 and aims to collect an integrated luminosity of  $50\,{\rm ab}^{-1}$ by 2024.
This will provide a dataset that is about
a factor of 50 times larger than that collected by BaBar and Belle together.
The Belle II experiment is expected to measure the branching fraction of the $B \to K^{(*)} \nu\bar{\nu}$ decay to a 30\% precision on the SM branching fraction.
Belle II will also be able  to measure the ${\rm BR}(\bar B \to X_s \gamma)$ to a relative uncertainty of better than 
$\sim 6\%$ (for $E_{\gamma} > $ 1.7 GeV), matching the precision of theoretical predictions,
and reach a precision on $R_{\rm K^{(*)}}$ of $\sim 2\%$.

Beyond LHCb and Belle II, the experimental rare $b$ decay programme would
benefit greatly from a future circular electron-positron collider collecting
several inverse attobarns of integrated luminosity at the $Z$ peak
\cite{d'Enterria:2016yqx}. Combined with a cleaner environment compared
to a hadron collider and a larger boost for $b$ hadrons compared to Belle II, this
would allow to improve measurements of decays with neutrinos in
the final state, such as $b\to q\tau^+\tau^-$ processes or the leptonic
charged-current decays $B^+\to\ell^+\nu_\ell$.

The challenge for the theory in the coming years will be to keep pace with the increased experimental accuracy.
While the uncertainties due to unknown higher-order \textit{perturbative} effects
have been reduced to the sub-percent level\footnote{An exception is the
Wilson coefficient $C_9$, where NLO electroweak corrections are not fully known
yet.} in recent years, the main challenge will be to reduce
hadronic uncertainties.
Progress is expected from LQCD that will determine decay constants and
form factors to higher precision. But also other tools like SCET, QCDF, or LCSR
will play an important role to reduce uncertainties due to hadronic effects like
the long-distance charm loop effect in $b \to s \ell^+ \ell^-$ decays.

In the coming decade, rare $b$ decays will continue to play a central role in the understanding of the underlying patterns of SM physics and
in setting up new directions in model building for non-SM contributions. The exploitation of the full datasets of the LHC experiments and Belle II
is a fantastic opportunity but also a big challenge for both the theoretical and experimental communities.

\section*{Acknowledgements} 

\noindent The authors would like to thank W.~Altmannshofer, C.~Bobeth, A.~Buras, T.~Gershon, C.~Langenbruch, and V.~Vagnoni for helpful discussion and valuable feedback on the manuscript. 
T.\,B. and G.\,L. would like to thank the members of the LHCb collaboration for their productive collaboration that led to some of the results discussed in this article. 
This work is supported in part by the Royal Society (T.\,B.), the INFN (G.\,L.),
and by the DFG cluster of excellence ``Origin and Structure of the Universe''
(D.\,S.).

\clearpage

\bibliography{bibliography}
\end{document}